\documentclass{report}

\usepackage[textsize=tiny]{todonotes} 

\usepackage{graphicx}
\usepackage{dirtytalk}
\usepackage{url}
\usepackage{hyperref}
\usepackage{amsthm}
\usepackage{amssymb}
\usepackage{authblk}

\theoremstyle{definition}
\newtheorem{definition}{Definition}

\theoremstyle{definition}
\newtheorem{example}{Example}
\newcommand{\chapterauthor}[1]{%
  {\parindent0pt\vspace*{-25pt}%
  \linespread{1.1}\large\scshape#1%
  \par\nobreak\vspace*{35pt}}
  
}
\graphicspath{ {./images/} }

\begin{document}
\begin{titlepage}

\newcommand{\HRule}{\rule{\linewidth}{0.5mm}} 

\center 
 

\includegraphics[scale=.3]{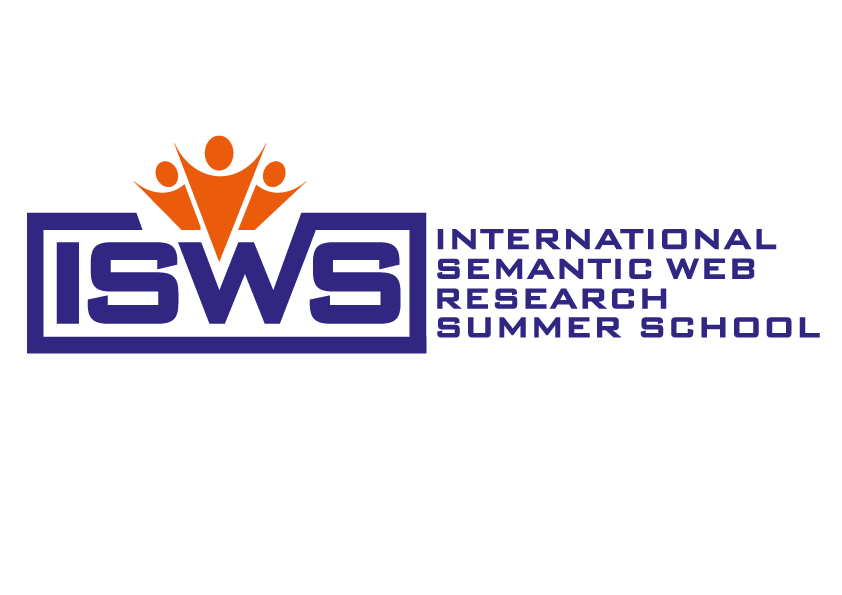}\\[1cm] 


\HRule \\[0.4cm]
{ \huge \bfseries Linked Open Data Validity}\\[0.2cm] A Technical Report from ISWS 2018\\[0.4cm]
\HRule \\[1.5cm]
 




{\large \today}\\[1cm] 
\large{Bertinoro, Italy}
\vfill 

\end{titlepage}

%
%
%
%
%

\chapter*{Authors}
\textbf{Main Editors}\\
Mehwish Alam, Semantic Technology Lab, ISTC-CNR, Rome, Italy\\
Russa Biswas, FIZ, Karlsruhe Institute of Technology, AIFB\\



\noindent \textbf{Supervisors}\\
Claudia d'Amato, University of Bari, Italy\\
Michael Cochez, Fraunhofer FIT, Germany\\
John Domingue, KMi, Open University and 
President of STI International, UK\\
Marieke van Erp, 
DHLab, KNAW Humanities Cluster, Netherlands\\
Aldo Gangemi, University of Bologna and STLab, ISTC-CNR, Rome, Italy\\
Valentina Presutti, Semantic Technology Lab, ISTC-CNR, Rome, Italy\\
Sebastian Rudolph, TU Dresden, Germany\\
Harald Sack, FIZ, Karlsruhe Institute of Technology, AIFB\\
Ruben Verborgh, IDLab Ghent University/IMEC, Belgium\\
Maria-Esther Vidal, Leibniz Information Centre For Science and Technology University Library, Germany, and Universidad Simon Bolivar, Venezuela\\

\noindent \textbf{Students}\\
Tayeb Abderrahmani Ghorfi,	IRSTEA	Catherine Roussey (IRSTEA)\\
Esha	Agrawal,	University Of Koblenz\\
Omar	Alqawasmeh,	Jean Monnet University - University of Lyon, Laboratoire Hubert Curien\\
Amina	ANNANE,	University of Montpellier France\\	
Amr	Azzam,	WU\\
Andrew	Berezovskyi,	KTH Royal Institute of Technology\\
Russa	Biswas,	FIZ, Karlsruhe Institute of Technology, AIFB\\
Mathias	Bonduel,	KU Leuven (Technology Campus Ghent)\\	
Quentin	Brabant,	Université de Lorraine, LORIA\\
Cristina-Iulia	Bucur,	Vrije Univesiteit Amsterdam, The Netherlands\\
Elena	Camossi,	Science and Technology Organization, Centre for Maritime Research and Experimentation\\	
Valentina Anita	Carriero,	ISTC-CNR (STLab)\\
Shruthi	Chari,	Rensselaer Polytechnic Institute\\	
David	Chaves Fraga,	Ontology Engineering Group - Universidad Politécnica de Madrid\\
Fiorela	Ciroku,	University of Koblenz-Landau\\
Vincenzo	Cutrona,	University of Milano-Bicocca\\	
Rahma	DANDAN,	Paris University 13\\
Pedro	del Pozo Jiménez,	Ontology Engineering Group, Universidad Politécnica de Madrid\\
Danilo	Dessì,	University of Cagliari\\
Valerio	Di Carlo,	BUP Solutions\\		
Ahmed El Amine	DJEBRI,	WIMMICS/Inria\\
Faiq Miftakhul	Falakh\\
Alba	Fernández Izquierdo,	Ontology Engineering Group (UPM)\\
Giuseppe	Futia,	Nexa Center for Internet \& Society (Politecnico di Torino)\\
Simone	Gasperoni,	Whitehall Reply Srl, Reply Spa\\
Arnaud	GRALL,	LS2N - University Of Nantes, GFI Informatique\\
Lars	Heling,	Karlsruhe Institute of Technology (KIT)\\
Noura	HERRADI,	Conservatoire National des Arts et Métiers (CNAM) Paris\\	
Subhi	Issa,	Conservatoire National des Arts et Metiers, CNAM-CEDRIC\\	
Samaneh	Jozashoori,	L3S Research Center (Leibniz Universitaet Hannover)\\	
Nyoman	Juniarta,	Université de Lorraine, CNRS, Inria, LORIA\\	
Lucie-Aimée	Kaffee,	University of Southampton\\	
Ilkcan	Keles,	Aalborg University\\
Prashant	Khare	Knowledge Media Institute, The Open University, UK\\
Viktor	Kovtun,	Leibniz University Hannover, L3S Research Center\\		
Valentina	Leone,	CIRSFID (University of Bologna), Computer Science Department (University of Turin)\\
Siying	LI,	Sorbonne universities, Université de technologie de Compiègne\\
Sven	Lieber,	Ghent University\\
Pasquale	Lisena,	EURECOM	Raphaël Troncy - EURECOM\\	
Tatiana	Makhalova,	INRIA Nancy - Grand Est (France), NRU HSE (Russia)\\
Ludovica	Marinucci,	ISTC-CNR\\		
Thomas	Minier,	University of Nantes\\
Benjamin	MOREAU,	LS2N Nantes, OpenDataSoft Nantes\\
Alberto	Moya Loustaunau,	University of Chile\\	
Durgesh	Nandini,	University of Trento, Italy\\	
Sylwia	Ozdowska,	SimSoft Industry\\		
Amanda	Pacini de Moura,	Solidaridad Network	Violaine Laurens, at Solidaridad Network\\	
Swati	Padhee,	Kno.e.sis Research Center, Wright State University, Dayton, Ohio, USA\\ 
Guillermo	Palma,	L3S Research Center\\
Pierre-Henri,	Paris	Conservatoire National des Arts et Métiers (CNAM)\\
Roberto	Reda,	University of Bologna\\
Ettore	Rizza,	Université Libre de Bruxelles\\	
Henry	Rosales-Méndez,	University of Chile\\		
Luca	Sciullo,	Università di Bologna\\	
Humasak	Simanjuntak,	Organisations, Information and Knowledge Research Group, Department of Computer Science, The University of Sheffield\\	
Carlo	Stomeo,	Alma Mater Studiorum, CNR\\
Thiviyan	Thanapalasingam,	Knowledge Media Institute, The Open University\\	
Tabea Tietz, FIZ, Karlsruhe Institute of Technology, AIFB \\
Dalia	Varanka,	U.S. Geological Survey\\
Michael	Wolowyk,	SpringerNature\\
Maximilian	Zocholl,	CMRE
\chapter*{Abstract}
Linked Open Data (LOD) is the publicly available RDF data in the Web. Each LOD entity is identified by a URI and accessible via HTTP. LOD encodes global-scale knowledge potentially available to any human as well as artificial intelligence that may want to benefit from it as background knowledge for supporting their tasks. 
LOD has emerged as the backbone of applications in diverse fields such as Natural Language Processing, Information Retrieval, Computer Vision, Speech Recognition, and many more. Nevertheless, regardless of the specific tasks that LOD-based tools aim to address, the reuse of such knowledge may be challenging for diverse reasons, e.g. semantic heterogeneity, provenance, and data quality. As aptly stated by Heath et al. \say{Linked Data might be outdated, imprecise, or simply wrong}: there arouses a necessity to investigate the problem of \emph{linked data validity}. This work reports a collaborative effort performed by nine teams of students, guided by an equal number of senior researchers, attending the International Semantic Web Research School (ISWS 2018) towards addressing such investigation from different perspectives coupled with different approaches to tackle the issue. 


\tableofcontents
\listoffigures
\listoftables

\chapter{Introduction}

In computer science, data validation is the process of ensuring data have undergone data cleansing to ensure they have data quality, that is, that they are both correct and useful. 
To this end, usually so-called validation rules or constraints are applied to check for correctness, meaningfulness, as well as for data security.
Linked Open Data are considered as interlinked, structured, and publicly available datasets encoded and accessible via W3C standard protocols. 
Fine grained access is enabled by utilizing IRIs (International Resource Identifiers) as universal address schema for each single data item.
Linked Open Data can be retrieved and manipulated via HTTP (Hypertext Transfer Protocol) using standard Web access (i.e. port 80).
Furthermore, Linked Open Data is encoded via RDF (Resource Description Framework) that structures data values in terms of simple triples (subject, property, object) thereby enabling the implementation of knowledge graphs by the interlinking of data items within a local repository, but also and in particular with external Linked Open Data sources.
To exploit the real potential of Linked Open Data, the SPARQL query language enables sophisticated federated queries among them.

Linked Open Data is based on the idea to realize the large-scale implementation of a lightweight Semantic Web. 
Due to its simplistic design principles, it has been possible to easily transfer large existing data repositories into a RDF representation. Furthermore, Linked Open Data has been created via automated analysis of natural language texts or other unstructured data. 
Thereby giving way to the introduction of errors, insufficencies, inaccuracies, ambiguities, misjudgements, etc.
Moreover, Linked Open Data has also been created directly from user inputs enabling further potential error including different levels of trustworthiness, reliability, and accuracy.
The very same holds for the introduction of interlinkings among different Linked Open Datasets.

The semantic backend of Linked Open Data is ensured via ontologies providing a formal, machine understandable definition of properties, classes, their relationship among each other including potential constraints, as well as axiomatic rules.
For Linked Open Data, those ontologies are often made available in terms of RDF vocabularies providing canonical terms to be used to name properties and classes.
The formal definition of the ontology based on description logics mostly remains hidden to the end user but is accessible for automated evaluation and validation.
Ontologies may be manually defined or also automatically created via knowledge mining techniques.
However, both possible approaches might again lead to logical or structural errors and other insufficiencies that prevent Linked Open Data to make use of its full potential.

The potential of Linked Open Data lies in its ability for large scale data integration accompanied by fully automated machine understanding.
Yet, this vision fails if data quality in terms of Linked Data Validity cannot be guaranteed.

This paper has the goal to shade light on different aspects of Linked Data Validity. It is a collection of nine differently focused contributions provided by students of the International Semantic Web Research Summer School (ISWS 2018) in Bertinoro, Italy.
Overall, five different approaches have been taken into account, which will be outlined in the subsequent Parts:

\begin{itemize}
    \item Part \ref{part1}: \textbf{Contextual Linked Data Validity}. First, the Natural Language Processing perspective is taken into account with a special emphasis on context derived from text (cf. Chapter \ref{sec:42s}), while subsequently contextual dimensions for assessing LOD validity are defined and implemented as  SPARQL templates  to assess existing Knowledge Graphs (cf. Chapter \ref{sec:ravenclaw}).
    \item Part \ref{part2}: \textbf{Data Quality Dimensions for Linked Data Validity}. Here, the different dimensions of data quality are taken into account to propose an approach for the general improvement of Linked Data validity (cf. Chapter \ref{sec:gryffinder}).
    \item Part \ref{part3}: \textbf{Embedding Based Approaches for Linked Data Validity.} First, a generalized framework for linked data validity based on knowledge graph embeddings is discussed (cf. Chapter \ref{sec:hobbits}), followed by its application to the important use case of validating LOD against Common Sense Knowledge (Chapter \ref{sec:deloreans}.
    \item Part \ref{part4}: \textbf{Logic-Based Approaches for Linked Data Validity.} Here, the application of description logics-based approaches to ensure linked data validity is described such as learning logical constraints (cf. Chapter \ref{sec:mordor}, and extending SHACL with restrictions (cf. Chapter \ref{sec:dragons}.
    \item Part \ref{part5}: \textbf{Distributed Approaches for Linked Data Validity.} For sake of efficiency the implementation of Linked Data Validity requires a distributed approach. First, a combination of BLockchains and Linked Data is introduced applied to the usecase of validating personal data (cf. Chapter \ref{sec:hufflepuff}. Furthermore, an approach to tackle Linked Data incompleteness is presented (cf. Chapter \ref{sec:jedis}.
\end{itemize}

\part{Contextual Linked Data Validity}
\label{part1}

\chapter{Finding validity in the space between and across text and structured data}
\label{sec:42s}
\chapterauthor{Amina Annane, Amr Azzam, Ilkcan Keles, Ludovica Marinucci, Amanda Pacini de Moura, Omar Qawasmeh, Roberto Reda, Tabea Tietz, Marieke van Erp}

\hfill

\noindent Research questions:
\begin{itemize}
    \item When you go from text to structured data how do assess validity on a piece of information? 
    \item How do you cope with imperfect systems that extract information from text to structured formats?
    \item How do you deal with contradicting or incomplete information? 
    \item How to deal with fluid definitions of concepts? For example the concept of an Event? This is different across many different domains (and LOD datasets) but may be expressed through the same class (for example (sem:Event). 
\end{itemize}

\begin{definition}[NLP Perspective]
Whenever an entity is extracted from a text and refers to an entity in a trusted Linked Data dataset and the entity’s properties, either extracted from text, or provided in the Linked Data resource, are aligned, then we assess the data element as valid.
\end{definition}

Even today, most of the content on the Web is available only in unstructured format, and in natural language text in particular. Also, as large volumes of non-electronic textual documents, such as books and manuscripts in libraries and archives, are being digitised, undergoing optical character recognition (OCR) and made available online, we are faced with a huge potential of unstructured data that could feed the growth of the Linked Data Cloud\footnote{Linked Open Data Cloud. \url{http://lod-cloud.net/}}

However, to actually integrate this content into the Web of Data, we need effective and efficient techniques to extract and capture the relevant data \cite{mccallum2005information}. Natural Language Processing (NLP) encompasses a variety of computational techniques for the automatic analysis and representation of human language. As such, NLP can arguably be used to produce structured datasets from unstructured textual documents, which in turn could be used to enrich, compare and/or match with existing Linked Data sets.

This raises two main issues for data validity: {\bf textual data validity}, which refers to the validity of data extracted from texts, and {\bf Linked Data validity}, which concerns the validity of structured datasets. We propose that structured data extracted from text through NLP is a fruitful approach to address both issues, depending on the case at hand: structured data from reliable sources could be used to validate data extracted with NLP, and reliable textual sources could be processed with NLP techniques to be used as a reference knowledge base to validate Linked Data sets. This leads us to our definition of Linked Data validity from an NLP perspective: whenever an entity is extracted from a text and refers to an entity in a trusted Linked Data dataset and the entity’s properties, either extracted from text, or provided in the Linked Data resource, are aligned, then we assess the data element as valid. Trust in this sense refers to metadata quality (e.g. precision and recall) as well as intrinsic data qualities \cite{ceolin2015linking}. 

In order to demonstrate this, we have performed initial processing and analysis on a corpus of Italian travel writings by native English speakers\footnote{\url{https://sites.google.com/view/travelwritingsonitaly/}} to extract data on locations, and then matched the extracted data with the two structured open data sets on geographic locations. To extract the textual data, we applied an NLP technique called Named Entity Recognition (NER), which identifies and extracts all mentions of named entities (nouns or noun phrases “serving as a name for something or someone” \cite{marrero2013named}) and categorizes them according to their types.

The corpus was selected due to four main factors, all which add to the chances of occurrences of contradicting and/or competing data, and thus to interesting cases for assessing data validity. First, the corpus spans a period of 75 years (1867 to 1932), so it potentially involves changes to names and attributes of locations over time. Second, it includes texts from several different authors, so even though it is one single corpus, it covers several different sources of information. Third, travel writings are a literary genre that, while not necessarily fictional, has no commitment with providing exclusively factual information. And fourth and last, all authors are foreign travelers, and so potentially unknowledgeable on the regions which they are describing.

We hope that our analysis and approach may define not only provide a definition what Linked Data validity may look like from a NLP perspective, but also show why this is an issue worth investigating further and which could be the main points of interest for future work.

\section{Related Work}

\cite{vasardani2013locating} offers a survey of the existing literature contributing to locating place names. The authors focus on the positional uncertainties and extent of vagueness frequently associated with the place names and with the differences between common users perception and the representation of places in gazetteers. In our work, we attempt to address the problem of uncertainty (or validity) of place names extracted from textual documents by exploiting existing knowledge resources -- structured Linked Open Data resources.

\cite{ceolin2011estimating} aims to address the uncertainty of categorical Web data by means of the Beta-Binomial, Dirichlet-Multinomial and Dirichlet Process models. The authors mainly focused on two validity issues: (i) {\bf multi-authoring} nature of the Web data, and (ii) {\bf the time variability}. Our work addresses the same Web data validity issues. However, in our approach, we propose to use existing structured linked datasets (i.e., GeoNames\footnote{GeoNames \url{http://www.geonames.org/}} and DBPedia\footnote{DBpedia \url{https://wiki.dbpedia.org/}}) to validate the information --place names-- extracted from textual documents. 

In \cite{shen2012linden}, a framework called LINDEN is presented to link named entities extracted from textual documents using a knowledge base, called YAGO, an open-domain ontology combining Wikipedia and WordNet \cite{suchanek2008yago}. To link a given pair of textual named entities (i.e., entities extracted from text), the authors proposed to identify equivalent entities in YAGO, then to derive a link between the textual named entities according to the link between the YAGO entities when it exists. 
Linking textual named entities to existing Web knowledge resources is a common task between our approach and that presented in \cite{shen2012linden}. However, \cite{shen2012linden} focuses on linking textual named entities, while our work focuses on validating textual named entities. Moreover, in  \cite{shen2012linden}, the authors exploited one knowledge base (i.e., YAGO), while in our work, we used two knowledge bases (i.e., GeoNames and DBPedia).

\cite{van2015georeferencing} propose an automatic approach for georeferencing of textual localities identified in a database of animal specimens, using GeoNames, Google Maps and the Global Biodiversity Information Facility. However, our approach takes a specific domain raw text as an input.  Our goal is not to georeference, but to validate the identification of these locations using GeoNames and DBpedia.

\cite{grover2010use} reports on the the use of  Edinburgh geoparser for georeferencing digitized historical collections, in particular the paper describes the work that was undertaken to configure the geoparser for the collections. The validity of data extracted is done by consulting lists of large places derived from GeoNames and Wikipedia and decisions are made based on a ranking system.  However, the authors don't make any assumptions about whether the data in GeoNames or the sources from which they extract information is valid or not.

\section{Resources}

The structured data can be in the form of an RDF dataset such as DBpedia and GeoNames and the unstructured data can be in any form of natural language text. We have chosen to work with a corpus of historical writings regarding travel itineraries named as “Two days we have passed with the ancients… Visions of Italy between XIX and XX century”\footnote{Italian Travel Writings Corpus \url{https://sites.google.com/view/travelwritingsonitaly/}}. We propose that this dataset provides rich use cases for addressing the textual data validity defined in Introduction section for 4 reasons:
\begin{itemize}
    \item It contains 30 books that correspond to the accounts written by travelers who are native English speakers traveling in Italy. 
    \item The corpus consists of the accounts of travelers who have visited Italy within the period of 1867 and 1932. These writings share a common genre, namely "travel writing". Therefore, we expect to extract location entities that are valid during the time of the travelling. However, given that the corpus covers a span of 75 years, it potentially includes cases of contradicting information due to various updates on geographical entities. 
    \item The corpus might also contain missing or invalid information due to the fact that the travelers included in the dataset are not Italian natives, and therefore we cannot assume that they are experts on the places they visited. 
    \item  The corpus also contains pieces of non-factual data, such as the traveler’s opinions and impressions. 
\end{itemize}

Since the dataset we select corresponds to the geographical data, we selected structured data sources that deal with the geographical data. In this project, we utilize GeoNames and  DBpedia. GeoNames is a database of geographical names that contains more than 10,000,000 entities. The project is initiated by the geographical information retrieval researchers and the core database is provided by official government sources and the users are able to update and improve the database by manually editing contained information. Ambassadors from all continents contribute to the GeoNames dataset with their specific expertise. Thus, we assume that the data included in GeoNames is of sufficient quality. In addition, we select DBpedia as a reliable structured database since it is based on Wikipedia, that provides the volunteers with methods to enter new information and to update inconsistent or wrong information. Therefore, we assume that it is a reliable source of information regarding the geographical entities. The current version of DBpedia contains around 735,000 places. Information in DBpedia is not updated live, but around twice a year, thus, it is not sensitive for live information, e.g. an earthquake in a certain location or a sudden political conflict between states. However, since working with historical data and not with live events, we propose that it is valid to include geographical information from DBpedia. 

\section{Proposed Approach}

As mentioned in the Introduction section, NLP can be utilized to assess two different issues of validity, textual data validity and Linked Data validity. 

\paragraph{\textbf{\textit{Textual data validity}}} refers to the validity of the information that is extracted from documents of a given corpus. In our work, we use the named entities obtained by the NLP pipeline to achieve this goal. Our proposed method consists of 5 steps:
\begin{itemize}
    \item Sentence Tokenization: This corresponds to determining sentences from the input corpus. 
    \item Word Tokenization: This corresponds to the determining words within each sentence identified in the sentence tokenization step.
    \item PoS Tagging: This step annonates the tokenized sentences with part of speech (PoS) tags.  
    \item Named Entity Recognition (NER): This step identifies different types of entities employing the output of PoS tagging. In the NLP literature, the recognized entities can either belong to one class (named entity) or a set of classes (place, organization, location). For the textual data validity problem, the choice of a single class or a set of classes depends on the use case.
    \item Named Entity Linking (NEL): This step links the named entities obtained by the previous step to the structured datasets. In our method, this corresponds to linking entities to the linked open data sources. Since the underlying assumption is that the structured datasets are reliable, we can conclude that the entities that have been linked are valid entities.
\end{itemize}

\paragraph{Example 1.} Consider the sentence “For though all over Italy traces of the miracle are apparent, Florence was its very home and still can point to the greatest number of its achievements.”. The outputs obtained at the end of the steps are provided below.
\begin{itemize}
    \item Word Tokenization: {For, though, all, over, Italy, traces, of, the, miracle, are, apparent, Florence, was, its, very, home, and, still, can, point, to, the, greatest, number, of, its, achievements}
    \item PoS Tagging: {(For, IN), (though, IN), (all, DT), (over, IN), (Italy, NNP), (traces, NNS), (of, IN), (the, DT), (miracle, NN), (are, VBP), (apparent, JJ), (Florence, NNP), (was, VBD), (its, PRP\$), (very, RB), (home, NN), (and, CC), (still, RB), (can, MD), (point, VB), (to, TO), (the, DT), (greatest, JJS), (number, NN), (of, IN), (its, PRP\$), (achievements, NNS)} 
    \item NER: {(Italy, location), (Florence, location)}
    \item NEL: {(Bertinoro, location, 2343, bertinoro\_URI), (Italy, location, 585, italy\_URI)}
\end{itemize}

\paragraph{Linked Data validity} refers to the validation of Linked Data using the information extracted from trusted textual sources. In order to identify whether a given RDF triple is valid or not, we also propose an approach based on the NLP pipeline. This approach goes deeper into the text, as it also tries to identify relations after the NER step, to generate <subject> <predicate> <object> triples. These triples can then be matched to the RDF triples whose validity we aim to assess. If the information is consistent between the input and extracted relations, we conclude that the RDF triple is valid according to the textual data. Moreover, the proposed method can also be employed in order to find out the missing information related to the entities that are part of the structured data set. Due to time constraints, this approach is yet to be implemented. 

\paragraph{Example 2.} Let us assume that a structured dataset contains an RDF triple {\tt (dbr:Istanbul dbo:populationMetro 11,174,200)}. However, we have a document that is published recently that has a statement “The population of Istanbul is 14,657,434 as of 31.12.2015”. The last step of the algorithm should be able to identify the RDF triple {\tt (dbr:Istanbul, dbo:populationMetro, 14,657,434)}. Then, we can conclude that the input RDF triple is not valid. 

\subsection{Evaluation and Results: Use case/Proof of concept - Experiments}
\begin{figure}[]
\includegraphics[width=1.0\textwidth]{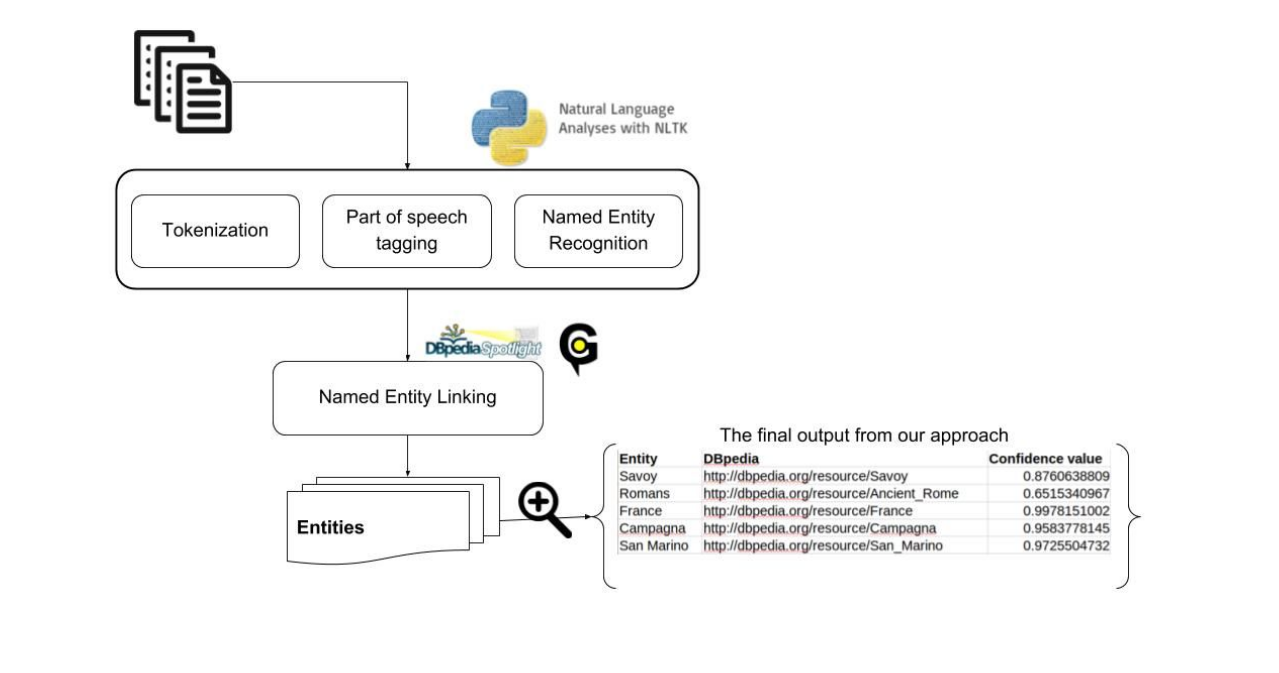}
\caption{Natural Language Processing workflow}
\label{fig:42s_fig1}
\end{figure}

As explained in the Resources section, a corpus consisting of travel diaries of English-speaking travelers in Italy between 1867 and 1932 was used. Furthermore, DBpedia and GeoNames were selected as the connecting structured databases since they contain geographical entities. We present our experimental workflow in Figure~\ref{fig:42s_fig1}. 

In order to complete tokenization, Part-of-Speech tagging, and NER, we used the Natural Language Toolkit library (NLTK)\footnote{\url{https://www.nltk.org/}} \cite{bird2009natural}. NLTK offers an easy-to-use interface and it has a built-in classifier for NER. We extracted all named entities belonging to the Person, Location and Organization categories, and then focused only on Location entities. Then, we used GeoNames and DBpedia for NEL. In order to enhance the matching quality, we have used the exact matching method. We used 29 documents out of 30 documents for our analysis, since one of the documents had an unicode encoding error. 

In total, we have identified 16,037 named location entities in 29 documents. Linking with GeoNames produced 8181 linked entities, and with DBpedia, 8,762. We were thus able to validate more than 50\% of the entities with either one of the structured data sets.

For the next step of our analysis, we selected only the linked entities from GeoNames.
\begin{figure}[]
\includegraphics[width=0.8\textwidth]{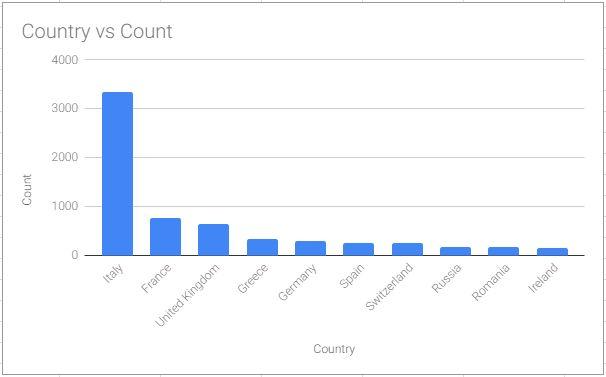}
\caption{Top-10 Countries in Linked Location Entities}
\label{fig:42s_fig2}
\end{figure}
First, we checked country information for these entities. Figure ~\ref{fig:42s_fig2} presents the top-10 countries where the linked location entities are. As expected, most of them are located in Italy. This suggests that GeoNames database has a good coverage of geographical entities in Italy. We have also entities from other countries. This might be due to several reasons. First of all, the name of the current location might be different than its name in the time of the author’s visit to Italy. Second, there might be some locations that are now part of a different country. Third, there may exist geographical entities with the same name in other countries. 

\begin{figure}[]
\includegraphics[width=0.8\textwidth]{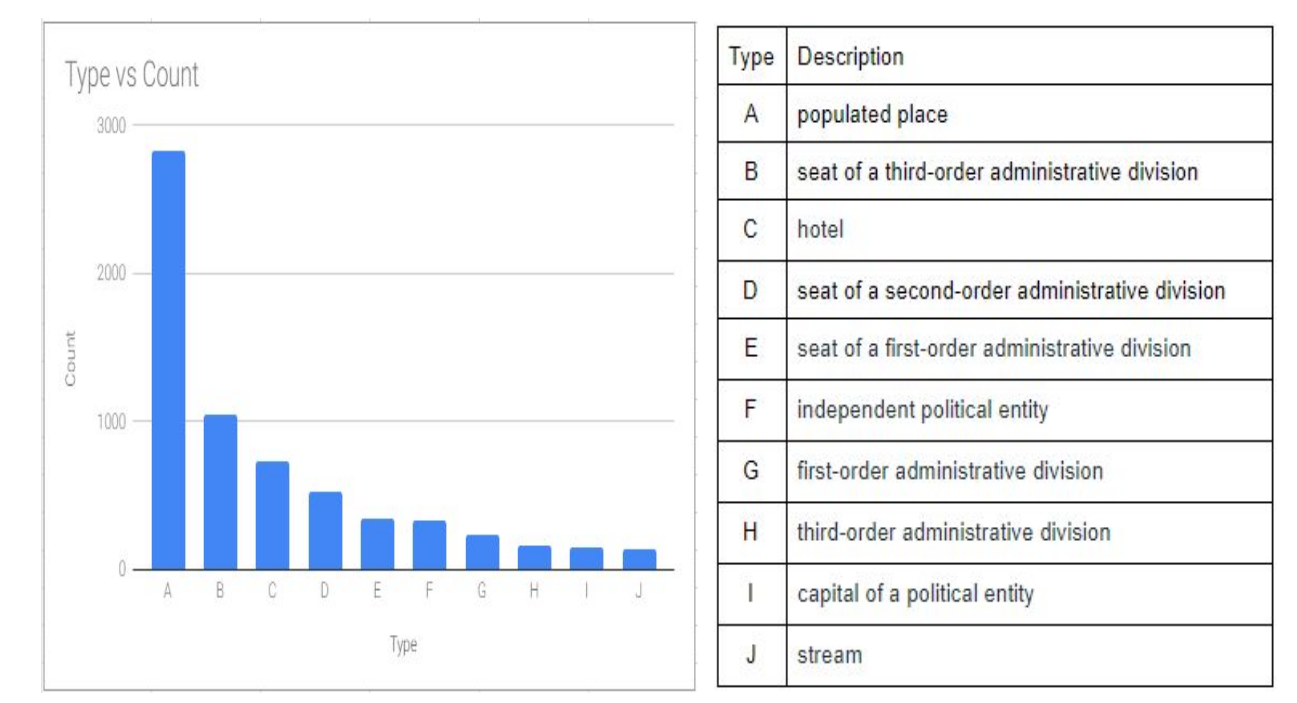}
\caption{Top-10 Location Types in Linked Location Entities}
\label{fig:42s_fig3}
\end{figure}

Figure~\ref{fig:42s_fig3} presents the top-10 types of the linked entities. As expected, the named entities are generally populated areas and administrative areas. However, the third most frequent location type is hotel. This probably corresponds to some problems regarding entity linking since the selected corpus consists of historical travel documents dated between 1867 and 1932. The reason of entities being linked to hotels would be the contemporary hotels with historical names. In future work issue needs to be checked in further detail. 

\begin{figure}[]
\includegraphics[width=1.0\textwidth]{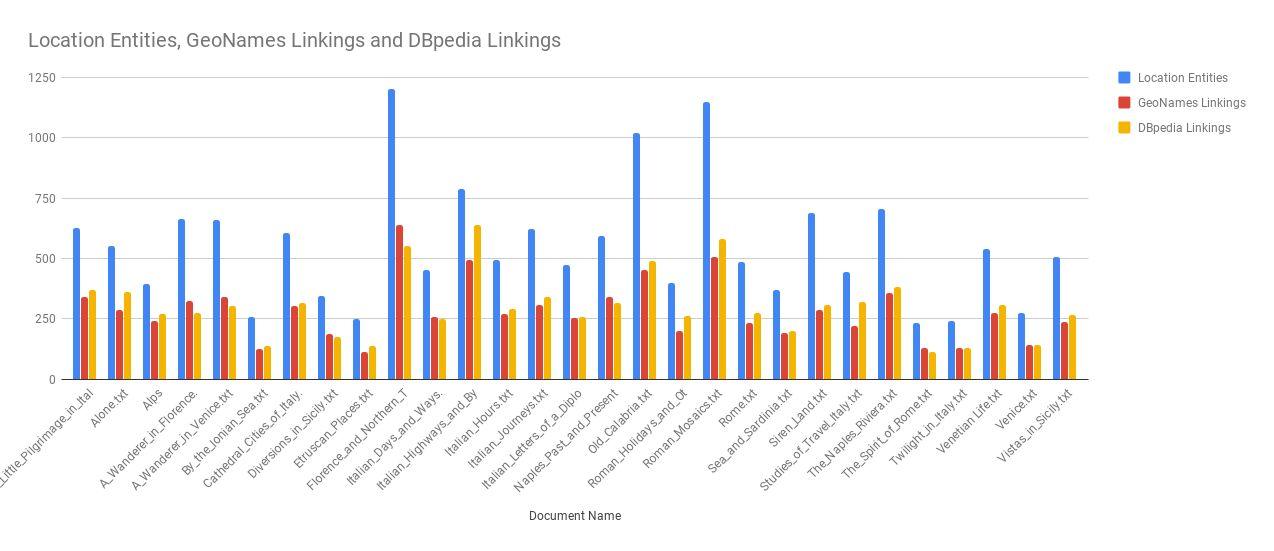}
\caption{Number of Entities and Entity Linkings from GeoNames and DBpedia}
\label{fig:42s_fig4}
\end{figure}

Figure~\ref{fig:42s_fig4} displays the number of location entities, the number of entities linked using GeoNames and the number of entities linked using DBpedia for each file. The text under each column-group corresponds to the title of the document. As can be seen, the number of entity linkings from GeoNames and DBpedia is quite dependent on the content of the document. In half of the documents GeoNames performs slightly better than DBpedia and vice versa. The figure shows that it cannot be clearly stated that one of the selected structured database works better than the other one for the textual data validity of documents regarding geographical entities. However, we have found an example corresponding to a name change of a location in Sicily. The previous name was Monte San Giuliano and now it is called as Erice. When we lookup the name of Monte San Giuliano from GeoNames, we managed to find the contemporary location entity due to the fact that GeoNames contains the information regarding old names. However, it was not possible to locate this entity in DBpedia. For this reason, if the entities are extracted from documents corresponding to historical information, it would be better to utilize GeoNames database. 

\section{Discussion and Conclusion}

Textual documents are a rich source of knowledge that, due to their unstructured nature, is currently unavailable in the Linked Data cloud. NLP techniques and tools are specially developed to extract the information encoded in text so that it can be structured and analyzed in a systematic manner. Until now, the opportunities for intersection between NLP and Linked Data have not received much attention from either the NLP or the Semantic Web community, even though there is an unexplored potential for investigation and application to real-world problems.

We proposed an approach to explore this intersection, based on two definitions of validity: textual data validity and Linked Data validity. We selected a textual corpus of travel writings from the 19th and 20th centuries, and applied NLP-based methods to extract location entities. Then, we linked those entities to the structured Linked Data from DBpedia and GeoNames in order to validate the extracted data. 

The contributions of this paper include:
\begin{itemize}
    \item A definition of Linked Data validity in the context of Natural Language Processing;
    \item The combination of two trusted knowledge sources to validate the entities extracted from text;
    \item The execution of experiments on a corpus of original travel writings by native English speakers;
    \item A proposition of a generic approach which may be easily reproduced in other contexts.
\end{itemize}

Our approach has the following strengths: 

\begin{itemize}
    \item We use knowledge from different types of sources (i.e. extracted through NLP and from Linked Data) 
    \item Our prototype uses off-the-shelf tools, providing an easy entry-point into assessing Linked Data validity from the NLP perspective. 
\end{itemize}

Naturally, there are also some weaknesses to our approach: 
\begin{itemize}
\item Assumption that dbpedia/geonames are reliable sources for validating the data
\item NLP tools are not adapted to the historical travel writings domain and thus may make more mistakes than optimised resources.
\end{itemize}

In our work we addressed the issues of Textual data validity and Linked Data validity.
We showed that structured data extracted from text through NLP is a promising approach to address both the issues. Structured data from reliable sources could be used to validate data extracted with NLP, and reliable textual sources could be processed with NLP techniques to be used as a reference knowledge base to validate Linked Data sets.

In this research report, we focused on the first aspect of Linked Data validity from an NLP perspective, namely checking the output of an NLP system against a Linked Data resource. In future work, we will also address the second aspect, namely checking the validity of a Linked Data resource using NLP output extracted from a reliable text source. We will connect to research on trust and provenance on the semantic web, to assess and model trust and reliability. 

Furthermore, we plan to extend our experiments by enlarging the dataset, consider more knowledge bases to compare with and include other domains. We plan to extract more properties, attributes, and historical information about the extracted locations as such a list of properties might further automate the validation process. Finally, for those entities that are not found in the different knowledge bases, we plan to have an automatic system to add them there with the different extracted properties. For example, in case of extracting a piece of historical information as we saw in the case of the old names of Erice as Monte San Giuliano, we can update this new information to the required knowledge base such as DBpedia.

\chapter{Validity and Context}
\label{sec:ravenclaw}
\chapterauthor{Esha Agrawal, Valerio Di Carlo, Sven Lieber, Pasquale Lisena,
Durgesh Nandini, Pierre-Henri Paris, Harald Sack}

Following are the research questions targeted in this study:
\begin{itemize}
\item Is Linked Data Validity always the same...and will it stay that way?
\item If it seems valid to you, is it also valid to me? 
\item If it has been valid 10 years ago, is it still valid...and will it stay that way? 
\item What is the intended use of some particular Linked Data and how does the intended use influence LOD validity?
\end{itemize}
Following two context for for LOD Validity have been considered: (i) What are contextual dimensions/factors relevant for LOD validity? and
(ii) How to determine, analyze, and leverage context (and pragmatics) relevant for deciding on LOD validity?
This section also discusses LOD Validity Evolution over Time meaning that time is a special context dimension for LOD validity which leads to the following problems: (i) How does LOD validity change (evolve) over time?  (ii) How can we model that and How can we make use of it?

\begin{definition}[Linked Data Validity] Validity of queried data is subjective and based on contextual information of the data and the user who queries the data.
\end{definition}

Linked Open Data (LOD) is open data, released on the Web under an open license, which does not impede its reuse for free, and is linked to other data on the Web \cite{hartig2010publishing,linkeddataweb}. Since everyone can upload linked data to the Web, validity becomes important. 

According to the Oxford Dictionary, validity is “the quality of being logically or factually sound; soundness or cogency”\cite{validity}. However, the validity of a LOD dataset is subject to various contexts and might hold true only for a certain timespan or under certain circumstances. For example, Barack Obama was only the president of the United States from 2009 until 2017, while Homer Simpson is a Nuclear Safety Inspector only in the context of the TV show “The Simpsons”. 

In general, any part of knowledge is potentially affected by the context in which it has been created and in which it will be used. Differences in time, space, and intentions produce different impacts on the human experience, from which the knowledge is generated. Additionally, even the most trusted scientific certainty is valid until it is replaced by a new one that makes it outdated, affecting also the whole world surrounding it. 

Linked Data is context dependent, but context information is usually not specified explicitly at all, mixed with other data, or only available implicitly, as e.g. encoded in a natural language text string which is only understood by humans \cite{homola2010modeling}. Our research is therefore limited to the {\bf contextual dimensions} that can be used to determine the validity of a LOD dataset or a part of that. From this point of view, the {\bf context} of a LOD dataset can be considered as a set of dimensions that might differ between the creation and the usage of that particular dataset and might even cause a change in the information.

Invalidity of data based on context occurs when, on the change of one of these dimensions, the described information becomes incorrect. In other words, LOD validity context is a set of attributes implicitly or explicitly surrounding knowledge data that allow us to define to establish the validity of the data. This context is important for the creation (authorship context) of data as well as for its usage (user context). In both the cases, the interpretation, the point of view, belief and background information are important.

In particular, we have two contributions: firstly, we identify four contextual dimensions for LOD validity and analyze to which extent current LOD datasets provide this information. Secondly, we provide SPARQL query templates to help users to query temporal data without any previous knowledge of the time management within dataset.

The remaining of the report is organized as follows. First we cover related work regarding contextual information in Linked Data. Then we introduce a working definition for LOD Validity and context and then define contextual dimensions. We surveyed existing datasets regarding our identified dimensions and according to our findings proposed the usage of an ontology. We also proposed templates that should be provided with metadata to facilitate temporal query writing for user.

\section{Related Work}\label{sec:ravenclaw_related_work}

While research of contexts has been extensively discussed in AI \cite{akman1996steps}, there are still no comprehensive studies on the formal representation of contexts and its application on Semantic Web. Guha et al. \cite{guha2004contexts} already highlighted the obstacles posed by differences in data context: for example, two datasets may provide their data using the same data model and the same vocabulary. However, subtle differences in data context pose additional challenges for aggregation: these datasets may be related to different topics, they may have been created in different time or from different points of view.

Information about the context is often not explicitly specified in the available Semantic Web resources, and even if so, it often does not follow a formally defined representation model, even inside the same resource. A few number of extensions to Semantic Web languages have been already proposed with the aim to handle context \cite{udrea2010annotated,khriyenko2006framework,serafini2011contextual,serafini2012contextualized}:

Both Annotated RDF \cite{udrea2010annotated} and Context Description Framework \cite{khriyenko2006framework} extend RDF triples with an n-tuple of attributes with partially ordered domains. The additional components can be used to represent the provenance of an RDF triple or it could also be used for directly attaching other kind of meta-facts like context information.

Serafini, L. et al. \cite{serafini2011contextual,serafini2012contextualized} proposed a different approach, called Contextualized Knowledge Repository (CKR), build on top of the description logic OWL 2 \cite{w3c2009owl}. Contextual information is assigned to contexts in form of dimensional attributes that specify the boundaries within which the knowledge base is assumed to be true. The context formalization is sufficiently expressive but at the same time more complex.
The presented approaches vary widely and a broadly accepted consensus has not yet been reached so far. Moreover, all of them require extensive work to adapt the existing knowledge bases to the proposed new formalism. As opposite, we propose an approach that:
\begin{itemize}
    \item makes easier to extend the existing knowledge resources with context information;
    \item allows to access them considering the user context.
\end{itemize}

Another important issue is, which definition of a context is reasonable to use within Semantic Web: there is, as yet, no universally accepted definition nor any comprehensive understanding of how to represent context in the area of knowledge base system. An overview of existing interpretations of context can be found at \cite{jansen1993context}.

\section{Proposed Approach}

Contextual information is important for LOD validity. Existing work requires an adaption of existing knowledge bases. In the following, we i.a. define different context dimensions and how they can be used to describe meta information of datasets, which doesn’t involve an adaption of existing knowledge bases.

\paragraph{Overview}

As previously stated in Section~\ref{sec:ravenclaw_related_work}, there is not yet a widely accepted definition of context in the field of Semantic Web. To formulate our definition, we choose to start from a relatively general one, extracted from the American Heritage Dictionary \cite{mifflin2000american}:

\say{1. The part of a text or statement that surrounds a particular word or passage and determines its meaning. 2. The circumstances in which an event occurs; a setting.}

The first definition is largely applied in the field of Natural Language Processing when dealing with textual data, while the second one has been already applied in many AI fields, for example Intelligent Information Retrieval [2]. Based on the second definition, we can identify at least three different levels at which it can be applied in RDF:
\begin{enumerate}
    \item {\bf Dataset Level:} This is the external context surrounding an entire dataset. It reflects the circumstances in which the dataset has been created (e.g. information about the source, time of creation, purpose of the dataset, name of the author and much more).
    \item {\bf Entity Level:} This is the internal context surrounding an entity of a graph. It reflects the circumstances in which the concept represented by the entity “lives” or occurs.
    \item {\bf Triple Level:} This is the specific context surrounding a single triple in a graph. It reflects the circumstances in which the relation between the subject and the object holds.
\end{enumerate}

The approaches of \cite{khriyenko2006framework,guha2004contexts} follows the third definition, while the one of \cite{serafini2011contextual,serafini2012contextualized} is based on the first one. As explained in \cite{bozzato2012context},  approaches based on the triple definition makes the knowledge difficult to be shared, encapsulated and easily identified. For this reason, on our approach, we rely only on the Dataset and the Entity Level definitions. The definition of the user context we adopt follows the widely used definition employed in AI systems \cite{akman1996steps}: the circumstances in which the user queries the knowledge resources (e.g. geo-location, language, interests, purposes etc.).

Based on the previous definitions of context, we define {\bf LOD Validity} in the following way:

\say{Given the context of the knowledge resource and the context of the user, the validity of the retrieved data is a function of the similarity between the two contexts.}

\paragraph{Dimensions and metrics}
Context is not an absolute and independent measure. Several dimensions with their respective metrics can have an influence on the context.
We have identified three different contextual dimensions (i) spatio-temporal, (ii) purpose/intention, and (iii) knowledge base population. All the three contextual dimensions apply to the knowledge base; the first two dimensions additionally apply to the user, who is querying the knowledge base (see Figure 1).
The first and maybe the most important dimension is composed by {\bf spatio-temporal contextual factors}. Several related metrics can influence the context:
\begin{itemize}
    \item Time at a triple level: a fact can become invalid with time, therefore there is a need for time information such as start, end, duration, last update.
    \item Time at an entity level: properties and values of an entity can change over time.
    \item Time at dataset level: an event that appends after the update or creation of a dataset can not be found in this dataset, this is the reason why creation date and last update time are important information to have.
    \item Geographic, political and cultural at a dataset level: a political belief, or the native language, or the location can influence the answer one would expect. Is an adult someone who is older than 18 years or 21 years?
\end{itemize}

The second dimension is the {\bf purpose or intention of the dataset}. A dataset might be created for a certain purpose that could be modeled with a list of topics. A dataset that does not contain the topics required to answer a query will not be able to provide the expected answer to this specific query. Thus, both user and dataset intentions must match or at least overlap. For example, the President of the U.S.A. can differ between a dataset about politics and a dataset about fictional characters.

The third dimension is {\bf Knowledge Base population context}. What is the provenance of the data? How many sources are there? What are the methods and/or algorithms used to populate the KB? A user may, for example, prefer to have human generated data like Wikidata over programmatically generated data like DBpedia. Also, when creating a dataset from a source dataset, approximations or wrong information may be propagated from the source dataset to the new dataset.

\section{Survey of Resources}

\paragraph{Generalistic datasets}
This category includes as e.g. Wikidata and DBpedia. Even if {\bf context metadata} are not expressed among their triples, it is well know what the purpose of the dataset is and how they have been generated.

As contextual information (meta-information about the validity about a single atomic information), Wikidata provides property qualifiers\footnote{\url{https://www.wikidata.org/wiki/Wikidata:SPARQL_tutorial\#Qualifiers} (accessed on the 06/07/2018)} that have the capability to declare (amongst others) the start and end time of the validity of a statement (e.g. “USA has president Obama”)

\begin{verbatim}
?statement1 pq:P580 2009. # qualifier: start time: 2009 (pseudo-syntax)    
\end{verbatim}

\paragraph{Domain or application specific datasets}

This category includes the less popular datasets that cover a specific domain or application. These datasets often include more descriptive metadata about themselves, frequently following the Dublin Core standard, so that they easier to parse and to include in other dataset collections (i.e. in the LOD cloud). A survey reveals that the authors of these datasets are in part conscious of the importance of expliciting context, even if with different outcomes.

Table 1 presents a brief survey made on the provided datasets. Temporal context is the most commonly expressed via properties such as {\tt dct:created}, {\tt dcterms:issued} or {\tt prov:startedAtTime}. The purpose of the dataset is moat times contained in the documentation or in a free-text description. Geo-political or methods contextual metadata are not provided.

A positive example of the context of generation of each entity is provided by ArCo, where a specific {\tt MetaInfo} object is directly linked to the subject entity that is described, specifying the time of generation and the exact source of the information. Always in ArCo, some information is directly represented as temporal-dependent, such as the location of a cultural object\footnote{E.g. \url{http://wit.istc.cnr.it/lodview/resource/TimeIndexedQualifiedLocation/0100200684-alternative-1.html} (accessed on the 06/07/2018) }. 

\section{The Provenance Ontology}

The need for metadata describing the context of the data generation is not new in the LOD environment, and different ways of modelling it have been proposed. One existing solution is the Provenance Vocabulary Core Ontology\footnote{\url{http://trdf.sourceforge.net/provenance/ns.html\#} (accessed on the 06/07/2018)} \cite{hartig2010publishing}. Extending the W3C PROV Ontology (commonly known as PROV-O) \cite{missier2013w3c}, this vocabulary defines the {\tt DataCreation} event, to which it is possible to directly link a set of properties that cover our newly introduced contextual dimensions:
\begin{itemize}
\item {\tt prov:atLocation} (geo-spatial)
\item {\tt prov:atTime} (time)
\item {\tt prv:usedData} (kb population, source)
\item {\tt prv:performedBy} + {\tt prov:SoftwareAgent} (kb population, methods)
\item {\tt prv:performedBy} + {\tt prv:HumanAgent} (kb population, author)
\end{itemize}

The {\tt DataCreation} can be linked through {\tt prv:createdBy} to the dataset or to any entity, giving the possibility of expliciting the context at different granularities.
Figure 2 shows an example of how to model the {\tt DataCreation} for a generic dataset. Provenance ({\tt prv:} or {\tt prov:} for PROV-O original properties) is used for most of the dimensions, while Dublin Core\footnote{\url{http://purl.org/dc/elements/1.1/subject} (accessed on the 06/07/2018)} ({\tt dc:}) is used for the purpose definition.

\subsubsection{Time Handling Templates}

The goal of the presented time handling templates is to facilitate data usage by giving intelligible hints to the user about how data can be temporally queried. This removes the need for a time-consuming study of the entire dataset structure from the user side.
For each dataset, example(s) SPARQL queries should be provided by the data owner in the form of metadata. Thus, any data user could quickly be able to write a temporal query.

{\bf DBpedia} handles duration (or time periods) in several ways (the following list may not be exhaustive):
\begin{itemize}
\item By using an {\bf instance} of the {\tt dbo:TimePeriod} class (or one of its subclasses). 
\begin{itemize}
    \item specific datatype properties might indicate the duration of the considered time period. For example, we can consider a time window of the career of the football player Paul Pogba. {\tt dbr:Paul\_Pogba\_\_3} is an instance of a subclass of {\tt dbo:TimePeriod} and has the property {\tt dbo:years} indicating the year of this period of time.
    
\noindent Template:

\begin{verbatim}
SELECT *
WHERE {
  [SUBJECT] a dbo:TimePeriod ;
        dbo:[DATATYPE\_PROPERTY\_WITH\_TIME\_RANGE] ?timeValue
}    
\end{verbatim}

\item specifying the considered time period directly in the type. For example, the Julian year 1003 is represented by the resource {\bf dbr:1003} whose type is {\tt dbo:Year}.

\end{itemize}

\item By using specific (couples of) datatype properties. The type of the time measurement (e.g. year) is specified in the name and more formally in the range. The differentiation between starting and ending events is encoded in the name of the property. For example, {\tt dbo:activeYearsStartYear} and {\tt dbo:activeYearsEndYear} or {\tt dbo:activeYearsStartDate} and {\tt dbo:activeYearsEndDate}.

\noindent Template:

\begin{verbatim}
SELECT *
WHERE {
  ?subject dbo:[PropertyName][Start|End][TimeType]
}    
\end{verbatim}

However, since the semantics of the properties is not explicitly provided, its interpretation requires the manual effort of the data creator.

\end{itemize}

{\bf Wikidata} on the other side uses the concept of qualifiers to express additional facts and constraints about a triple (by using the specific prefixes: p, ps and pq for alternative namespaces to distinguish qualifiers from regular properties). For example, the assertion “Crimean Peninsula is a disputed territory since 2014” is expressed by the statements $s_1 =  <Crimean, is a, disputed territory>$ and $s_2=<s_1, start time, 2014>$. Wikidata template:

\begin{verbatim}
SELECT *
WHERE {  
  ?subject p:[PROPERTY\_ID] ?statement.
  ?statement pq:[TIME\_PROPERTY\_ID] ?timeInformation .
}    
\end{verbatim}

\section{Conclusion}

As stated in the introduction, knowledge is created by humans. Anyway, humans have their own beliefs, which might introduce a bias. We argue that this belief is an important contextual dimension for LOD validity as well. For example, Wikipedia is an online encyclopedia which is curated by multiple users and therefore might contain less bias than a dataset created and curated by a single person. However, these beliefs are manifold and possibly implicit, which makes it hard to express them explicitly, both, formally and informally. Therefore we didn’t include a contextual personal belief dimension.

LOD contains contextual information on the dataset level in the form of meta information and within the dataset in the form of data. Based on examples, we have shown that contextual information are an important part of LOD Validity. For example, data may vary over time at multiple levels, or user's expectation may depend on her cultural context. 
In this work, we have provided a set of dimensions that can influence either dataset and user contexts. We demonstrate the importance, for both user and dataset owner, to provide this information in the form of metadata using the Provenance Ontology. We also provide a way to add, in metadata, templates to show to users how to use temporal data in the dataset without time-consuming study of the data.

We proposed to reuse existing vocabularies to describe contextual meta information of datasets. Future work can investigate the usage of statistical data (and their semantic representation) regarding contextual dataset data, to facilitate the selection of a dataset fitting the purpose of the user.

\begin{table}[htp!]
\centering
\scriptsize{
\begin{tabular}{|c|c|c|c|c|c|c|} \hline

DIMENSION           &  Scholarly Data  &  Data.cnr.it  &  ArCo    &  Pubmed  &  Food     &  Food (subdatasets) \\ \hline  
time                &  DATASET         &               &  ENTITY  &  ENTITY  &  DATASET  &  DATASET             \\ \hline 
geo-political       &                  &               &          &          &           &                  \\ \hline    
purpose/intention   &  UI              &               &          &          &  DATASET  &                  \\ \hline    
author              &  DATASET         &  DATASET      &  ENTITY  &          &  DATASET  &                  \\ \hline    
source              &  DATASET         &               &          &  ENTITY  &  ENTITY   &  ENTITY           \\ \hline  
methods              &          &               &          &    &    &           \\ \hline  
\end{tabular}
}
\caption{A survey on the contextual information in the datasets provided for that report with the help of a SPARQL endpoint. Green indicates that information is explicit and machine-readable. Yellow indicates that information is present as plain natural language text only to be further interpreted. Otherwise, no information is provided.} \label{tab:ravenclaw_1}
\end{table}


\begin{figure}[]
\includegraphics[width=0.8\textwidth]{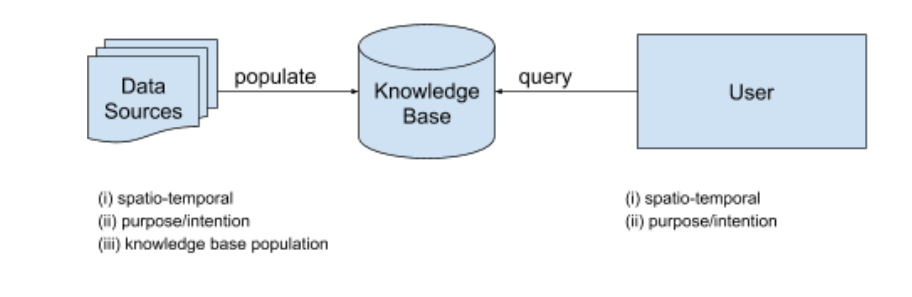}
\caption{The three identified contextual dimensions for LOD Validity. All three are relevant for
the dataset and two of them are relevant for the user.}
\label{fig:ravenclaw_fig1}
\end{figure}

\begin{figure}[]
\includegraphics[height=0.6\textwidth]{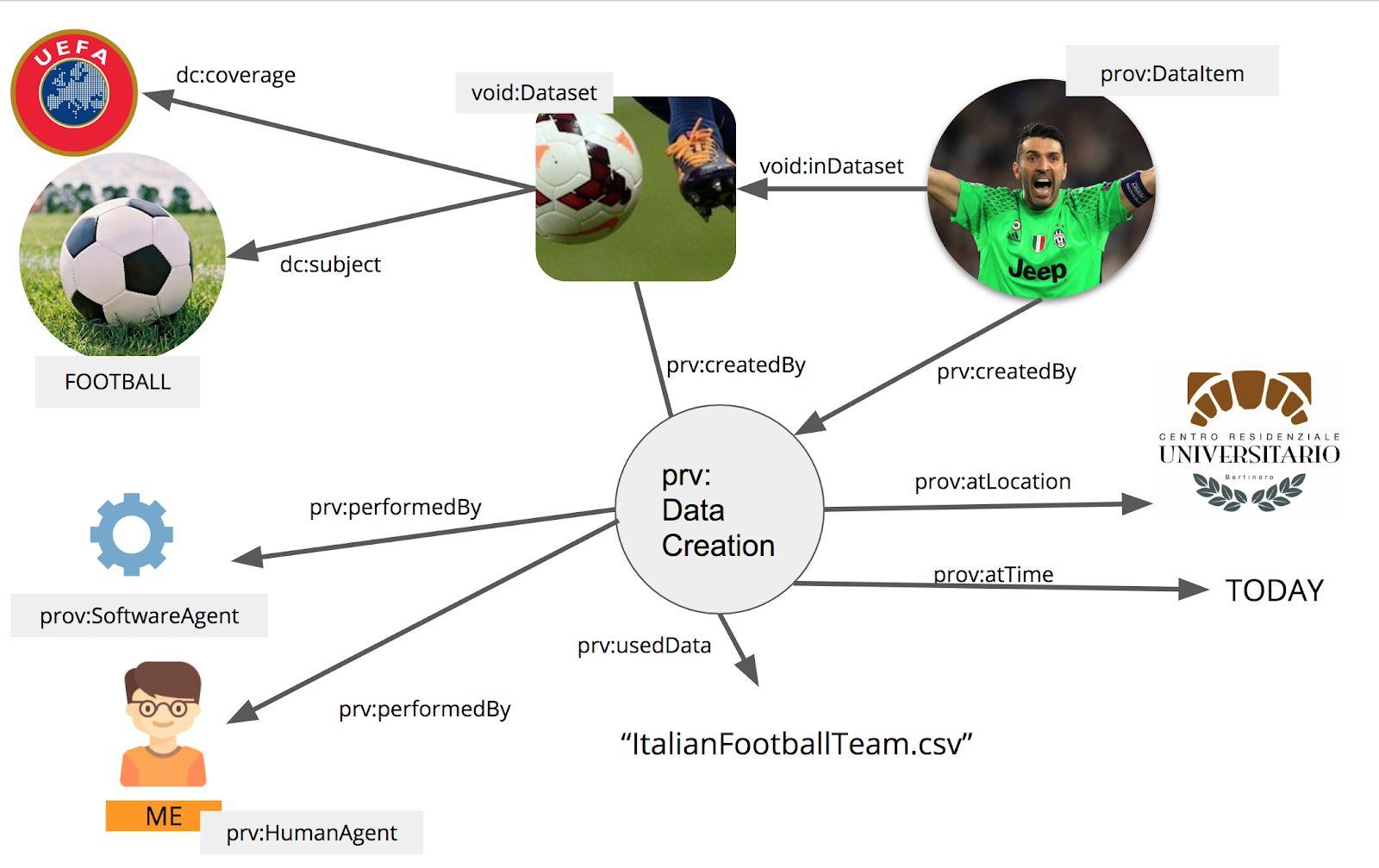}
\caption{The contextual metadata for a Football dataset realised by the authors today in
Bertinoro.}
\label{fig:ravenclaw_fig2}
\end{figure}

\part{Data Quality Dimensions for Linked Data Validity}
\label{part2}
\chapter{A Framework for LOD Validity using Data Quality Dimensions}
\label{sec:gryffinder}
\chapterauthor{ Mathias Bonduel, Rahma Dandan, Valentina Leone, Giuseppe Futia, Henry Rosales-Méndez, Sylwia Ozdowska, Guillermo Palma, Aldo Gangemi}
Research Questions:
\begin{itemize}
    \item What are exemplary use cases for LOD validity?
    \item How to establish validity metrics that are sensible both/either to structure (internal), as well as to tasks, existing knowledge and sustainability (external)? 
    \item What is a typical LOD unit to be checked for validity?
\end{itemize}

\begin{definition}[Data Quality Dimensions]
The notion of validity, in our specific case, is related to two different perspectives: (i) an internal perspective and (ii) an external perspective. The internal perspective is built on data quality dimensions such as: accuracy, completeness, consistency, and novelty. These dimensions involve, on the one side, the data itself (A-box statements) and, on the other side, the ontologies that describes data (T-Box statements). The external perspective is driven by typical issues and contents related to a specific domain of the data. In our paper, we focus on the internal perspective, while we mention issues related to the external perspective in the discussion. 
\end{definition}


Linked Data (LD) represent the backbone for systems that exploit domain-specific or domain-independent structured data published on the Web. The capacity of such systems to retrieve valuable knowledge from LD is strictly related to the validity dimension of the available data. 

Regarding a relevant use case for LD validity, we can discuss the case when someone new in a certain domain wants to collect some basic information about a subject. This person typically starts with googling and/or looking for a general database as Wikipedia to get some general introduction, before reading more detailed information. A similar approach could be valid for DBpedia (general KB) and an expert KB related to the domain. The validity of the general KB can be important as the general KB will be used by non-experts who cannot directly see if some information is (in)valid.

The notion of validity, in our specific case, is related to two different perspectives: (i) an internal perspective and (ii) an external perspective. The internal perspective is built on data quality dimensions such as: accuracy, completeness, consistency, and novelty. These dimensions involve, on the one side, the data itself (A-box statements) and, on the other side, the ontologies that describes data (T-Box statements). The external perspective is driven by typical issues and contents related to a specific domain of the data, that in some cases could bring different results in terms of validity compared to the internal one. To better understand both perspectives, consider the following assertion: “ex:book ex:isWrittenIn 2054”. From an internal point of view, this assertion is not valid, because a book can be written in the past and not in the future. Nevertheless, if this assertion models a scenario related to a science fiction set in the future, this statement is probably valid. Our intuition is that both perspectives should be considered and evaluated to effectively establish LD validity.

For the internal perspective, our approach is based on a comparison between a Ground-Truth Knowledge Graph (GT-KG) that plays the role of oracle in our evaluation and a Test-Set Knowledge Graph (TS-KG) that should be evaluated according to GT-KG. Our method exploits SPARQL queries and ontology patterns to measure accuracy, completeness, and consistency data quality dimensions mapped on precision, recall, and F1 metrics. We have decided to use ArCo as GT-KG and DBpedia as TS-KG. In another step, we translate Competency Questions (CQs) of a domain expert in SPARQL queries and ontology patterns on ArCo in order to detect validity issues based on the external perspective. In this case, we compute precision, recall, and F1-measure according to a human oracle or a natural-language oracle, an authoritative resource that covers the analysis domain. For the last case, results could be less accurate due to the automatic process of statements extraction from text. In our paper we deeply discuss the internal perspective, and we report some reflections related to the external perspective.
  
The paper is structured as follows: Section 2 shows related works, Section 3 describes data sources that we have exploited in our analysis, Section 4 provides details on the adopted method, Section 5 illustrates results and evaluation, and finally Section 6 reports conclusions and propose a discussion about our research work.

\section{Related Work}

Our research work is strictly related to the Linked Data Quality (LDQ) field, because we consider dimension like accuracy, completeness, consistency, and novelty in order to compute validity. In the field of LDQ, we identify three different type of contributions: (i) works focused on the definition of quality in LD, (ii) approaches to detect issues and improve quality according to such definitions, (iii) implementation of tools and platforms based on this approaches. For the first type of contribution, we remark the work of \cite{zaveri2016quality} that discusses with a systematic literature review many works on data quality assessment. For the second type of contributions focused on the approach, we mention the work of \cite{bizer2009quality} that propose to apply filters on all available data to preserve high-quality information. For the third kind of contribution, related to the implementation, we report the work of ~\cite{kontokostas2014test} that present a tool inspired by test-driven software development techniques, to detect quality problems in LOD. In particular, they define test to detect data quality problems, based on the semi-automated instantiation of a set of predefined patterns expressed in SPARQL language. Our research can be counted between works related to the approaches developed to identify quality issues, but focused on the dimension of validity.

As mentioned in the previous Section, we can also define CQs in order to establish the validity of an ontology (or a KG) for specific tasks. Traditionally, CQs are used for ontology development in specific use cases gathering functional user requirements \cite{jacobson1993object}, and ensuring that all relevant information is encoded. Other works are more focused on specific methodologies to use CQs. For instance \cite{couto2014application} proposed an approach to transform use cases descriptions expressed in a Controlled Natural Language into an ontology expressed in the Web Ontology Language (OWL), allowing the discovery of requirement patterns formulating queries over OWL datasets. In other cases CQs consist in a set of questions that an ontology should be able to answer correctly according to a given use case scenario \cite{gruninger1995role}. A wide spectrum of CQs, their usefulness in ontology authoring and possible integration into authoring tools have been investigated \cite{dennis2017computing,hofer2015semi,ren2014towards}. Unlike such research works, our approach does not focus on the construction of ontologies, but on their validation for the achievement of specific purposes within a well-defined domain.

Finally, for the data preparation stage for validity evaluation, we can mention works related to link discovery. Such works try to identify semantically equivalent objects in different LOD sources. Most of the existing approaches reduce the link discovery problem to a similarity computation problem adopting some similarity criteria and the corresponding measures in order to evaluate similarities among resources\cite{nentwig2017survey}. The selected criteria could involve both the properties and the semantic context of resources. However, all these approaches focus their attention in finding similarities among  LOD sources which belong to the same domain. On the contrary, in our project we tried to discover similarities among general and domain-specific LOD knowledge bases. Other techniques based on entity-linking like DBpedia Spotlight \cite{mendes2011dbpedia} and TellMeFirst \cite{rocha2015semantic} can be exploited for link discovery starting from natural language description of the entities.

\section{Resources}
A mentioned in the first Section, our approach requires at least one Ground-Truth Knowledge Graph (GTK) that plays the role of oracle in our evaluation and a Test-Set Knowledge Graph (TS-KG) that should be evaluated according to GT-KG. Several KGs have been proposed in the literature, many of them specialized in a particular domain, while general KGs commonly focus the attention real-world entities and its relations. 

Expert KGs focus on a specific domain which contains deep and detailed information about a particular area of knowledge. With this characteristics we can highlight DRUGS, a KG that includes a valuable information of drugs, since a bioinformatics and cheminformatics point of view. BIO2RDF is other expert KG that deal data for the Life Sciences. As we decided to focus our attention on the Cultural Heritage field, we chose ArCo as Ground-Truth Knowledge Graph. ArCo is a recent project, started in November 2017 by the Istituto Centrale per il Catalogo e la Documentazione (ICCD) and the Istituto di Scienze e Tecnologie della Cognizione (ISTC). Its aim is to enhance the Italian cultural value creating a network of ontologies which model the knowledge about the cultural heritage domain. From the modelling point of view, ArCo tries to apply good practices concerning both the ontology engineering field and the fulfillment of the users requirements. 

In particular, ArCo is a project oriented to the re-use and the alignment of existing ontologies through the adoption of ontology design patterns. Moreover, following an incremental development approach, it tries to fulfil in every stage the user requirements which are provided by a group of early adopters. Some examples of early adopters could be a firm, a public institution or a citizen. Their contribute to the development of the project testing the preliminary versions of the system and providing real use cases to the team of developers.

On the other side, one of the most popular general KG is DBpedia \cite{bizer2009quality}, which is automatically created from the Wikipedia editions, considering only the title, abstract and its semi-structured information (e.g., infobox fields, categories, page links, etc). In this way, the quality of the DBpedia data depends directly on the Wikipedia data, which is important because Wikipedia is a large and valuable source of entities, but its quality is questionable because anyone can contribute. This problem also affects the cross-language information of DBpedia. For instance, if we go to the page of Bologna in the English and Italian version of DBpedia, it will not be the equivalent information.

In order to homogenize the description of the information in DBpedia, the community has devoted efforts to develop an ontology scheme, which gathers specific information such as the properties of the Wikipedia infobox. This Ontology was  manually created, and currently consist in 685 classes which form a subsumption hierarchy and are described by 2,795 different properties. With this schema, the DBpedia ontology contains 4,233,000 instances where those that belong to the Person (1,450,000 instances) and Place (735,000 instances) class predominate.

\section{Proposed Approach}

Our approach is based on the general methodology on the general LDQ Assessment pipeline presented by \cite{rula2014methodology}. This methodology comprises four different stages: (i) Preparing the Input Data, (ii) Requirement Validation (iii) Linked Data Validation Analysis (iv) Linked Data Improvement. Figure 1 shows the pipeline of the methodology for the validation of Linked Data. The following sections describe the phases of our methodology. In the next paragraph we described from an high level point of view stage (i) and stage (iv) because our contribution is particularly focused on stage (ii) and Stage (iii).

\begin{figure}[t!]
\includegraphics[width=1.0\textwidth]{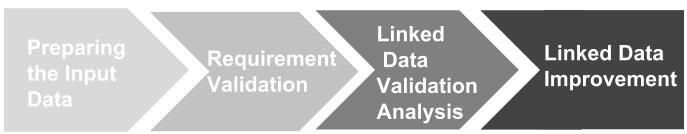}
\caption{Pipeline of the Methodology proposed for Linked Data Validity}
\label{fig:fig1}
\end{figure}

\paragraph{Stage i - Preparing the Input Data} 
After choosing the GT-KG and TS-KG, respectively ArCo and DBpedia and our specific case, we build a bridge between the two KG exploiting ontology matching and entity alignment techniques exploited manual and automatic tool to accomplish this task (see Related Work section on link discovery for more details). In this way, we create the conditions to compare a set of statements i.e. a subgraph, for the validation process. As we report in the Use Case Section we will start from relevant classes, properties, an entities linked in this stage.

\paragraph{Stage ii - Requirement Validation}
In our approach, we have defined data quality dimensions for the internal perspective as accuracy, completeness, consistency, and novelty. The dimensions are based on the quality assessment for linked data presented by Zaveri et al. \cite{zaveri2016quality}.

The {\bf accuracy} is related to the degree according to which one or more statements reported in the GT-KG are correctly represented in the TS-KG. The metrics identified for the validation of LD statements are the detection of inaccurate values, annotations, labellings and classifications by comparison with respect to a ground truth dataset. 

The {\bf completeness} validation of an entity in the TS-KG corresponds to the degree according to which information contained in the GT-KG is present in the TS-KG. This can be done be looking to specific statements and the mapped properties. Additionally, besides mapped properties, also Linked Data patterns of one of the KGs mapped to LD patterns of the other KG can be analysed to check the completeness dimension.

The {\bf consistency} validation means that the Linked Data statements should be free of contradictions w.r.t. defined constraints. The consistency can be represented at schema and data levels. Consistency at the schema level indicates that schema of a dataset should be free of contradictions, and consistency at the data level relies on the absence of inconsistencies in the A-Box in combination with its corresponding Tbox. 

The {\bf novelty} of Linked Data is defined as the set of relevant Linked Data statements that are in the dataset and that are not represented in the ground truth dataset. These Linked Data statements correspond to new predictions that should be validated.

\paragraph{Stage 3 - Linked Data Validation Analysis.}
The goal of this phase is to perform the validation specifying metrics that correspond to the 4 dimensions specified in the previous stage.

The accuracy degree of a group of LD statements can be determined, computing the Precision, Recall and F1 score of the number of the statements validated and not validated. 

$$Recall = \frac{No.~of~Linked~Data~statements~validated}{total~no.~of~linked~data~statements}$$

$$Precision = \frac{No.~of~Linked~Data~statements~validated}{No.~Linked~Data~statements~validated + No.~of~Linked~Data~statements~not~validated}$$

The completeness degree can be computed as follows:

$$Completeness = \frac{No.~of~real-world~entities~contained~in~the~Linked~data~statements}{Total~no.~of~real-world~entities~in~the~ground~truth~dataset}$$

The metric used for computation of the consistency is the number of inconsistent statements in the knowledge graph:
$$Consistency  = \frac{No.~of~inconsistent~values} {Total~no.~of~real-world~entities~in~the~ground~truth~dataset}$$

The novelty can be computed as follows: $$Novelty = No.~of~Linked~Data~statements~not~included~in~the~ground~truth~dataset$$

\paragraph{Stage 4: Linked Data Improvement}
In this stage, strategies to address the problems with the invalided statements are implemented. One strategy that can be the implementation of an automatic or semi-automatic system with recommendations for the invalid LD statements.

\section{Evaluation and Results: Use case/Proof of concept - Experiments}

In this Section we present a use case that exploits ARCO as GT-KG and DBpedia as TS-KG.  During the stage of “Preparing the input data” mentioned in previous Section, we perform the Ontology Matching and Similar Entity linking between ARCO and DBpedia. In this way we are able to obtain an entity matching between instances of ARCO and instances of DBpedia. For instance, we are able to state that the entity identified in ARCO as Colosseum\footnote{\url{http://dati.beniculturali.it/lodview/mibact/luoghi/resource/CulturalInstituteOrSite/20734l}} is identified to be probably the same entity in DBpedia \footnote{\url{http://it.dbpedia.org/resource/Colosseo/l}}. Each Arco instance can be related to multiple DBpedia instances and each DBpedia instance can be related to multiple Arco instances. We assume that this relatiness is stored in a separate graph. 

We start identifying the most common properties of ARCO classes. As mentioned in the previous paragraph, in our case we focus on the ARCO class\footnote{\url{http://dati.beniculturali.it//cis/CulturalInstituteOrSite}}, counting the most common proporties with the following SPARQL query.

\begin{verbatim}
SELECT DISTINCT ?class ?p (COUNT(?p) AS ?numberOfProperties)
WHERE{
	?class a owl:Class .
	?inst a ?class ;
		?p ?o .
	# classes cannot be blank nodes + no owl:Thing and owl:Nothing
	FILTER (?class != owl:Nothing)
	FILTER (?class != owl:Thing)
	FILTER (!isBlank(?class))
} 
GROUP BY ?class ?p
ORDER BY ?class    
\end{verbatim}

According to the results obtained through this query we have chosen properties and values reported in the Table 1 to compute accuracy, completeness, and novelty. The results of this query can be used as a weighing factor for the different validity measures related to properties. This table shows an example of the LD validation of several relevant properties of the real-world entity Colosseum. 


\begin{figure}[]
\includegraphics[width=1.0\textwidth]{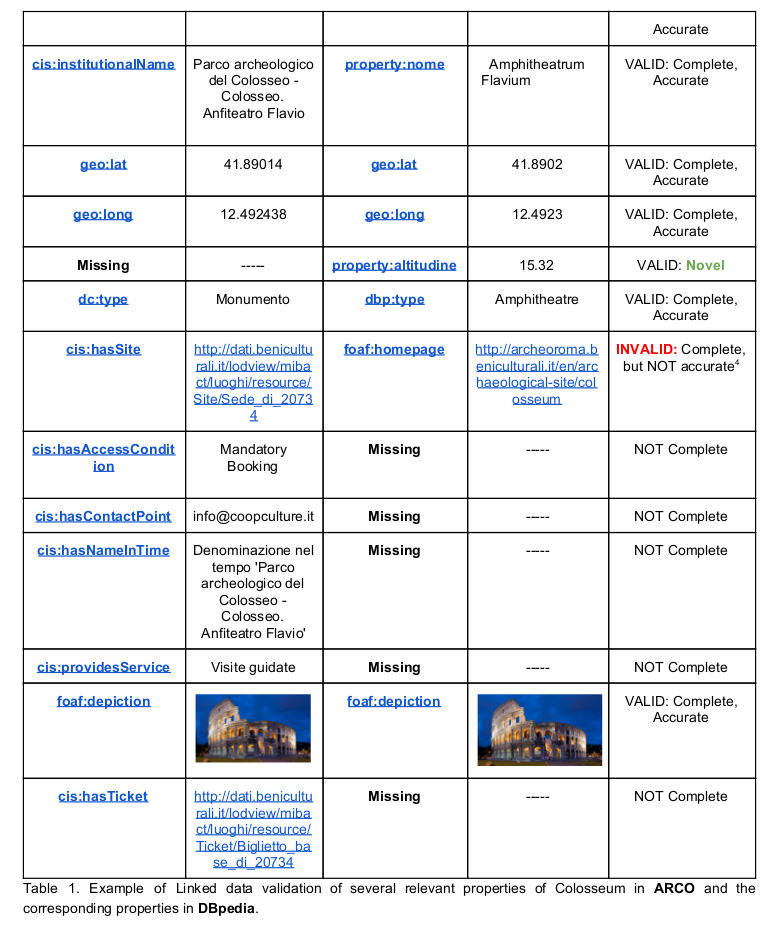}
\caption{}
\label{fig:gryffindor_table1}
\end{figure}

Determining the consistency of matched entities using {\tt owl:sameAs} in combination with an ontology alignment of both the TB-KG and the GT-KG (include restrictions), can be described with the following example:

\begin{verbatim}
@prefix cis: <http://dati.beniculturali.it/cis/> .
@prefix core: <https://w3id.org/arco/core/> .
@prefix arco: <http://dati.beniculturali.it/mibact/luoghi/resource/CulturalInstituteOrSite/> .
@prefix dbo: <http://dbpedia.org/ontology/> .
@prefix yag: <http://dbpedia.org/class/yago/> .
@prefix dbr: <http://dbpedia.org/resource/> .

#Arco Tbox
cis:CulturalInstituteOrSite a owl:Class .
core:AgentRole a owl:Class .
cis:CulturalInstituteOrSite owl:disjointWith core:AgentRole .

#Arco Abox
arco:20734 a cis:CulturalInstituteOrSite .

#DBpedia Tbox
dbo:Venue a owl:Class .
yag:YagoLegalActorGeo a owl:Class .

#DBpedia Abox
dbr:Colosseum a dbo:Venue , yag:YagoLegalActorGeo .

#Arco-DBpedia ontology mapping
core:AgentRole owl:equivalentClass yag:YagoLegalActorGeo .

#Arco-DBpedia entity linking
arco:20734 owl:sameAs dbr:Colosseum .

\end{verbatim}

If these graphs are analysed by a reasoning engine, it will come across an inconsistency as the owl:disjointWith restriction is violated. Debugging systems and their heuristic methods can be used by a machine to determine which triples might be causing the inconsistency. In the above case, there might be three triples that could be considered relating to the ontology mapping, the entity linking or a wrongly asserted triple in the TB-KG Abox:

\begin{verbatim}
#Arco-DBpedia ontology mapping
core:AgentRole owl:equivalentClass yag:YagoLegalActorGeo .
#Arco-DBpedia entity linking
arco:20734 owl:sameAs dbr:Colosseum .
#DBpedia Abox
dbr:Colosseum a yag:YagoLegalActorGeo .    
\end{verbatim}

For a human interpreter, it is quite obvious that that the inconsistency is caused by the wrongly asserted triple in the TB-KG, but machines cannot easily deal with it. We assume there are ten statements on the entity in the TB-KG.

The final computation of the metrics corresponding to four dimensions of Linked Data Validity is as follows:

$$Precision = \frac{6}{6+1} = 0.86$$
$$Recall = \frac{6}{13} = 0.46$$
$$ F_1 score = 0.60$$
$$Completeness  = \frac{7}{12}=0.58$$
$$ Consistency  = 0.1$$
$$Novelty = 1$$

\section{Conclusion and Discussion}

The paper presents an approach to establish the validity of Linked Data according to an internal perspective considering specific dimensions related to data quality domain, in particular: accuracy, completeness, consistency and novelty. In some cases like cis:Description for ArCo and ontology:Abstract in DBpedia we compare the two statements on the according to such dimension. 

In other cases we also focus on ontology patterns. Considering a simple example of geographic information. In the Arco ontology we detect properties like geo:lat and geo:long associated to a specific entity like Colosseum. In DBpedia we can have the concept Point, that specified latitude and longitude, associated to the Colosseum entity. Therefore, to compare such data, we can exploit this kind of pattern.

In some cases we have considered some statements valid according to the internal perspectives instead of external perspective. For instance, we have noticed that geolocated information about the Colosseum are slightly different in case of ArCo and DBpedia. Such statements can be considered valid for establishing a point on a map, but we can need more accuracy if a robot should perform a job in that area. For this specific case we can define a CQ that establish validity for such specific purpose.

Finally, about the novelty dimension, we state in rough way that a statement is novel (and valid) if it appears in DBpedia and not in ArCo. Nevertheless, as future works, we should perform much more analysis on this novel statement in order to establish the validity of this novel statement.


\part{Embedding Based Approaches for Linked Data Validity}
\label{part3}

\chapter{
Validating Knowledge Graphs with the help of Embedding
}
\label{sec:hobbits}
\chapterauthor{Vincenzo Cutrona, Simone Gasperoni, Nyoman Juniarta, Prashant Khare,    Siying LI, Benjamin Moreau, Swati Padhee, Michael Cochez}

\begin{figure}
\includegraphics[width=1.0\textwidth]{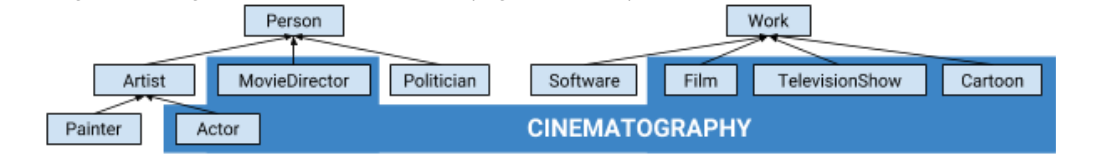}
\caption{Relevant DBpedia classes for the Cinematography topic}
\label{fig:hobbits_fig1}
\end{figure}

A huge volume of data is being curated and added to generic Knowledge Graphs (KGs) with every passing day.
The web of data has grown from 12 datasets in 2007 to more than 1160 datasets now.
The English version of DBpedia released in 2016 has more than 6.6 million entities and 1.7 billion triples.
Domain-specific applications sometimes need additional information represented in external KGs in order to enrich their existing information. 
Assuming this information has been collected already, this task could be addressed by: (a) obtaining a specific-domain KG, or (b) by extracting a subgraphs from a generic KG.
Option (a) is straightforward, since it requires downloading the whole specific-domain KGs (when available), while the solution of (b) is still a challenging problem.
Let's assume that an application needs data about a specific topic, such as cinematography (which includes Film, TV series, Cartoons, Actors, \dots). 
Thus, we can consider a specific topic as a subgraph of a KG that contains only instances that are related to that topic.
Considering a generic KG, in many cases the information is organized based on a taxonomy that does not reflect our needs, i.e., it is not organized by ``the topics'' (cf. fig.~\ref{fig:hobbits_fig1}). 
For example, in DBpedia the concept of cinematography is not represented directly, and Films, TV Shows and Cartoons are grouped together with other classes (e.g., Software) into the broader concept Work.

Validity could be defined in several ways, depending on the specific scenario.
There is no generally accepted definition of validity in the literature, and likely this is also not possible as is observable from the many viewpoints presented in this report.
Among other choices, validity can be defined in terms of relevancy with respect to a domain, schema-level consistency (i.e, properties and classes are used according to the ontology), and temporal validity.
\textbf{In this chapter, we focus on validity in terms of relevancy to a domain.
Thus, we define the validity as the relevance of a (sub)graph with respect to a specific domain.}
More in detail, a graph is valid when all properties and entities are relevant, with respect to a specific domain.
Based on this definition, we propose a methodology to extend an existing KG using properties/entities that are relevant to the selected domain (which we will also call the \emph{topic}). 

To summarize, the main contributions of this work are the following:
\begin{itemize}
\item Identifying the most relevant subgraph with respect to a topic from a generic KG;
\item Using knowledge graph embedding techniques aiming at topic-relevant subgraph identification;
\item Identifying the nature of predicates being relevant to a topic of interest.
\end{itemize}

\section{Related Work}

\paragraph{Domain-specific subgraph relevancy}

A variety of domain-specific subgraph extraction works have addressed the issue of validity in terms of relevancy.
These methods usually employ the relatedness of associated concepts to the domain of interest \cite{lalithsena2016harnessing}. 
The work by Lalithsena et al. \cite{lalithsena2017domain} considers that the relevancy of a concept to a particular domain can be determined through the type and lexical semantics of the category label associated with that concept.
Furthermore, Perozzi et al. \cite{perozzi2014focused} proposed a graph clustering with user preference, i.e. the finding a subgraph with regard to the users interest.
As opposed to that work, where relevant nodes are determined using the Euclidean distance between nodes, we propose an approach to identify the most relevant subgraph by combining  spatial and contextual semantics of nodes at the same time.
Our proposed contextual similarity (via Topic Modeling) augmented with KG embedding based approach contributes to identify the nature of predicates, whether they are more responsive to cross-domain or inter-domain relations.

\paragraph{Knowledge Validity}

One of the prominent works in automatic KG construction and prediction of the correctness of facts is by Dong et.al \cite{dong2014knowledge}. 
In that work, instead of focusing on text-based extraction, they combined extractions from Web content with some prior knowledge.
Bhatia et.al \cite{bhatia2018tell} also designed an approach to complement the validity of facts in automatic KGs curation by taking into consideration the descriptive explanations about these facts.
Bhatia and Vishwakarma \cite{bhatia2018know} have shown the significance of context in studying the entities of interest while searching huge KGs.
However, we propose to extend the context by complementing the spatial neighborhood of entities with the context of predicates (edges) connecting these entities.  

\paragraph{Topic modeling}

In this report, we also apply the task of topic modeling \cite{wallach2006topic}. In this task, given a dataset of documents, where each document is a text, we try to obtain a set of \emph{topics} that are present among the documents.
The most important step is the grouping of articles, where an article can be present in more than one grouping.
Each group corresponds to a topic.
Then, in order to map a topic to a label (e.g. sport, health, politics), we look at the frequent words among the articles in that group.
One basic way to perform this grouping is by applying Latent Dirichlet Allocation (LDA) \cite{blei2012probabilistic}, which allocates articles into different topics.

\paragraph{Knowledge graph embedding}

The purpose of knowledge graph embedding is to embed KG into a low dimensional space by preserving some properties of it.
This allows graph algorithm to be computed efficiently.
Yao, \cite{yao2017incorporating} proposes a knowledge graph embedding algorithm to achieve topic modeling.
However, that work does not take into account property values which contain an essential part of the knowledge.
Numerous KG embedding techniques have been proposed \cite{cai2018comprehensive}. 
In this report, we focus on node embedding algorithms that preserve node position in the graph and thus, graph topology such as Laplacian eigenmaps \cite{belkin2002laplacian}, Random walk \cite{fouss2007random}, DeepWalk \cite{perozzi2014focused} and Node2Vec \cite{grover2016node2vec}.

\section{Resources}

In our approach, we focus on identifying a specific-domain subgraph, given a generic graph.
Thus, in general, we can select any generic graph as our input.
However, since our approach heavily relies on descriptions of entities for the topic modeling, we need a KG that provides descriptions about entities.
Looking at the widely studied generic KGs, we see that DBpedia provides long abstracts.
In addition, most of the reviewed approaches use this graph for experimental evaluation, so this choice also enables comparative experimentation.

We are mostly interested in computing the relevance of properties that allow us to enrich an existing dataset with external information.
Thus, considering DBpedia, we can identify two kinds of properties:

\begin{description}
	\item[Hierarchical Predicates] are used for structuring the knowledge and include predicates indicating broader concepts, subclass relations, disjointedness, etc.
	Often, these predicates will only be used on more abstract entities.
	To determine the relevance of entities connected with these predicates it is crucial to investigate the entities themselves.
	\item[Non-hierarchical Predicates] are typically more context specific and could include predicates like directedBy, writer, actedIn, etc. 
	For these predicates, it is usually not needed to scrutinize each entity separately.
	Rather, once it is established that the predicate is relevant for the domain, then all nodes connected by it are relevant as well.
\end{description}

\begin{figure}[t!]
\includegraphics[width=1.0\textwidth]{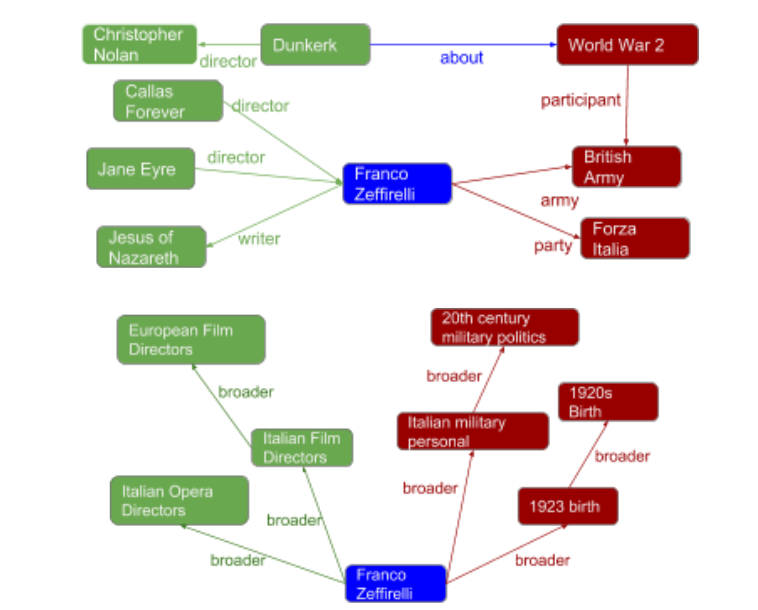}
\caption{Non-hierarchical predicates (top) vs hierarchical predicates (bottom)}
\label{fig:hobbits_fig2}
\end{figure}

In the cinematography use case, an example of predicates related to this domain is shown in Figure~\ref{fig:hobbits_fig2} (example by Lalithsena et.al \cite{lalithsena2017domain}).

\section{Proposed Concept}

We are interested in enriching an existing KG (which we assume to be valid) with information represented in DBpedia.
With reference to our definition of validity (the \emph{Topic}), we want to find  properties within the DBpedia KG that are relevant for our specific domain.
For example if our existing KG represents Scientists, we are probably interested in properties such as \texttt{\small dbo:doctoralAdvisor} or \texttt{\small dbo:almaMater}.


\begin{figure}[t]
\includegraphics[width=1.0\textwidth]{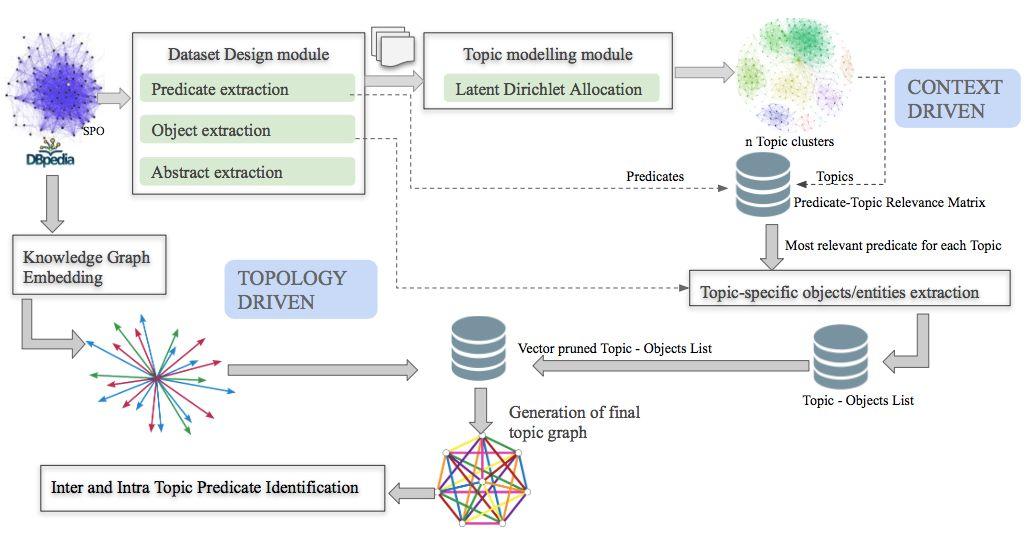}
\caption{Architecture describing the overall pipeline}
\label{fig:hobbits_fig3}
\end{figure}

In this report, we start the investigation of a new approach to find relevant information with reference to a given context, based on topic modeling and graph embedding.
Figure~\ref{fig:hobbits_fig3} depicts the pipeline.
The first step of the pipeline is to find topics represented by the KG. 
To find properties related to the domain, we instantiate a typical topic modeling task as follows:

\begin{itemize}
    \item We select the set $P$ of all properties $p$ in the graph.
    \item For each property $p_i$ we then collect the set $O_i$ of all entities $o$ that appear as object of the property.
    \item Given $p_i$ and $O_i$, we create the document $p_i(O_i)$ containing the concatenation of the abstract (i.e., the textual description) of all entities in $O_i$
    \item We run a topic modeling task over all documents $p_i(O_i)$. The number of clusters is set manually.
    \begin{itemize}
        \item As a result, we obtain a matrix $M$ where the row $i$ corresponds to $p_i$, while each column $j$ represents a topic.
        These topics can be labeled manually by looking at words that are contained in each cluster, i.e., a cluster that contains the words city, lake, neighborhood, and capital could be labeled as Location. 
        \item  A cell value $M_{ij}$ represent the probability that property $i$ belongs to topic $j$
    \end{itemize}

    \item Based on the above matrix, we are now able to fetch relevant information from DBpedia by selecting properties which have a probability higher than a set threshold $t$. 
    \item Note that this pipeline, does not give a clear indication the relevance of the values (i.e., objects) of these properties; even if we are able to fetch the correct information (because we know the right property).
\end{itemize}

After the topic modeling step, the set with properties $Pr_j$ closest related to each topic can be identified.
Then we can find the objects for each of these properties.
This will result in a collection of objects/entities $L_j$ that is strongly oriented towards the chosen topic/cluster $j$.
 
Next, we use graph embeddings to further narrow down the domain oriented list $L_j$, to create a more cohesive network, based on spatial topology of the nodes in the graph (since the contextuality has already been taken care of). 
We can do this by representing nodes as vectors in a space using a graph embedding algorithm that preserves the topological structure of the graph (e.g. DeepWalk). 
Then, we look up nodes of $G_j$ in the embedded graph and compute outliers of the embedded space.
Once the outliers are identified, we can remove isolated objects in $L_j$ and remove them; and then recreate a graph $G_j$ with the remaining objects.

Now, for each subgraph $G_j$, we analyze the properties of each object.
We analyze how often each property has a property path with nodes that are not a part of this subgraph; we do this for all the properties of every object in the graph.
Then we normalize this score in the range $0-1$, which given an indication how often a predicate takes us out of the domain.

This process is repeated for all topics generated during the topic modeling phase.
In the end we can determine whether the behavior of properties has a pattern throughout different topics.
This can help us in determining if there are certain properties that have a tendency to take us out of the domain while some may not.

This approach can help us to be selective with our approach while expanding the semantics for the data in a given scenario.
We can accordingly choose the properties to expand the semantics depending on whether it requires more cross-domain knowledge or retain the scope of semantics to be within the domain.

\section{Proof of concept and Evaluation Framework}

In this section we describe a methodology to test the proposed approach.
Possible metrics for evaluating different aspects of our work can be grouped as follows:
\begin{description}
    \item[Graph reduction] these metrics give an indication of the capability of our approach to reduce a generic graph to a smaller specific-domain subgraph.
    \item [Impact on accuracy and recall] these metrics demonstrate the performance of our approach in terms of accuracy (i.e., relevance of the retrieved entities, non-relevant retrieved entities, missed entities, etc.)
    \item [Impact on run-time] we have to measure how much time we can save using the proposed approach, instead of running ad-hoc queries in order to retrieved entities related to manually selected properties.
    \item [Application based evaluation] in the end, the data collected by the approach would be used as part of another application(e.g., a recommender systems). An investigation would measure how our approach is able to ease the enrichment phase in different application domains.
\end{description}

To get an impression of the feasibility of what we propose, we already did some initial experiments.
We performed an n-hop expansion of hierarchical categories in DBpedia.
We traversed the DBpedia categories connected by the \texttt{\small skos:broader} relation starting from the root node of four topics (Databases, Datamining, Machine\_Learning, and Information\_Retrieval). Table \ref{tab:hobbit_1} shows our results using n-hop expansion technique.

\begin{table}[htp!]
\centering
\begin{tabular}{|c|c|c|} \hline
Root category &Number of hops & Number of subcategories extracted\\ \hline
Databases& 8 & 880\\ \hline
Datamining & 8 & 15 \\ \hline
Machine Learning & 8 & 2193\\ \hline
Information Retrieval & 8 & 8557 \\ \hline

\end{tabular}
\caption{Analysis of different topics subgraph sizes with the same number of hops traversed.} \label{tab:hobbit_1}
\end{table}

It is evident that for the same number of hops selected (8 in this case) we obtained varying amounts of subcategories using the n-hop expansion technique.
Our approach is supposed to automatically extract the most relevant subgraph irrelevant of the number of hops traversed.
We also present an initial analysis on the effect of number of hops traversal with respect to number of subcategories extracted for a particular domain Film in Table~\ref{tab:hobbit_2} below.

\begin{table}[htp!]
\centering
\begin{tabular}{|c|c|c|} \hline
Number of hops & Number of subcategories extracted\\ \hline
 20 & 1048799 \\ \hline
 10 & 220311 \\ \hline
 5 & 25425 \\ \hline

\end{tabular}
\caption{Analysis of the number of hops expansion for a particular domain.} \label{tab:hobbit_2}
\end{table}

Table~\ref{tab:hobbit_2} shows that given a particular domain of interest, n-hop expansion subgraph extraction can provide a diverse size of subcategories.
The selection of the most relevant subgraph here depends on manual selection with the performance of the graph for intended applications.
However, we propose to evaluate our automatic topic driven approach with respect the most relevant n-hop expansion subgraph.

Precision, recall, execution time, and comparison with topic modelling approaches and knowledge graph embedding approaches.
As they are generally available and widely used in research, we suggest to evaluate our approach using DBpedia and Wikidata.
To conduct this evaluation, on would:
\begin{itemize}
    \item Select multiple topics
    \item Manually extract specific domain subraphs from DBpedia and Wikidata (which are then used as a gold standard)
    \item For each topic, generate the specific topic subgraph from DBpedia and Wikidata using our approach and state of the art in knowledge graph embedding and topic modelling.
    \item Measure the execution time.
    \item For each execution and identified subgraphs, compute precision and recall.
    \item Compare this approach with results of others.
\end{itemize}

We predict that our approach would be able to obtain a higher precision and recall, but worse execution time than the state of the art.

\section{Conclusion and Discussion}

In this report, we analyzed the concept of Linked Data validity from a specific perspective, namely the problem of enriching a domain-specific subgraph considering relevancy of a property or an entity to the domain from generic KGs.
Then, we suggested an approach based on topic modeling and knowledge graph embedding.
We also have designed a preliminary experiment in order to evaluate our proposed approach.

Instead of a specific topic, our approach can be applied to any other topic.
This is interesting since there can be many topics in a generic KG. Moreover, the topic is obtained both spatially and contextually.
We have also analyzed the tendency of how often a property takes us out of (has paths to entities outside of) the domain.

As a future work, a complete evlaution of this method would be needed. 
Besides, more sophisticated methods could be applied.

\chapter{LOD Validity, perspective of Common Sense Knowledge}
\label{sec:deloreans}
\chapterauthor{Russa Biswas, Elena Camossi, Shruthi Chari, Viktor Kovtun, Luca Sciullo, Humasak Simanjuntak, Valentina Presutti}

This section targets the following research questions:

\begin{itemize}
    \item What is a definition of LOD validity?
    \item What is a proper model for representing LOD validity?
    \item What measures allow a fair assessment of LOD validity? And what are their associated metrics?
    \item How can one compute such metrics in a distributed environment such as LOD?
    \item Is there any pattern in LOD that allows us to distinguish a generally valid statement (e.g. common sense fact) from a context dependent one? If yes, why?
\end{itemize}

Commonsense knowledge is knowledge shared by all people about the world e.g. the sun rises, a human can walk, etc. while domain- or application-dependent knowledge models objects in order to address specific requirements. In the last case, the same objects may be modelled in very different ways, sometimes incompatible with each other. As far as LOD validity is concerned, we would like to investigate the question whether there is any emerging pattern in LOD that suggests the presence of facts that are “valid” as commonsense knowledge.

\begin{definition}[Common Sense Knowledge]
Linked Data Validity has been defined by Heath “Linked Data might be outdated, imprecise, or simply wrong.”[4] However, it is a complex task to validate LOD from the perspective of common sense knowledge. The validity of Linked Data for common sense consists in assessing whether a Linked Data triple (or set of triples) expresses knowledge that humans need to understand situations, text, dialogues, etc. and that does not derive from special competences and expertise.
\end{definition}

Since its introduction, the Linked Open Data (LOD) cloud has been constantly increasing in the number of datasets, which are part of it. The available LOD now cover extensive sources of general knowledge such as DBpedia, but also more specialised sources such as lexical and linguistic corpora used in Natural Language Processing Framester, or result of scientific projects such as ConceptNet, a crowdsourced resource to investigate common sense reasoning.  These globally available huge knowledge bases are often federated and are the backbone of many applications in the field of data mining, information retrieval, natural language processing as well as many intelligent systems. 
In this era of pervasive Artificial Intelligence, the research on robotic assistants and smart houses has been at the focal point to helping users in daily life. In this emerging domain of research, grounding of common sense into user interface systems plays an integral part.
Common sense knowledge is intuitively defined as knowledge shared by all people about the world. It is an inherent form of knowledge that can be extended to wide variety of skills humans possess. Commonsense knowledge broadly comprises of inherent knowledge, knowledge shared by a larger community i.e. some globally accepted fact or knowledge acquired through day to day life experiences or certain conditions etc. However, common sense knowledge possesses different attitude depending on the domain of the discourse, and is societal and application dependent. Said differently, a fact can be interpreted differently depending on its context. It can be broadly classified into different domains such as psychology, physical reasoning, planning, understanding the language etc.. Common sense knowledge is squeezed out from our experience and  can vary from the simplest of actions involved in our daily life such as opening a door to complex actions such as driving cars.  Hence, for designing wide variety of intelligent systems responsible for doing usual household chores to driving autonomous cars, there emerges a huge necessity of commonsense knowledge. 
Over time, the open knowledge base ConceptNet has emerged as one of the prominent backbone of commonsense knowledge for the intelligent systems. ConceptNet triples are generated as an amalgamation of the information contributed by humans in all the different properties of the entity. However, due to human intervention in the construction of triple set, huge amount of triples in ConceptNet are incorrect, incomplete and inconsistent. For instance, the entity Baseball in ConceptNet contains a fact   associating baseball to Barack Obama   

\begin{verbatim}
https://w3id.org/framester/conceptnet5/data/en/baseball
https://w3id.org/framester/conceptnet5/schema/IsA
https://w3id.org/framester/conceptnet5/data/en/barack\_obama
\end{verbatim}

Also, commonsense knowledge infers that run is a form an exercise and not a device as specified in ConceptNet 

\begin{verbatim}
https://w3id.org/framester/conceptnet5/data/en/run
https://w3id.org/framester/conceptnet5/schema/UsedFor
https://w3id.org/framester/conceptnet5/data/en/device_that_work
\end{verbatim}

Hence, there arises a necessity of validating the facts present in the ConceptNet with a perspective of commonsense knowledge.
In this work, we design and validate with a proof of concept the application of machine learning approaches for the automatic annotation of common sense, to distinguish common sense facts in knowledge graphs from general knowledge. To create an annotated knowledge base to be used for training, we design and apply a crowd sourcing experiments to validate, against common sense, facts selected from ConceptNet in the domain of human actions. 
The main contributions for this work are:

\begin{itemize}
    \item a systematic novel and reusable approach of validating facts related to human actions derived from ConceptNet using supervised classification algorithm followed by validation using crowd sourcing methods;
    \item The design of a novel approach to generate reusable vectors for the triples leveraging graph embedding techniques.
\end{itemize}

\section{Related Work}
The aim of this work is to design a systematic approach to investigate patterns in LOD and validate the facts in LOD in the perspective of commonsense knowledge within the context of human’s actions. A few recent studies focus on semantically enriching knowledge graphs with commonsense knowledge. Common sense is elicited either from language features and structures or by inherent notions formalised in foundational ontologies trough semantic alignment. Recent approaches apply machine learning. Very recent works try application of deep learning to infer common sense knowledge from large corpora of texts.
\paragraph{\textbf{\textit{Classification-Based Approaches.}}} Asprino et al. \cite{Asprino2018} focus on the assessment of foundational distinctions over LOD entities, hypothesizing they can be validated against common sense. They aim at distinguishing and formally asserting whether an LOD entity refers to a class or an individual, or whether the entity is a physical object or not, foundational notions that are assumed to match common sense. They design and execute a set of experiments to extract these foundational notions from LOD, comparing two approaches. They first transform the problem into a supervised classification problem, exploiting entities features extracted from the DBpedia knowledge base; namely: the entity abstract, its URI and the incoming and outgoing entity properties. Then, the authors compare this method with an unsupervised alignment-based classification that exploits the alignments between DBpedia entities and WordNet, Wiktionary and OmegaWiki, linked data encoding lexical and linguistic knowledge. The authors run the final experiment to validate the results against common sense using first crowdsourcing and expert-based evaluation. Our contribution is inspired from this prior work and we intend to extend the work and design a classification process for actions related to human beings according to common sense.
\paragraph{\textbf{\textit{Alignment-Based Approaches.}}} Other works exploit foundational ontology-based semantic annotation of lexical resources that can be used to support commonsense reasoning, i.e. to make inferences based on common sense knowledge. Gangemi et al. \cite{Gangemi2003} made a first attempt to align WordNet upper level synsets and the foundational ontology DOLCE, extended by Silva et al. \cite{Silva2016} verbs in order to support also common sense reasoning on events, actions, states and processes.
\paragraph{\textbf{\textit{Deep-Learning-Based Approaches.}}} Other works assume that contextual common sense knowledge is captured by language and try to infer it from a part of the discourse or from text corpora for question answering. Recently, Neural-Based Language Models trained on big text corpora have been applied to improve natural language applications, suggesting that these models may be able to learn common sense information. Larz Kunze et al. \cite{kunze} presented a system, that converts commonsense knowledge from the large Open Mind Indoor Common Sense database from natural language into a Description Logic representation, that allows for automated reasoning and for relating it to other sources of knowledge. Additionally, Trinh and Le et. al \cite{Trinh2018} focus on commonsense reasoning based on deep learning. The authors use an array of large RNN language models that operate at word or character level on LM-1-Billion, CommonCrawl, SQuAD, Gutenberg Books, and a customized corpus for this task and show that diversity of training data plays an important role in test performance. Their method skipped the usage of annotated knowledge bases. However, in this work our aim is focussed on the validation of LOD facts from the common sense perspective. In order to identify the actions, this work can be extended.
\paragraph{\textbf{\textit{Commonsense Knowledge Bases.}}} Another notable work in the commonsense knowledge domain is the crowdsourced machine readable knowledge graph ConceptNet.. OpenCyc represents one of the early works of commonsense knowledge, which includes an ontology and uses a proprietary representation language. As a result, the direct usage of both these commonsense knowledge bases as a backbone for applications related to intelligent systems. As already mentioned, in this work we intend to validate the triples from the ConceptNet according to common sense.

\section{Resources} 
In this section we introduce the resources used in this work.
Framester is a hub between FrameNet, WordNet, VerbNet, BabelNet, DBpedia, Yago, DOLCE-Zero and ConceptNet as well as other resources. Framester does not simply create a strongly connected knowledge graph, but also applies a rigorous formal treatment for Fillmore's frame semantics, enabling full-fledged OWL querying and reasoning on the created joint frame-based knowledge graph. ConceptNet, originated from the crowdsourcing project Open Mind Common Sense, is a freely-available semantic network, designed to help computers understand the meanings of words that people use.  Since the focus for this work is to validate the facts in the LOD from the perspective of common sense, therefore ConceptNet has been used as the primary dataset. However, since both DBpedia and ConceptNet is contained within Framester, the link between them has also been leveraged.
In order to identify only the types of actions performed by human beings we considered other two image datasets, namely ‘UCF101: a Dataset of 101 Human Actions Classes From Videos in The Wild’ \cite{ucf} and ‘Stanford 40 Actions’ \cite{standard40} as background knowledge. UCF101 is currently the largest dataset of human actions. It consists of 101 action classes, over 13k clips and 27 hours of video data. The database consists of realistic user uploaded videos containing camera motion and cluttered background. On the other hand, the Stanford 40 Action Dataset contains images of humans performing 40 actions. In each image, we provide a bounding box of the person, who is performing the action, indicated by the filename of the image. There are 9532 images in total with 180-300 images per action class. Only the labels of actions from these two datasets are extracted to identify the possible types of actions, which could be performed by human beings.
Therefore, the data collection can be viewed upon as a two step process on a broad level:
\begin{itemize}
    \item Identify the types of actions that could be performed by human beings from UCF101 and Stanford 40 Actions dataset.
    \item Find triples from the ConceptNet, which are related to these type of actions.
\end{itemize}
\subsection{Proposed approach}
As already mentioned, the goal of our work is to propose the model as classification problem for the validation of the triples from the LOD from the perspective of commonsense knowledge. We intend to classify the triples into two classes: commonsense knowledge and not commonsense knowledge. The approach can be defined as a 4-step process:
\begin{itemize}
    \item Select triples from the knowledge base.
    \item Annotate the triples using crowd sourcing approach.
    \item Generate vectors for each triple using graph embedding.
    \item Classify the vectors using supervised classifiers.
\end{itemize}
After collecting the triples from the knowledge bases, we annotated the data collected as a proof of concept for the design using crowd sourcing approach. The features for the classification problem are generated using the RDF2Vec \cite{ristoski2016rdf2vec} graph embedding algorithm. 
\paragraph{\textbf{\textit{RDF2Vec}}}\cite{ristoski2016rdf2vec} is an approach of latent representations of entities of a knowledge graph into a lower dimensional feature space with the property that semantically similar entities appear closer to each other in the feature space. Similar to the word2vec word vectors, these vectors are generated by learning a distributed representation of the entities and their properties in the underlying knowledge graph. The vector length can be restricted to 200 features. In this embedding approach, the RDF graph is first converted to a sequence of entities which can be considered as sentences. The generation of these sequence of entities is done by choosing node subgraph depth d. This depth d, is the number of hops from the starting node. Using the hops, the connection between the ConceptNet and DBpedia can be leveraged since these 2 knowledge graphs do not share a direct common link. However, it is to be noted in this case all the prefixes from both the knowledge graphs are kept intact in their namespace to identify the properties with same labels in the vector space. 
\paragraph{\textbf{\textit{Classification Process}}}The vectors generated from RDF2Vec can be directly used for the classification process. For the validation of the LOD facts, we design a binary classifier, in which we intend to classify triples as commonsense knowledge and not commonsense knowledge. The classifiers used for the purpose are Random Forest and SVM.
\textbf{\textit{Random Forest.}} Random forests or random decision forests are an ensemble learning method for classification, regression and other tasks, that operates by constructing a multitude of decision trees at training time and outputting the class, that is the mode of the classes (classification) or mean prediction (regression) of the individual trees.
\textbf{\textit{Support Vector Machine (SVM).}} An SVM model is a representation of the examples as points in space, mapped so that the examples of the separate categories are divided by a clear gap that is as wide as possible. New examples are then mapped into that same space and predicted to belong to a category, based on which side of the gap they fall.

\section{Experimental Setup}

\paragraph{\textbf{\textit{Data Collection for Proof of Concept.}}} A preliminary set of candidate common sense triples has been selected from ConceptNet, focusing triples related to human actions. Exploring these triples, and in particular the triples properties, we performed a manual alignment towards other knowledge graphs, such as DBpedia to search for connected facts, and selected candidates triples for domain knowledge or general knowledge. As a proof of concept, triples have been manually selected, using the ConceptNet GUI, but the approach can be automated by querying the ConceptNet SPARQL endpoint, exploring connected triples in ConceptNet, and align with other knowledge graphs. 

\paragraph{\textbf{\textit{Automated Data Collection.}}} The triples for the task can be extracted automatically in a couple of possible ways. Since the data comprises of actions involved by human beings, taking into consideration verbs and finding the triples surrounding the verbs could be useful. Also, selection of proper frames from Framester could possibly lead us to generate appropriate triples for the task. Moreover, the word embedding vectors, which are available for ConceptNet, could be used to identify the triples from the knowledge base, by taking into account the vectors which are in close proximity in the vector space.

\begin{figure}[]
\centering
\includegraphics[width=0.8\textwidth]{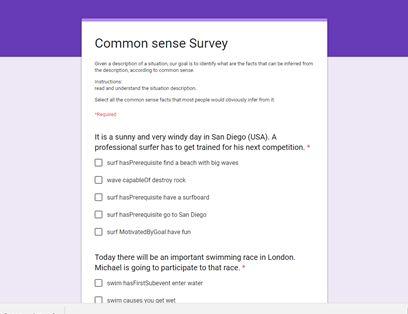}
\caption{Use of crowdsourcing for triple annotation}
\label{fig:deloreans_fig1}
\end{figure}

\paragraph{\textbf{\textit{Annotation by crowdsourcing.}}} The dataset created as proof of concept has been annotated, to distinguish common sense from general knowledge, using crowdsourcing. A preliminary set of demonstrative multiple-choice questions has been prepared at this scope, uploaded on Google Forms and proposed to some ISWS 2018 attendants. In Figure ~\ref{fig:deloreans_fig1}, we show the interface and an example of question to distinguish common sense and domain knowledge facts related to surfing. An important aspect of common sense is the context of the situation, which needs to be taken into account to distinguish common sense from general knowledge. Indeed, common sense reasoning is not meant to derive all possible knowledge, that is usually not formalised in knowledge graph, but only the one, that is relevant for the situation. Indeed, a fact could be a common sense for a specific situation, but not in another one. 

According to the agreed crowdsourcing methodological process, after an initial test run with volunteers, the questionnaire has been revised and adapted, taking into consideration the comments received from the participants. In particular, the task description and the preliminary considerations have been improved, including an example of what common sense is.  

This work can be extended with additional questions and extending the list of question choices, by exploring automatically the knowledge graph associated to each selected activity. 

\begin{figure}[]
\includegraphics[width=0.8\textwidth]{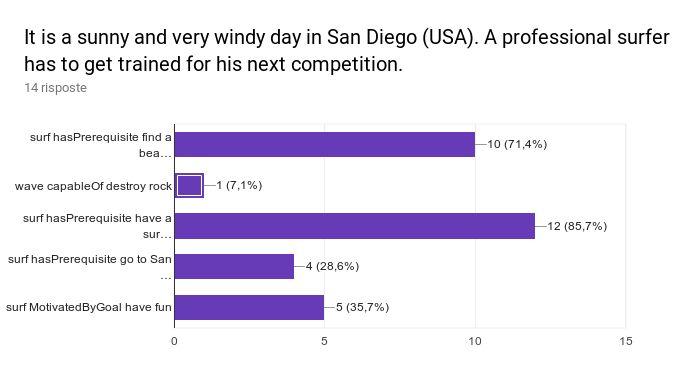}
\caption{Example of question in the survey}
\label{fig:deloreans_fig2}
\end{figure}

Figure ~\ref{fig:deloreans_fig2} shows data collected from one of the 5 questions we asked to 14 people. All the data
are available at online\footnote{\url{https://docs.google.com/forms/d/1E9dpMcTBz27KjBq9ZoKQxrD8RWOi3t4E4tneCmgXLg0/viewanalytics}}

While the results for some triples confirm the association with common sense, like for instance the one claiming that you need a surfboard in order to surf, some of them seem to represent some ambiguity. For instance, it is no clear why surf has as a prerequisite the fact that you have to go to San Francisco. In this case, we expected to obtain very homogeneous results, since everyone knows that San Francisco is a wonderful place for surfing, but not the only one in the world. Hence, ambiguity represents one of the most important results we need to discuss.

\begin{table}[]
\centering
\begin{tabular}{|c|c|c|c|} \hline
Subject & Predicate & Object  &  Validity \\ \hline
swim  & usedFor  & exercise  &  1 \\ \hline
run  &  causes  &  breathlessness  &  1 \\ \hline
disease  &  causes &  virus & 1 \\ \hline
shower & UsedFor & Clean your Tooth & 0 \\ \hline
eat & causes & death & 0 \\ \hline
climb & usedFor & go up & 1 \\ \hline
smoking & hasPrerequisite & cigarette & 1 \\ \hline
\end{tabular}
\caption{Results from the crowdsourcing annotation.} \label{tab:deloreans_1}
\end{table}

\paragraph{Discussion on the Crowdsourced annotated data.}
As previously discuss, results showed a certain degree of ambiguity. We identified three main possible reasons.
First, there could be users with a low reliability. There are several strategies to detect them, for instance by collecting a statistical-meaningful set of results or by using some golden questions. We put some of them, and we will use and extend them in future investigations. Golden questions can be used in the crowdflower platform, that implements some automatic mechanism for computing reliability and trust score for workers.

Second, ambiguity can be strictly related to language itself or just by some misunderstanding of the question, maybe because of the user’s cultural background.

Third, ambiguity can simply represent those concepts in the middle between common sense and what we consider general knowledge. Actually, these results can contain some important information [6] about the users that participated, like, for instance, if their knowledge or common sense is cultural-biased. In order to retrieve this kind of information, we will try to make clusters of data basing on the geographical region of each person that will make the survey.

\section{Discussion and Conclusion}
In this work we have investigated potential approaches for common sense annotation of LOD facts, to distinguish common sense from general knowledge in the context of a discourse. We propose an approach addressing the following research question: 

Is there any pattern in LOD that allows us to distinguish a generally valid statement (e.g. common sense fact) from a context dependent one? If yes, why?

We partially address also the following question: What is a proper model for representing LOD validity?

Specifically, this work is a contribution to the areas of common sense reasoning and semantic web. The automatic tagging of common sense facts could help enlarge existing knowledge graphs with additional facts, which can be inferred on the basis of common sense knowledge.  The approach, described herein, is inspired from the current trends in the literature, and proposes the application of supervised classification, to distinguish commonsense knowledge from domain knowledge. The proof of concept described in this work leverage existing sources of common sense, specifically ConceptNet and Framester, and expanded to other knowledge graphs through alignment to expand the domain of the discourse. Frames, in particular, look promising to identify the sets of facts, potentially related to common sense. A crowdsourcing experiment has been also designed and run as a proof of concept, demonstrating crowdsourcing may be used to produce annotated datasets useful for training a classifier.

The demonstrative proof of concept described in this paper may be evolved in an automatic approach where SPARQL queries are used to construct the knowledge base used for training, and LOD properties are used to expand the knowledge base starting from initial seeds. In our experiments we initially considered a list of human actions and started analysing common sense related to these actions to define the approach. Analogously, other seeds may be identified considering other potential topics of common sense. 

This study has identified potential future lines of investigation. In particular, the dependency of common sense from the context, which has been underlined in this work as the context of the discourse for the crowdsourcing annotation step, could be expanded to consider also the effect of the cultural bias, which affects the perception of what common sense is. Clearly, if common sense is knowledge that is acquired on the basis of experience, the learning environment is an important aspect to be taken into account. On the same line, also time, age and sex of the people, involved in the discourse, may potentially bias the distinction between common sense and general knowledge. In some contexts, there could also be no clear distinction between stereotypes and common sense.

Other potential directions of investigation could explore alternative machine learning techniques, including deep learning. Despite the results obtained using unsupervised machine learning approaches are promising, the choice of the corpora used for learning clearly affect the quality. This shortcoming could be mitigated by the fast expansion of LOD; however, robust statistical approaches such as Bayesian Deep Learning can be investigated. The investigation of the relationships between stereotypes and jokes and common sense could be interesting from a social science perspective and could also help prepare cleaner datasets to be used for training.          

\part{Logic-Based Approaches for Linked Data Validity}
\label{part4}
\chapter{Assessing Linked Data Validity}\label{sec:mordor}
\chapterauthor{Danilo Dess{\'i}, Faiq Miftakhul Falakh, Noura Herradi, Pedro del Pozo Jim{\'e}nez, Lucie-Aim{\'e}e Kaffee, Carlo Stomeo, Claudia d'Amato}

The attention towards Knowledge Graphs (KGs) is increasing in the last few years, for instance by developing applications exploiting 
KGs, such as those grounded on the exploitation of Linked Data (LD). However, an issue may occur when  using KGs and LD in particular: it is not always possible to assess the validity of the data/information therein. 
This is particularly important in the  perspective of reusing (portions of) KGs, since invalid statements maybe involuntarily reused. Hence, assessing the validity of a (portion of) KG results a key issue to be solved. 

In this chapter, we focus on defining the notion of \emph{validity} for a (RDF) statement and on the problem of assessing the validity of a given statement. The validity of a KG will be regarded, by extension, as the problem of assessing the validity of all the statements composing the KG. 

Informally, a statement is valid if it complies with 
a set of constraints, that are possibly formally defined. 
Depending on the type of constraints, it is possible to distinguish between a notion of validity for a statement that is context/domain dependent, that is the validity of a statement depends  
on the context to which it belongs e.g. constraints concerning the common sense of a certain domain; and a notion of validity for a statement that is context/domain independent, that is it applies independently to the particular context/domain the statement belongs to. Constraints belonging to this second category may be expressed as logical rules. 
 

Constraints, and most of all domain independent constraints, may be known in advance, but more often they can be encapsulated within the data/information available within the KGs, e.g. constraints may change over the time because data within the KG are evolving; or in presence of very large KGs there could not be enough knowledge available for pre-defined constraints. As such, being able to 
somehow learn constraints (a model) from the data itself results a key point for assessing the validity of a statement. Once 
the learned model is obtained, new/additional constraints may be defined. Hence, the validity of the KG can be assessed with respect to the whole collection of constraints by checking 
every statement with respect to the given constraints. 

In this work, we focus on learning logical constraints. 
Specifically, 
the following DL like constraints could be learned: 
\begin{itemize}
    \item {\bf domain and range constraints} for a property; 
    \item {\bf functional property restriction} $(\top \sqsubseteq \leq 1r)$; 
    \item {\bf maximum cardinality restriction} $\top (\sqsubseteq \leq nr, r \in N$);
    \item {\bf class constraint} $(\top \sqsubseteq \forall R.C.)$; 
    \item {\bf datatype constraint}.
\end{itemize}
 
As for the representation language for the learned model, we adopt SHACL (Shape Constraint Language)\footnote{https://www.w3.org/TR/shacl/}, the latest W3C standard for validation in RDF knowledge graph since it currently results as a promising language that is receiving a lot of attentions. 

{\bf Use Case.} In order to make more concrete our proposal, we briefly illustrate a use case in the cultural heritage domain by particularly using the ArCo data collection, which 
contains a collection of resources belonging to Italian cultural heritage. More details concerning this dataset are reported 
in Sect.~\ref{sec:appendix}. In the following, we show examples aiming to clarify 
three types of constraints: 
\begin{itemize}
    \item The {\bf maximum cardinality restriction}: 
    it stands for the maximum number of triples for one subject, e.g. Example (1). The property hasAgentRole can have 2 
    instantiation  in its range for the same resource \textsf{moneta RIC 219} at its domain. 
        \begin{example}
            \begin{itemize}
                \item   
                \item moneta RIC 219 w3id:hasAgentRole w3id.org:AgentRole/0600152253-cataloguing-agency
                \item moneta RIC 219 w3id:hasAgentRole w3id:AgentRole/0600152253-heritage-protection-agency
            \end{itemize}
        \end{example}
    \item {\bf Class constraint}: limits 
    the type for a given property. E.g. in Example (2),  hasConservationStatus refers to an object of type \textsf{w3id:ConservationStatus} that also represents 
    the range of the property. 
        \begin{example}
            \begin{itemize}
                \item - 
                \item moneta RIC 219 w3id.org:hasConservationStatus w3id.org:0600152253-stato-conservazione-1
                \item w3id.org:0600152253-stato-conservazione-1 rdfs:type w3id:ConservationStatus
            \end{itemize}
        \end{example}
    \item {\bf Datatype constraints}: used for specifing typed attributed as objects, 
    as in Example (3), where the property \textsf{rdfs:comment} refers to a string attribute value. 
        \begin{example} 
                moneta RIC 219 rdfs:comment ''moneta, RIC 219, AE2, Romana imperiale'' 
        \end{example}
\end{itemize}

\section{Related Work}

The problem LD validity and more in general KGs validity 
has not been largely investigated in the literature. An aspect somehow related and investigated in the literature is instead the assessment of 
 LD quality. In this section, we first briefly explore the main state of the art concerning LD quality, hence the literature concerning constraint representation and extraction is surveyed. 

{\bf Data Quality.} LD quality is a widely explored field in the semantic web research, in which context validity is included at times. Clark and Parsia defined validity in terms of data correctness and integrity. Zaveri et al.\cite{zaveri2016quality} create a famework to explore linked data quality where validity is seen as 
one dimension of LD quality. In this survey, Zaveri et al. classify data quality dimensions under 4 main categories: accessibility, intrinsic properties, contextual and representational dimensions.
In a more application-oriented work~\cite{gayolinked}, validation is proposed 
through the usage of Shape Expression Language. 
Both cited works lack of a clear definition of validity that we aim at providing in this work. 

{\bf Constraint representation.} There are many ways to represent constraints in RDF graphs. Tao et al.~\cite{Tao:2010:ICO:2898607.2898837} 
propose Integrity Constraint (IC), a constraint representation using an OWL ontology and specifically by using OWL syntax and structure. 
Fischer et al.\cite{Fischer15} introduce RDF Data Description to represent constraints in a RDF graph. In their approach, an RDF dataset is called valid (or consistent) if every constraint can be entailed by the graph.

{\bf Constraints Extraction.} Related works concerning learning rules from KGs can be found in the literature. 
One way of doing so is  mostly by exploiting rule mining methods~\cite{d2016evolutionary,d2016ontology}. 
Here rules are automatically discovered from KGs and  represented in SWRL 
whilst, we propose to use 
SHACL for representing constraints that are learned from KGs and that are ultimately used for validating (possibly also new) statements of a KG. 
Another solution 
for mining logical rules from a KG is represented by 
AIME system~\cite{galarraga2013amie} and its upgrade  AIME+\cite{galarraga2015fast} where, by exploiting Inductive Logical Programming solutions, a method for reducing the incompleteness of a KG while taking into account taking into account the Open World Assumption is proposed, differently from our goal aiming at on validating statements of a KG. 

\section{Proposed Approach}\label{sec:approach}

The problem we want to solve is learning/finding 
constraints for a RDF KG by exploiting the evidence coming from the data therein, 
hence apply the discovered constraints on (potentially new) triples in order to validate them. 
In this section, we specifically draft a solution for learning three types of constraints (see end of  Sect.~\ref{sec:LD_Validity} for details on them) reported in the following, hence we briefly present the validation process once constraints are available. 
\begin{itemize}
    \item Cardinality constraints.
    \item Class constraints.
    \item Datatype constraints.
\end{itemize}
{\bf Cardinality constraints.}  
Cardinality constraints can be detected through the usage of existing statistical solutions. 
Specifically, given a set T of triples for a given property $p$, maximum (risp. minimum) cardinality constraints (under some considerations about the domain of interest) could be assessed by statistically inspecting the number of triples available for the considered KG. 
An example of definition of a maximum cardinality expressed by SHACL is reported in the following, where by statistically inspecting the available data, the conclusion that us learned is that each person can have at most one birth date.

\begin{verbatim}
ex:MaxCountExampleShape
	a sh:NodeShape ;
	sh:targetNode ex:Bob ;
	sh:property [
		sh:path ex:birthDate ;
		sh:maxCount 1 ;
	] .    
\end{verbatim}

{\bf Class constraints.} Class constraints require that individuals that participate in a predicate should be instances of certain class types. To find this kind of constraints a straightforward way to go 
could be querying  
the KG in order to find the classes to which individuals participating in the predicate belong to, and then 
assuming that all retrieved classes are valid for the property. However, 
KGs may contain noisy data and, therefore, some classes should not been considered for the class constraint of 
the property. An alternative way for approaching the problem could be exploiting 
ML approaches, and specifically concept learning approaches~\cite{FanizzidE08,Lehmann09,BuhmannLW16,RizzodFE17} for assessing the concept description actually describing the collection of individuals participating in the predicate. Concept learning approaches are indeed more noisy tolerant and as such they would be more suitable for the described scenario. 
In the following, an example of 
class constraint expressed by SHACL is reported.  

\begin{verbatim}
ex:ClassExampleShape
	a sh:NodeShape ;
	sh:targetNode ex:Bob, ex:Alice, ex:Carol ;
	sh:property [
		sh:path ex:address ;
		sh:class ex:PostalAddress ;
	] .
\end{verbatim}

{\bf Datatype constraints.} Datatype constraints require that individuals that participate in a predicate should be instances of certain Literal types (numeric, String, etc). Here 
we assume that for a considered 
property there could be only one datatype. 
We focus on two kinds datatypes: numeric (integer) and string, but other datatypes can be further investigated. The approach that is envisioned is described in the following. Given a property $p$, 
the set of all objects that are related to $p$ are collected. Then, based on the datatype occurrences related to $p$, a majority voting criterion is applied defining 
the most common datatype value.
Alternative approaches could be also considered. Specifically:

\begin{itemize}
    \item the exploitation of methods for performing regression tasks 
    if the datatype is Integer.
    \item the exploitation of embeddings methods for performing similarity-based solutions between values, if the type is string.
\end{itemize}

String embeddings can be computed by using algorithms at the state of the art as Google word2vec~\cite{Mikolov13}, Glove~\cite{CochezRPP17} and so on.
In the following, an example of Datatype constraint in SHACL is reported. 

\begin{verbatim}
ex:DatatypeExampleShape
	a sh:NodeShape ;
	sh:targetNode ex:Alice, ex:Bob, ex:Carol ;
	sh:property [
		sh:path ex:age ;
		sh:datatype xsd:integer ;
	] .
\end{verbatim}

\paragraph{Matching the SHACL constraints to RDF dataset.} 
The SHACL instance graph identifies the nodes in the data graph selected through targets and filters and that will be compared against the defined constraints. The data graph nodes that are identified by the targets and filters are called "focus nodes". Specifically, focus nodes are all nodes in the graph that:
\begin{itemize}
    \item match any of the targets, and
    \item pass all of the filter Shapes. 
\end{itemize}

SHACL can be used for documenting data structures or the input and output of processes, driving user interface generation or navigation because these processes all require testing some nodes in a graph against shapes. The process is called "validation" and the result is called a "validation result". The validation 
fails if validating any test against each "focus node" returns fail, otherwise the 
validation is passed.

\section{Evaluation}

We want validate a KG by assessing the validity of 
triples in the KG with respect to a set of constraints. As illustrated in the previous section, our hypothesis is that (some) constraints may be learned from the data. As such, the aim of this section is to set up an evaluation protocol for 
assessing the effectiveness of the constraints that are learned from the data. 
Formally, We hypothesize (H1) Our approach is able to learn constraints expressed in SHACL to be used for identifying 
valid triples. Given H1, we evaluate our approach on the following research questions:
\begin{itemize}
    \item RQ1 Can we cover a majority of triples in the KG with our constraints? 
    \item RQ2 Are the constraints 
    contradicting?
    \item RQ3 Are the triples identified as valid plausible to a human?
\end{itemize}

\begin{table}[htp!]
\centering
\begin{tabular}{|p{4cm}|p{4cm}|p{4cm}|} \hline
Research Questions & Evaluation  & Results   \\ \hline
RQ1 Can we cover a majority of triples in the KG with our constraints? 
& automatic, number of triples covered by the constraints 
& percentage of triples covered (the higher, the better)   \\ \hline
RQ2 Are the constraints 
contradicting? & look at all extracted constraints, 
evaluate contradictions & no constraints 
should be contradicting  \\ \hline
RQ3 Are the triples identified as valid / plausible to a human? & expert experiment, ask experts to evaluate plausibility & all (or a high percentage) of validated triples should be plausible to humans  \\ \hline

\end{tabular}
\caption{Research questions for evaluation and how they are applied.} 
\label{tab:mordor_1}
\end{table}


RQ1 looks into how many triples can be covered by the constraints, to get an idea of how comprehensive the extracted rules are. 
The metric to be used for the purpose is based on counting the number of triples that are validated by the constraints, as well as the number of triples that are not valid according to the constraints. 
This evaluation can provide 
an insight into how comprehensive the learned constrains 
are, and how much of the KG can be somehow covered. 

The goal of RQ2 is either to ensure that no contradicting constraints are learned or alternatively to assess the impact of contradicting constraints with respect to all constraints that are learned. 
Furthermore understanding the reason for having contradicting constraints would be very important in order to improve the design of the proposed solution so that limiting such an undesired effect. 

Finally, with RQ3, we want to assess 
whether the triples identified as valid by the learned constraints are plausible to a human. 
For the purpose a survey with Semantic Web experts is envisioned. It could be conducted as follows. First of all, all valid triples coming from the discovered constraints are collected. 
Hence a 
sample of the randomly selected valid triples is obtained. The cardinality of the sample should 
depends on the number of triples validated by each type of constraint. 
The selected sample is provided to 
a group of experts 
jointly with the instruction to mark which triples are considered as valid, invalid, and/or that seem plausible, i.e. where the content might be wrong but it could possible in the real world. For example, Barack Obama married\_to Angelina Jolie is not correct, but somehow possible. 

\section{Discussions and Conclusions}

We introduced an approach to discover/learn 
constraints from a Knowledge Graph. Our approach relies on a mixture of statistical 
and machine learning methods and on SHACL as representation language.
We focused on three constraints, namely 
cardinality constraints, class constraints, and datatype constraints. 
We also presented an evaluation protocol for our proposed solution. 
While our approach is limited to the discussed constraints, it can be seen as a good starting point for further investigations of the topic.

\section{Appendix}\label{sec:appendix}
This section is aimed to show a proof of concept for the proposed solution. The adopted data collection for the purpose is 
ArCo, containing 
a plethora of resources belonging to Italian cultural heritage. The dataset has been examined by exploiting SPARQL queries. Some of them have been 
reported in the following. 

\begin{figure}[t]
\centering
\includegraphics[height=0.5\textwidth]{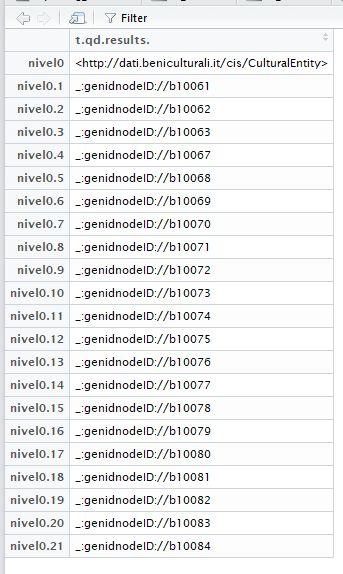}
\caption{Main roots found in ontology.}
\label{fig:mordor_fig1}
\end{figure}

The dataset contains 
around 154 classes. 
We 
focused 
on 
ArCo:CulturalEntity concept acting as the root of our exploration for a total of about 
20 classes inspected. As for the rest of the main reachable concepts, we found 
unknown names, only numeric identifiers, as it is shown 
in Figure ~\ref{fig:mordor_fig1} where results have been collected by 
using Query 1. 

\begin{figure}[]
\includegraphics[height=0.5\textwidth]{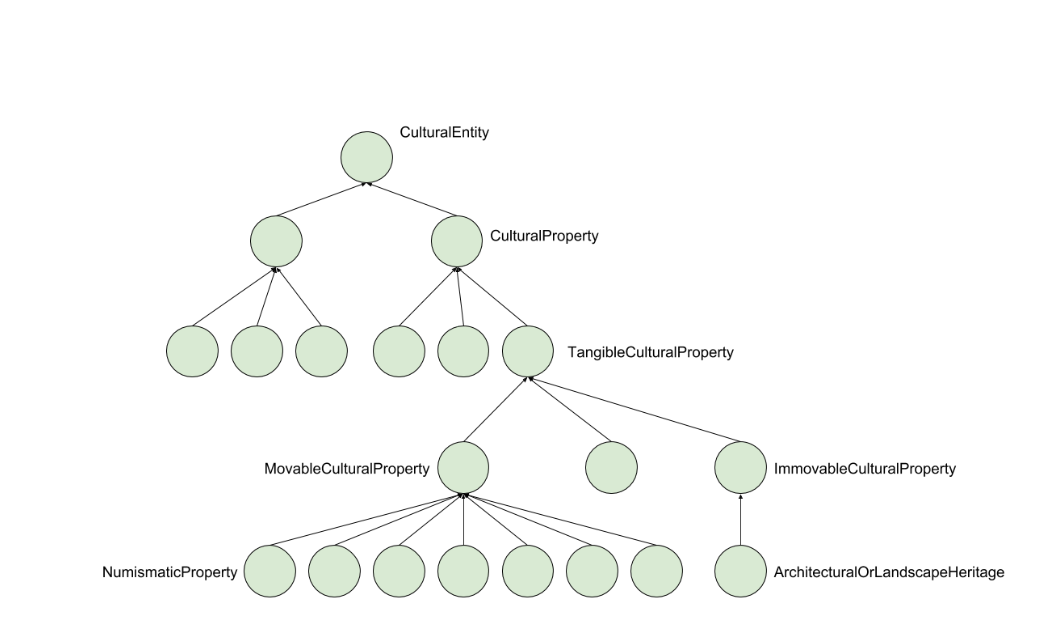}
\caption{Distribution of subclasses of CulturalEntity. NumismaticProperty is subclassOf MovableCulturalProperty, which is subClassOf TangibleCulturalProperty, which is subClassOf CulturalProperty, and this, is subClassOf CulturalEntity.
}
\label{fig:mordor_fig2}
\end{figure}

As for 
ArCo:CulturalEntity, several subclasses can be discovered. 
Queries 2, 3, 4 and 5 show how to obtain this information. 
Figure~\ref{fig:mordor_fig2} depicts 
a diagram of the concept hierarchy and 
relationships among classes, 
for instance, ArCo:NumismaticProperty.

In order to exemplify each type of constraint explained in Sect.~\ref{sec:approach}, 
we focus on a resources belonging to the class NumismaticProperty. The list of properties that are involved with NumismaticProperty is found by using 
Query 6.  
Hence the resource 
\url{<https://w3id.org/arco/resource/NumismaticProperty/0600152253>}, which name is “moneta RIC 219”, is selected. 
Query 7 is used for retrieving the resource related information 
that can be used as a case of possible constraints learnt from the data.

\noindent Cardinality constraints: 
\begin{verbatim}
https://w3id.org/arco/core/hasAgentRole
https://w3id.org/arco/resource/AgentRole/0600152253-cataloguing-agency
https://w3id.org/arco/core/hasAgentRole
https://w3id.org/arco/resource/AgentRole/0600152253-heritage-protection-agency   
\end{verbatim}

The property hasAgentRole can have 2 different possibilities in its range given 
the same resource at its domain.\newline 

\noindent Classes constraints:
\begin{verbatim}
https://w3id.org/arco/objective/hasConservationStatus
https://w3id.org/arco/resource/ConservationStatus/0600152253-stato-conservazione-1
\end{verbatim}

\noindent related data, reachable in Query 8:
\begin{verbatim}
http://www.w3.org/1999/02/22-rdf-syntax-ns#type
https://w3id.org/arco/objective/ConservationStatus
https://w3id.org/arco/objective/hasConservationStatusType
https://w3id.org/arco/objective/intero
    
\end{verbatim}

The property hasConservationStatus points to a resource belonging to ConservationStatus class. Thus, the class expected in the range of this property should be ConservationStatus.

\noindent Datatype constraints:
\begin{verbatim}
http://www.w3.org/2000/01/rdf-schema#comment moneta, RIC 219, AE2, Romana imperiale    
\end{verbatim}

In this case, the property rdfs:comment should has a string at the range part.

\noindent Query 1: Classes that acts as roots
\begin{verbatim}
Select distinct ?nivel0
Where {
?nivel1 rdfs:subClassOf ?nivel0 .
?nivel2 rdfs:subClassOf ?nivel1 .
?nivel3 rdfs:subClassOf ?nivel2 .
?nivel4 rdfs:subClassOf ?nivel3 .
}    
\end{verbatim}

\noindent Query 2: First level subclasses from {\tt ArCo:CulturalEntity}  

\begin{verbatim}
Select distinct (<http://dati.beniculturali.it/cis/CulturalEntity>) as ?level0 ?level1
where {
?level1 rdfs:subClassOf <http://dati.beniculturali.it/cis/CulturalEntity>
} 

level1:
https://w3id.org/arco/core/CulturalProperty
https://w3id.org/arco/core/CulturalPropertyPart
\end{verbatim}

\noindent Query 3: Second level subclasses from {\tt ArCo:CultrualEntity}
\begin{verbatim}
Select distinct (<http://dati.beniculturali.it/cis/CulturalEntity) as ?level0 ?level1 ?level2
Where {
?level1 rdfs:subClassOf <http://dati.beniculturali.it/cis/CulturalEntity> .
?level2 rdfs:subClassOf ?level1 .
} 

level2:
https://w3id.org/arco/core/DemoEthnoAnthropologicalHeritage
https://w3id.org/arco/core/IntangibleCulturalProperty
https://w3id.org/arco/core/TangibleCulturalProperty
https://w3id.org/arco/core/CulturalPropertyComponent
https://w3id.org/arco/core/CulturalPropertyResidual
https://w3id.org/arco/core/SomeCulturalPropertyResiduals
\end{verbatim}

\noindent Query 4: Third level subclasses from {\tt ArCo:CultrualEntity}

\begin{verbatim}
Select distinct (<http://dati.beniculturali.it/cis/CulturalEntity>) as ?level0 ?level1 ?level2 ?level3
where {
?level1 rdfs:subClassOf <http://dati.beniculturali.it/cis/CulturalEntity> .
?level2 rdfs:subClassOf ?level1 .
?level3 rdfs:subClassOf ?level2 .
} 

level3:
https://w3id.org/arco/core/ArchaeologicalProperty
https://w3id.org/arco/core/ImmovableCulturalProperty
https://w3id.org/arco/core/MovableCulturalProperty
\end{verbatim}

\noindent Query 5: Fourth level subclasses from {\tt ArCo:CultrualEntity}

\begin{verbatim}
Select distinct (<http://dati.beniculturali.it/cis/CulturalEntity>) as ?level0 ?level1 ?level2 ?level3 ?level4
Where {
?level1 rdfs:subClassOf <http://dati.beniculturali.it/cis/CulturalEntity> .
?level2 rdfs:subClassOf ?level1 .
?level3 rdfs:subClassOf ?level2 .
?level4 rdfs:subClassOf ?level3 .
}    

level4:
https://w3id.org/arco/core/ArchitecturalOrLandscapeHeritage
https://w3id.org/arco/core/HistoricOrArtisticProperty
https://w3id.org/arco/core/MusicHeritage
https://w3id.org/arco/core/NaturalHeritage
https://w3id.org/arco/core/NumismaticProperty
https://w3id.org/arco/core/PhotographicHeritage
https://w3id.org/arco/core/ScientificOrTechnologicalHeritage
\end{verbatim}

\noindent Query 6: Properties about resources belonging to {\tt NumismaticProperty}

\begin{verbatim}
Select distinct ?p
Where {
?s ?p1 <http://www.w3id.org/arco/core/NumismaticProperty> .
?s ?p ?o .
}
p                   http://www.w3.org/1999/02/22-rdf-syntax-ns#type
p.1                 http://www.w3.org/2000/01/rdf-schema#label
p.2                 http://www.w3.org/2000/01/rdf-schema#comment
p.3                 https://w3id.org/arco/catalogue/isDescribedBy
p.4                 https://w3id.org/arco/core/hasAgentRole
p.5                 https://w3id.org/arco/core/hasCataloguingAgency
p.6                 https://w3id.org/arco/core/hasHeritageProtectionAgency
p.7                 https://w3id.org/arco/core/iccdNumber
p.8                 https://w3id.org/arco/core/regionIdentifier
p.9                 https://w3id.org/arco/core/uniqueIdentifier
p.10                https://w3id.org/arco/location/hasTimeIndexedQualifiedLocation
p.11                https://w3id.org/arco/objective/hasConservationStatus
p.12                https://w3id.org/arco/subjective/hasAuthorshipAttribution
p.13                https://w3id.org/arco/subjective/hasDating
p.14                https://w3id.org/arco/location/hasCulturalPropertyAddress
p.15                https://w3id.org/arco/objective/hasCulturalPropertyType
p.16                https://w3id.org/arco/objective/hasCommission
p.17                https://w3id.org/arco/core/suffix
p.18                https://w3id.org/arco/subjective/iconclassCode    
\end{verbatim}

\noindent Query 7: {\tt NumesmaticProperty} resource data example \url{<https://w3id.org/arco/resource/NumismaticProperty/0600152253>} , “moneta RIC 219”.
\begin{verbatim}
Select distinct ?p ?o
Where {
\url{<https://w3id.org/arco/resource/NumismaticProperty/0600152253>} ?p ?o .
}    
\end{verbatim}

\begin{figure}[]
\includegraphics[height=0.5\textwidth]{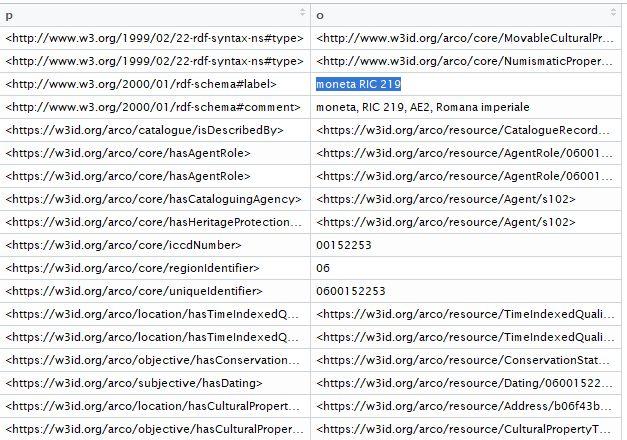}
\caption{Data related to the resource arco https://w3id.org/arco/resource/NumismaticProperty/:0600152253}
\label{fig:mordor_fig3}
\end{figure}

\noindent Query 8: related data to \url{<https://w3id.org/arco/resource/ConservationStatus/0600152253-stato-conservazione-1>}
linked to NumesmaticProperty resource selected by the property hasConservationStatus
\begin{verbatim}
Select distinct (<https://w3id.org/arco/resource/ConservationStatus/0600152253-stato-conservazione-1>)
as ?s ?p ?o
Where {
<https://w3id.org/arco/resource/ConservationStatus/0600152253-stato-conservazione-1> ?p ?o .
}    
\end{verbatim}

\chapter{Logical Validity}
\label{sec:dragons}
\chapterauthor{Andrew Berezovskyi, Quentin Brabant, Ahmed El Amine Djebri, Abderrahmani Ghorfi, Alba Fernández Izquierdo, Samaneh Jozashoori, Maximilian Zocholl, Sebastian Rudolph}



In this work, we consider linked data validity from a logical perspective, where we focus on the absence of inconsistencies. The latter may reshape according to the dimension of the data source, i.e., inside a single data source or between interlinked data sources.

Inconsistency can be assessed on two levels in the Semantic Web stack: Data Layer, and Schema Layer. In the first one, it is the existence of semantically contradictory values, e.g., the Eiffel Tower resource may have two different height values, while the height property should take only one value.
 
On the other hand, the schema layer may show several problems. The previous height constraint might be expressed as a functional property, while the violation of such a constraint is considered as an inconsistency. E.g., Height in different measurement units should be the same. However, if “1063 feets” and “324 meters” are linked to the Eiffel Tower resource with non contradictory, independent height properties, the result of the conversion of 1063 ft = 324,0024m and not 324 m. Sometimes, inconsistency in ontology refers to the unsatisfiability of classes, in other words, the existence of classes with no possible instances (as e.g. in the case where a class that inherits two complementary classes). 

From the LOD perspective, while assuming that each data source is consistent in its own context, it is needed to define a context for the larger one englobing their globality. The previously explained problems are to be extended towards the links between the sources.

In both cases, with or without model, or in a local or a global context, belief revision approaches \cite{lehmann2010ore} should be taken into account. Aiming to resolve the semantic issue by deleting some triples,  rejecting the updates or other aggregation procedures is to be specified.

For the purpose of this work, we are only focusing in inconsistencies in linked data sources having ontologies and employ the unique name assumption.

Certain attempts  have been made to use OWL for validation, most of them relying on the use of reasoners to detect errors as a sign of failed validation. However, the use of such an approach in business distributed linked data applications is problematic, as the reasoning stage would not happen until data is inserted into the triplestore, effectively breaking reasoning for all applications until the invalid data is removed. In addition, requests to such applications can be highly concurrent and made by multiple users, and if reasoning is not performed immediately, the application cannot trace back the request that caused a logical inconsistency in a knowledge base.

\begin{figure}[t!]
\includegraphics[width=1.0\textwidth]{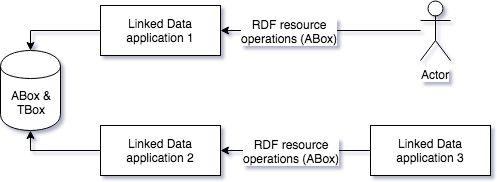}
\caption{Linked Data applications processing read \& write requests from the human users and machine
clients with linked data processing capability}
\label{fig:dragons_fig1}
\end{figure}

To illustrate the use-case with distributed linked data applications, we consider the system shown
in Figure ~\ref{fig:dragons_fig1}.

In the given system, the ontology underpinning the data is assumed to exist and to be consistent (according to the ontology consistency definition by \cite{bayoudhi2018repair}). We also assume the ontology engineers defined clearly their expectations of the ABox data, such as disjointness axioms and functional properties. However, we cannot assume the same of the external users interacting with the resources we manage over the HTTP/REST interface. For simplicity’s sake, we do not rely on any linked data specifications like W3C Linked Data Platform to be used in this case but simply assume the applications follow the 4 rules of Linked Data as laid out by Tim Berners-Lee\footnote{\url{https://www.w3.org/DesignIssues/LinkedData.html}}:
\begin{itemize}
    \item Use URIs as names for things.
    \item Use HTTP URIs so that people can look up those names.
    \item When someone looks up a URI, provide useful information, using the standards (RDF(S), SPARQL).
    \item Include links to other URIs. so that they can discover more things.
\end{itemize}

Most notably, these applications follow the third rule, which allows resources identified by a given URI to be retrieved, updated, and deleted using standard HTTP GET, PUT, and DELETE operations on that URI.

The main aim for this work is to enable software engineers to put reasoning on a frequently used “critical” path of the applications relying on reasoning over a knowledge base. At the same time, we aim at enabling ontology engineers to use the full power of the ontology languages and patterns without fear for logical inconsistencies caused by the erroneous RDF data added to the ABox. To do so, we propose the use of Shapes Constraint Language (SHACL) to ensure that datasets contain only data consistent with the original intentions of the ontology engineer.

\section{Related Work}

Until now, progress on the ontology consistency and using shapes to validate RDF data have been separate from each other.
\begin{itemize}
    \item How to Repair Inconsistency in OWL 2 DL Ontology Versions? \cite{bayoudhi2018repair}. In this paper, the authors have developed an a priori to checking ontology consistency. The used definition of consistency encompasses syntactical correctness, the absence of semantic contradictions and generic style constraints for OWL 2 DL. 
    \item ORE: A Tool for the enrichment, repair and validation of OWL based knowledge bases \cite{bayoudhi2018repair}. ORE uses OWL reasoning to detect inconsistencies in OWL based knowledge bases. It also uses the DL-Learner framework that can be used to detect potential problems  if instance data is available. However, relying only on reasoners is not  suitable for treating large knowledge graphs like LOD, due to scalability and due to the fact that reasoners cannot detect all inconsistencies in data.
    \item Using Description Logics for RDF Constraint Checking and Closed-World Recognition \cite{patel2015using}. There the authors discuss various approaches to validate data by enforcing certain constraints, including SPIN rules in TopQuadrant products, ICV (Integrity Constraint Violation) in Stardog, and OSLC, ShEx, and SHACL shapes. The authors note that shapes are most ``similar to determining whether an individual belongs to a Description Logic description''.
    \item TopBraid Composer \cite{TopQuadrant} allows to convert some OWL restrictions into a set of SHACL constraints. The downside of the presented approach is that the constraints are produced from the assumption that the ontology designers desired to apply a closed-world setting in their ontologies. For example, the ‘rdfs:range’ axiom from the OWL ontology will be naively translated into a ‘class’ or a ‘datatype’ constraint. Such translation will effectively prevent the case where the ontology engineer envisioned a case where a property ‘dad’ pointing to an instance of ‘Male’ would allow to infer that instance also to be a ‘Father’. Further, class disjointness and more complex axioms where values of ‘owl:allValuesFrom’, ‘owl:someValuesFrom’, ‘owl:hasValue’ or ‘owl:onClass’ are intersections of other restrictions are not handled.
\end{itemize}

\section{Resources}

In this work we are using two datasets to exemplify our approach: Bio2RDF and Wikidata.

\paragraph{Bio2RDF.} 
An open-source project that uses semantic web technologies to help the process of biomedical knowledge integration \cite{belleau2008bio2rdf}. It transforms a diverse set of heterogeneously formatted data from public biomedical and pharmaceutical databases such as KEGG, PDB, MGI, HGNC, NCBI, Drugbank, PubMed, dbSNP, and clinicaltrials.gov into a globally distributed network of Linked Data, through a unique URL, in the form of \url{http://bio2edf.org/namespace:id}. BioRDF with 11 billion triples across 35 datasets, provides the largest network of Linked Data for the Life Sciences applying Semanticscience Integrated Ontology which makes it a popular resource to help solve the problem of knowledge integration in bioinformatics.

\paragraph{Wikidata.} A central storage repository maintained by the Wikimedia Foundation. It aims to support other wikis by providing dynamic content with no need to be maintained in each individual wiki project. For example, statistics, dates, locations and other common data can be centralized in Wikidata.
Wikidata is one of the biggest and most cited source in the Web, with 49,243,630 data items that anyone can edit.

\section{Proposed Approach}\label{proposedapproach}

The proposed approach aims to check the “consistency” of Linked Open Data (LOD). Inconsistency happens when contradictory statements can be derived (from the data and ontology) by a reasoner. The use of OWL reasoners to detect such inconsistencies has a very high complexity, therefore, it might not scale to big datasets. Consequently, we propose the use of a validation mechanism, ensuring that data in an RDF base satisfy a given set of constraints, and preventing the reasoners from inferring unwanted relationships. Moreover, it enables the portability and reusability of the generated rules over other data sources. 

The definition of such constraints can come from several sources: (1) Knowledge engineer expertise, (2) ontologies, and (3) data. In this work we are only considering the definition of constraints through ontologies. Moreover, instead of expressing these SHACL constraints manually, we aim to automatically generate them from a set of ontology axioms. With this approach we allow a fast (although not necessarily complete) checking of the data consistency.

This approach shall include a proof (or an indication of a possibility of such proof) that the shape constraints derived from the ontology will be sound. By soundness, we mean that the set of the derived constraints will not prevent data, which would otherwise be valid and could be reasoned over without inconsistencies, from being inserted into the triplestore. Completeness (i.e. that every inconsistency-creating data would be detected by a SHACL shape and thus rejected) may not be achieved because the standard version of SHACL does not include a full OWL reasoner. Therefore, we cannot guarantee that any inconsistency arising after several steps of reasoning will be detected by SHACL shapes. An example of undetected inconsistency will be provided after the definition of the rules that we use to derive SHACL constraints from the ontology.

We below present the rules for deriving some of the description logic formulas into SHACL constraints. 

\begin{itemize}
    \item Rule 1: cardinality restriction. For every property R for which it is stated that R has cardinality at most n, we add the SHACL shape
    \begin{verbatim}
    ex:CardialityRestriction a sh:PropertyShape ;
		sh:path R ;
		sh:maxCount 1 .    
    \end{verbatim}

If the the same property also has an ‘rdfs:domain’ axiom with a class DC, a more specific shape may be added:

\begin{verbatim}
ex:SpecificCardialityRestriction a sh:NodeShape ;
    sh:targetClass DC ;
    sh:property [
      sh:path R ;
      sh:maxCount 1;
    ] .
    
\end{verbatim}

\item Rule 2: datatype restriction (range). For every property R whose range is the class C, and all class D which is explicitly disjoint from C, we add the SHACL shape
\begin{verbatim}
    DatataypeRestriction a sh:PropertyShape ;
    sh:path R ;
    sh:not [
   	 sh:datatype D ;
    ] .
\end{verbatim}

    \item  Rule 3: class disjointness. For any classes C and D that are explicitly stated to be disjoint, we add the SHACL shape
\begin{verbatim}
DisjointnessShape a sh:NodeShape ;
	sh:targetClass C ;
sh:property [
        		sh:path rdf:type ;
			sh:not [
        			sh:hasValue D ;
			]
        	].    
\end{verbatim}

\end{itemize}

Note that SHACL makes subclass inferences, so any instance that explicitly belongs to two classes C’ and D’ that are respective subclasses of C and D, would be rejected during SHACL validation. However, this SHACL shape does not cover all cases of inconsistencies arising from disjointness of classes. An example of such a case (involving self-cannibalism) is presented below.

\begin{figure}[t!]
\includegraphics[width=0.8\textwidth]{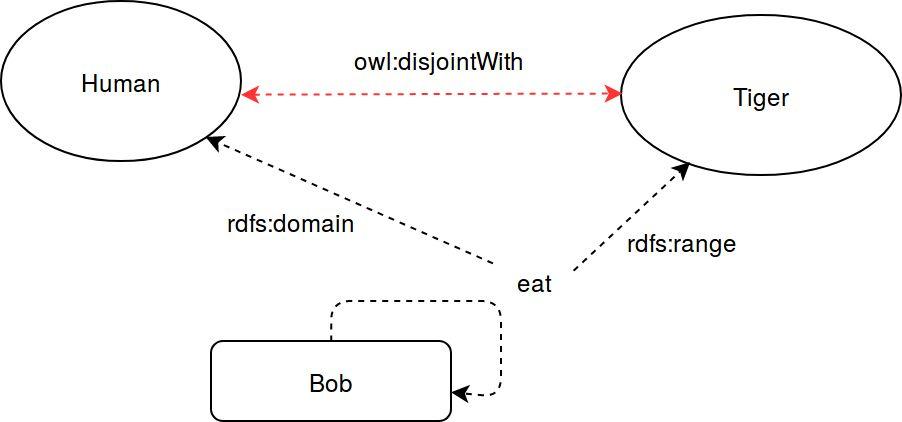}
\caption{Example of inconsistency detected by OWL reasoner}
\label{fig:dragons_fig2}
\end{figure}

As stated before, we cannot ensure that every inconsistency will be detected by the SHACL shapes generated by our rules. Figure \ref{fig:dragons_fig2}  depicts an example of inconsistency that arises after an inference that would be done by a full OWL reasoner but not by SHACL.

Since Bob eats himself, a reasoner would infer that Bob is a Human and a Tiger, while these two classes are supposed to be disjoint. Our SHACL shapes would not make any inference allowing to detect that Bob is a Human and a Tiger, and therefore would not detect the inconsistency. In order to detect such inconsistencies with SHACL shapes, it would be necessary to, either make some inference beforehand, or to derive a stronger set of SHACL rules from the ontology.

Now consider another example that the following triples already exist in bio2rdf:
@prefix bio2rdf:\url{<http://bio2rdf.org>}.

\begin{figure*}[h!]
\includegraphics[width=0.6\textwidth]{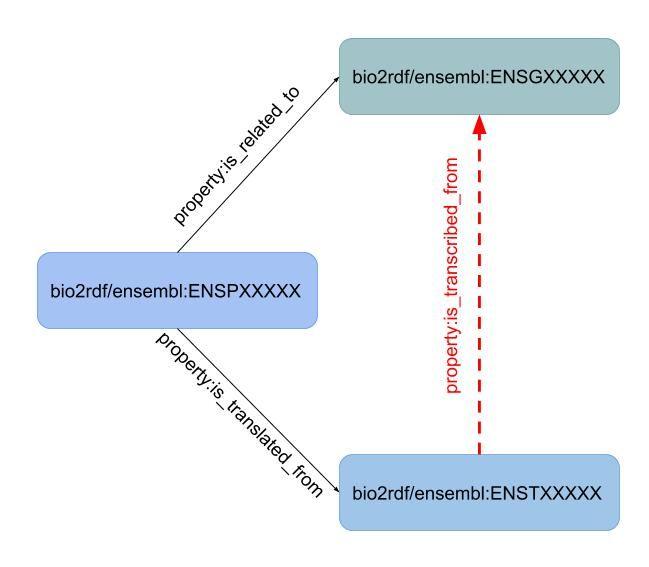}
\label{fig:dragons_fig3}
\end{figure*}
The properties in black are those that are explicitly mentioned in bio2rdf: The “Protein” with Ensembl id “ENSPXXXXX” is related to the “Gene” with Ensembl id “ENSGXXXXX” and is translated from the “transcript” with Ensembl id ENSTXXXXX. According to these two given triples, a domain expert can implicitly infer the property shown in red. 
Now assume the following triple to received and added to current data:

\begin{figure}[h!]
\includegraphics[height=0.5\textwidth]{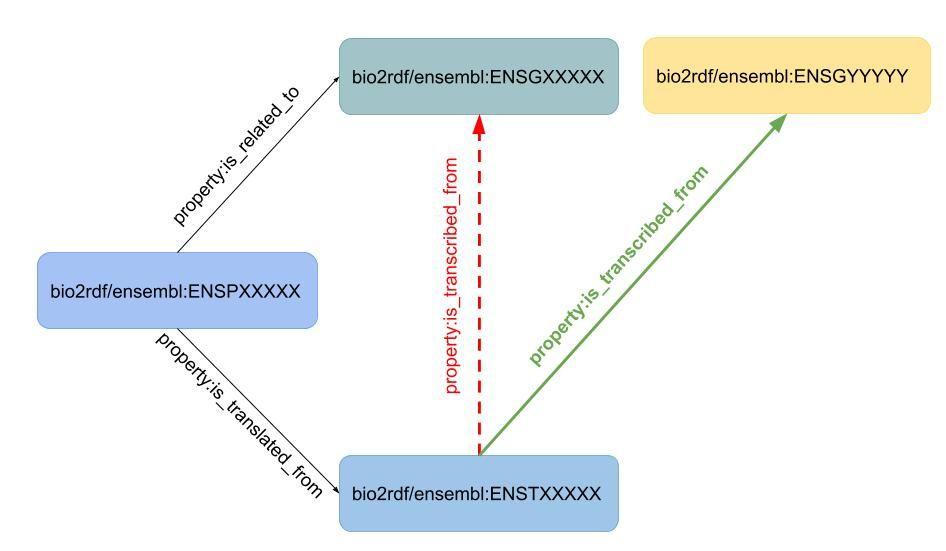}
\label{fig:dragons_fig4}
\end{figure}
The data that this triple expresses is in contrast with what is already derived from previously presented data. Therefore, to prevent “inconsistency”, all inferred data should also be considered in constraints. In this particular example, a SHACL shape may be used to restrict the cardinality of the ‘is\_transcribed\_from’ property to point to at most one Gene and a conjunctive constraint may be used to ensure that the ‘Transcript’ and the ‘Protein’ that was translated from it point to the very same Gene.

\section{Evaluation and Results}


In the following use-case, the Wikidata dataset is used and the following simplifications are made for the sake of readability and evaluation in a browser-based SHACL validator:
\begin{itemize}
    \item Wikidata entity ‘wde:Q515 for the City is referred to as ‘isw:City’
    \item Wikidata entity ‘wde:Q30185’ for the Mayor is referred to as ‘isw:Mayor’
    \item Wikidata entity ‘wde:Q146 for the Cat is referred to as ‘isw:Cat’
    \item The Mayor is declared to be a subclass of foaf:Person and the Cat is declared to be disjoint with a foaf:Person in order to exemplify the arising logical inconsistency when the information about Stubbs is added.
    \item The ‘hasMajor’ property is defined to directly link between City and Mayor class instances and the maximum cardinality restriction of 1 is defined. 
\end{itemize}
\begin{verbatim}
@prefix rdf:   <http://www.w3.org/1999/02/22-rdf-syntax-ns#> .
@prefix rdfs:  <http://www.w3.org/2000/01/rdf-schema#> .
@prefix owl:  <http://xmlns.com/foaf/0.1/> .
@prefix sh: <http://www.w3.org/ns/shacl#> .
@prefix foaf:  <http://xmlns.com/foaf/0.1/> .
@prefix isw: <http://isws.example.com/> .
@prefix wde: <http://www.wikidata.org/entity/> . 

isw:Mayor rdfs:subclassOf foaf:Person .
isw:Cat owl:disjointWith foaf:Person .
isw:hasMayor rdf:type owl:ObjectProperty ,
                      owl:FunctionalProperty ;
             rdfs:range :Mayor .
\end{verbatim}

\noindent
Rule 3 allows us to produce the shape with the following constraint:

\begin{verbatim}
    isw:MayorShape a sh:NodeShape ;
    sh:targetClass isw:Mayor ;
    sh:property [
            sh:path rdf:type ;
            sh:not [
                sh:hasValue isw:Cat
            ]
    ] .
\end{verbatim}

\noindent
Similarly, Rule 1 allows to derive a shape with the cardinality constraint:

\begin{verbatim}
isw:hasMayorShape a sh:PropertyShape ;
	sh:path isw:hasMayor ;
	sh:maxCount 1 .
\end{verbatim}

\noindent
Then, we validate the data prior to its insertion into the triplestore containing the knowledge base:

\begin{verbatim}
isw:jdoe   a           isw:Mayor, foaf:Person;
           foaf:name   "John Doe" .

isw:stubbs a           isw:Mayor, isw:Cat;
           foaf:name   "Stubbs".

isw:city1 a            isw:City;
          isw:hasMayor isw:jdoe, isw:stubbs .    
\end{verbatim}

\noindent
The validation against the set of the derived shapes produces the following validation report:
\begin{verbatim}
[
    a sh:ValidationResult;
    sh:focusNode isw:city1;
    sh:resultMessage "More than 1 values";
    sh:resultPath isw:hasMayor;
    sh:resultSeverity sh:Violation;
    sh:sourceConstraintComponent sh:MaxCountConstraintComponent;
    sh:sourceShape []
].
[
    a sh:ValidationResult;
    sh:focusNode isw:stubbs;
    sh:resultMessage "Value does have shape Blank node \_:n3625";
    sh:resultPath rdf:type;
    sh:resultSeverity sh:Violation;
    sh:sourceConstraintComponent sh:NotConstraintComponent;
    sh:sourceShape [];
    sh:value isw:Cat
].
\end{verbatim}

\noindent
The resulting report demonstrates how a set of shapes produced at the design time solely from the TBox part of the ontology allows to prevent the insertion of the resources that would cause a logical inconsistency for the reasoning process in the triplestore. As shown in the report there are two violations of the shapes. The first violation is triggered by the unmet cardinality restriction. The second violation derives from the disjointness axiom.

\section{Conclusion and Discussion}

Areas of Linked Open Data application are expanding beyond data dumps with the TBox immediately accompanied by the corresponding ABox. Linked Open Data technologies are used to power applications that allow concurrent modification of the ABox data in real-time and require scalability to handle Big Data. In this paper, we present an approach to ensure the logical consistency of the ontology at the runtime by checking the changes to the ABox against a set of statically generated SHACL shapes. These shapes were derived from the ontology TBox using a set of formal rules.

An ideal SHACL validation system would be sound and complete as described in Section~\ref{proposedapproach}. However, since OWL and SHACL rely on the open and closed world assumption respectively, soundness and completeness are difficult to achieve simultaneously (in other words SHACL shapes derived from the ontology tend to be too strong or too weak). We chose to ensure soundness, and left completeness for future work. A continuation of the work started in this document should contain proofs that soundness is indeed achieved, and further rules for SHACL shapes automated creation should be added in order to get closer to completeness.

Future work will also be directed to the extension of the approach  to support reasoning without unique name assumption. This extension  will require changes in some of the proposed shapes.


\part{Distributed Approaches for Linked Data Validity}
\label{part5}
\chapter{A Decentralized  Approach to Validating Personal Data Using a Combination of Blockchains and Linked Data}
\label{sec:hufflepuff}
\chapterauthor{ Cristina-Iulia Bucur, Fiorela Ciroku, Tatiana Makhalova, Ettore Rizza, Thiviyan Thanapalasingam, Dalia Varanka, Michael Wolowyk, John Domingue}

The objective of this study is to define a model of personal data validation in the context of decentralized systems. The distributed nature of Linked Data, through DBpedia, is integrated with Blockchain data storage in a conceptual model. This model is illustrated through multiple use cases that serve as proofs of concepts. We have constructed a set of rules for validating Linked Data and propose to implement them in smart contracts to implement a decentralised data validator. A part of the conceptual workflow is implemented through a web interface using Open BlockChain and DBpedia Spotlight. 

The current state of the World Wide Web is exposed to several issues caused by the over-centralisation of data: too few organisations yield too much power through their control of often private data. The Facebook and Cambridge Analytica scandal \cite{Rosenberg}  is a recent example. A related problem is that the ‘data poor’, citizens who suffer from insufficient data and a lack of control over it, are thus denied bank accounts, credit histories, and other facets of their identity that cause them to suffer financial hardship. Over 60 UK citizens of the ‘Windrush Generation’ have been erroneously deported because of a lack of citizenship data they once had that was lost due to government reorganization ~\cite{theweek}. Such reliance on central entities for validation means that consumers are passing control over their privacy and authenticity of personal information.  

This study examines the relation between decentralized data validation expressed through the integration of blockchain back-end data storage and Linked Data (LD). Decentralization is used to mean that no central authority has control over data and operations on this data. Blockchain technologies conform to the concept of decentralization: data is controlled and owned by the players in a neutral space/platform. Blockchains are useful for secure persistence, immutability, tracking and tracing all changes. Their main advantages is assessing the validity of how the data is used and keeping track of its usage. One of their disadvantage is that  the technique involves no indexing. Thus blockchains have issues with search solutions.  We propose to  put an LD layer over blockchain. The LD is needed when storing data and data can be heterogeneous. An LD layer might help with querying, reasoning and to add semantics to the data.

This study addresses the following broad research questions:
\begin{itemize}
    \item How does the concept of validity change in the context of a decentralized web?
    \item What does a decentralized approach to data validation look like?
    \item What benefits would accrue from a decentralized technology that supports validation in the context of LD?
\end{itemize}

The hypothesis of this research work examines whether Blockchains can provide a mechanism that can respond as a decentralised authentication platform to these questions. Blockchain is a distributed, public ledger that grows constantly as records of information exchange (transactions) are sequentially added to it in blocks \cite{pilkington2016blockchain}.

The problem of Linked Data (LD) validity is that even though LD access is decentralized, its publication is centralized. The data production is not transparent. The validity must be trusted based on the authoritative institution publishing the data whose ‘signature’ appears as an International Resource Identifier (IRI). IRIs are dereferencable, and can thus be dependably accessed with permissions or publicly.  The data is typically not encrypted.  Unauthorized access can allow the modification of information. The use of the data is not or is difficult to document and the processing of it is not transparent.  For example, a matchmaking site would produce a record on the Blockchain every time they process a user profile, making the user aware of how their user profile is used in. Furthermore, it provides a safer storage of information as it distributes them over a (large) network of computers, making it more resilient to data loss or corruption.

In computer science, data validation is generally considered as a “process that ensures the delivery of clean and clear data to the programs, applications and services using it" \cite{technopedia}. Beyond this definition that focuses on formal aspects of the data, the concept is also used in information science or data journalism as "the process of cross-checking the original data and obtaining further data from sources in order to enrich the available information." \cite{khosrow2005encyclopedia}. The term validation is also used in the blockchain context to describe the technical process that ensures that a transaction is validated by the network. In the context of this paper, data are said “valid” if one can assign to them a certain degree of trust and quality based on the validation of an authority or peer. It has two important aspects: validating data and how that data is used.

The motivation behind can be attributed to a number of well-known cases of misuse of trust by authorities who are in charge of centralised systems \cite{Lewin,Pepitone,Indiatoday}. The decentralised nature of Linked Data means that it is also prone to the aforementioned vulnerabilities. W3C’s Verifiable Claims Group aims to make the process of “expressing and exchanging credentials that have been verified by a third party easier and more secure on the Web” \cite{sporny2017}. This guideline would enable one to prove their claims, such as age for purchasing alcohol or credit-worthiness, without having to share any private data that will eventually be stored in a centralised platform. 

Our work is to present a solution for validating information on the Linked Data by leveraging the power of Blockchain technology. The selected data, DBpedia, is one of the recommended datasets by the summer school organizers (Dbpedia.com; ISWS 2018). The report presents a working demo that acts as a proof-of-concept and finally, we conclude the report by discussing the work required to maintain the sustainable growth of the network.

\section{Resources}

In our experiments we use the semantic annotation system DBpedia Spotlight. It allows for semantic queries in order to perform a range of NLP tasks. The tools can be assessed through a web application, as well as using a web Application Programming Interface (API).
\begin{itemize}
    \item The DBpedia knowledge base \cite{lehmann2015dbpedia} is the result of both collaborative and automated work that aims to extract of Wikipedia structured information in order to make them freely available on the Web, link them to other knowledge bases and allow them to be queried by computers \cite{auer2007dbpedia}.
    \item DBpedia Spotlight \cite{daiber2013improving} is a service of Named Entity Linking based on DBpedia that “looks for about 3.5M things of unknown or about 320 known types in text and tries to link them to their global unique identifiers in DBpedia.” The system uses context elements extracted from Wikipedia and keyword similarity measures to perform disambiguation. It can be downloaded and installed locally or queried with open APIs in ten languages. There are a variety of Linked Data that can be utilised to further evaluate the viability of our framework.
    \item We use Open Blockchain implemented by the Knowledge Media Institute to interact with a blockchain through four sets of API: User API, Store API, Util API and IPFS API. The first set of commands provides an authentication a user and managing its account. The sets of command Store API and Util API allow for fully interaction with the blockchain, including the requests for smart contracts stored in a blockchain and their hashes, registration of a new instance of the RDF store contract. IPFS API provides an assess to an IPFS storage.
\end{itemize}

\section{Proposed Approach}

\paragraph{Smart-Contract Response Principle}
Smart contract is an immutable self-executable code containing agreements that must be respected. For smart contracts, a set of rules is formulated as an executable code and the compliance with the rules is verified on the nodes.  User have access to smart contracts. Since we deal with two different types of users (trusted and untrusted) we propose to use two different validation models in our framework. The obtained responses (decisions) can be processed in two different ways (w.r.t. the type of users) in order to get consensus-based response.  We have constructed a general model that can be adapted for two different type of users.

The basis for the final decision is the majority vote. Let us consider how the majority vote model can be applied for the blockchain-based validation. On a query we get an infinite sequences of responses $r1, r_2,\dots,r_k$,  (one response for one claim). The claims can either be “accepted” or “rejected”, i.e. $r\in D , D = {0, 1}$, where 0 / 1 corresponds to ‘’reject” / “accept” responses, respectively. To take the final decision, we define the function $f: D \rightarrow D$:

\begin{equation}
    f(r_1,\dots,r_n) = [0.5 + \frac{\sum_i = 1,...,n r_i - 0.5}{n}]
\end{equation}

where $n\in N$ is a number of responses that are required for taking the final decision and $[.]$ is the floor function, i.e., it takes as input a real number and gives as output the greatest integer less than or equal to this number. The function takes n first responses and returns 0 / 1 in case where the final decision is to “accept” or “reject”, respectively.

\paragraph{Weak Validation Model}
When trusted users access smart contracts we use a weak validation model (for example, we consider a model where to get a Schengen Visa it is sufficient to obtain approvement or rejection only from one country). In this model, $n$ is a fixed value since all the responses are obtained from the reliable sources. In the simplest case, where $n=1$, to return the final decision the function takes the first received answer. Since responses are trusted their number is supposed to be small.

\paragraph{Strong Validation Model}

When untrusted users access smart contracts, we require more strict approvement rules. In other words, we require strong validation for untrusted responses (for example,. when ‘citizen’ assess to smart contracts the obtained responses should be verified carefully). We assume that the most of the users are trusted (or at least more than a half). In that case, the first model can be used when  n is large, i.e., to take a final decision a lot of responses are needed to be received. The weakness of the application when the model for untrusted users is the following. As the number of required responses should be large, to get the final decision can take a lot of time. We propose to use the difference-based model, where the final decision is taken when the number of “accept” or “reject” answer exceeds a chosen value, i.e., n is not fixed in advance, the number of responses that is needed to be received depends on the difference in the number of obtained “accept” and “reject” responses: 

$$n =argmin_{m\in N} [\mid  \sum_{i=1,..,m} I (r_i = 0) - I(r_i =1)\mid > Q ] $$

Where $I(.)$ is an indicator function, it takes 0 / 1, when the condition in the brackets is “false” / “true”, respectively.  Value $n$ is the minimal value when the difference between the number of “accept” and “reject” responses exceed the chosen threshold $Q$.

\begin{example}
Let us consider how the majority vote models work in practice.
\begin{enumerate}
    \item Case 1: $n$ is fixed. Let $n = 7$, i.e., to take a final decision $7$ responses are needed to be obtained. Assume the sequence of responses is 0101110.$$f(0,1,0,1,1,1,0)= [0.5 + \frac{4-0.5}{7}]=1 $$
    Thus, the final decision is "accept".
    \item Case 2: $n$ is not fixed and depends on the difference of the obtained responses. Let $Q = 2$. The and responses received are summarized in Table~\ref{tab:hufflepuff_1}.
\end{enumerate}

\begin{table}[htp!]
\centering
\begin{tabular}{|c|c|p{6cm}|} \hline
Sequence no. of responses &Response Value & Comments on the final decision \\ \hline
1 & 0 & Q = 1, the decision cannot be taken  \\ \hline
2 & 1 & Q = 0, the decision cannot be taken  \\ \hline
3 & 0 & Q = 1, the decision cannot be taken  \\ \hline
4 & 0 & Q = 2, the decision can be taken, n = 4,  $f(0,1,0,0)= [0.5 + \frac{1-0.5}{4}]=1 $ \\ \hline
\end{tabular}
\caption{The principle of decision making for non-fixed number of responses. Q is the difference in the number of responses.} \label{tab:hufflepuff_1}
\end{table}

\end{example}
The proposed models have the “limited-response” drawbacks. It means that in cases where only a few responses can be obtained, the response time for the final decision might be great. To avoid the time-lost problem, Q might be non-fixed in advance and a limit on maximal response time T is fixed. In that case, the requirement of the final decision can be relaxed to get the final result within the chosen dataframes.

\paragraph{Proof of Concept}

In our proof of concept we developed an application to test the Open Blockchain \cite{blockchaina}(2018a) infrastructure and API together with a Linked Open Data dataset. A brief demo can be found here: \url{https://hufflepuff-iswc.github.io}. The application has the complete workflow needed to store the data on blockchain and link it with linked open data functionalities. The Screenshot of the application is provided in the Figure 1 and description of each step is listed below:

\begin{itemize}
    \item In order to store the information in the blockchain the user should create an account and register himself with his credentials. In contradiction to the standard authentication methods - the user’s credentials are stored in encrypted decentralized way in the blockchain. After successful login the user gets an authentication token, which is then used for authorization of the next requests.
    \item In the next step the user has to create a new instance of the RDF store to put his data in using his authentication token. After the store is created it is put to a block and transaction number is returned back. Each transaction and block creation are visualised on top of the page.
    \item For the mining of a block some time is required. The user can check the status of the block mining by requesting the block receipt.
    \item Using authentication token and the transaction number the RDF store has to be registered in the blockchain. By registering the store the smart contract is created and the address of this contract is returned back to the user.
    \item At any time the user can check the RDF stores, which are associated with his account.
    \item In the next step the user has to load the file or data which has to be stored in the blockchain. The file/data is then automatically splitted to the validatable statements and semantic information in form of RDF triples is extracted from them.
    \item Finally, the extracted RDF data is stored in a transaction in the blockchain.
\end{itemize}

The user can choose which statements he wants to validate and select the trusted authorities suggestions provided by the system. Using the semantic information and fulfilling it with the contact information from the open sources the system will inform the authorities about the validation request. The validation request with the request status is stored in the user’s profile and can independently of other information be shared with the third parties.

In order validate the stored data the trusted authority signs the verifiable information using his private key. The data and the signature are put together to the blockchain application. 

By the verification of validation the organisation sends a request with the document to the system. The system retrieves the stored sentences, extracted RDF triples together with the signature of the trusted authorities and if the information could be validated successfully, puts a validation badge for each statement.

\begin{figure}[t!]
\centering
\includegraphics[height=1.0\textwidth]{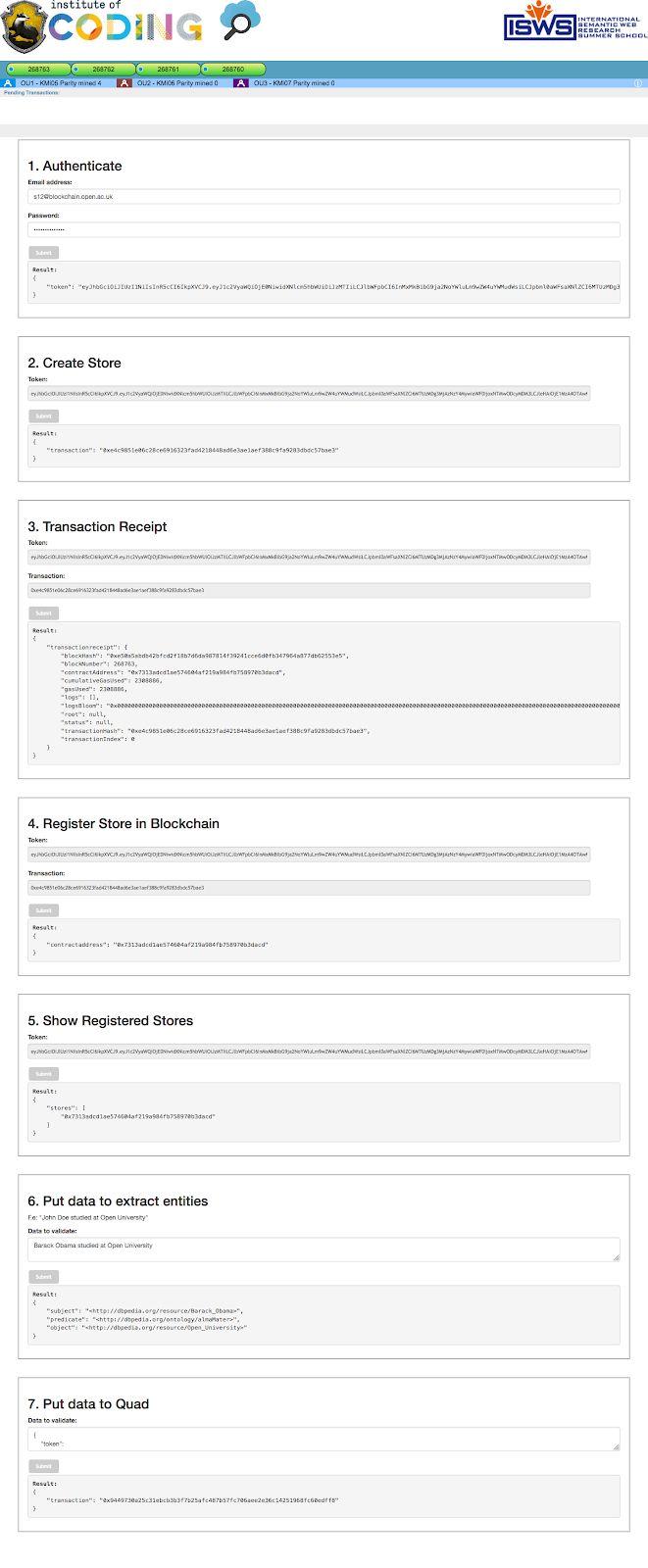}
\caption{Screenshot of the proof of concept application}
\label{fig:hufflepuff_fig1}
\end{figure}

\begin{figure}[]
\includegraphics[height=0.5\textwidth]{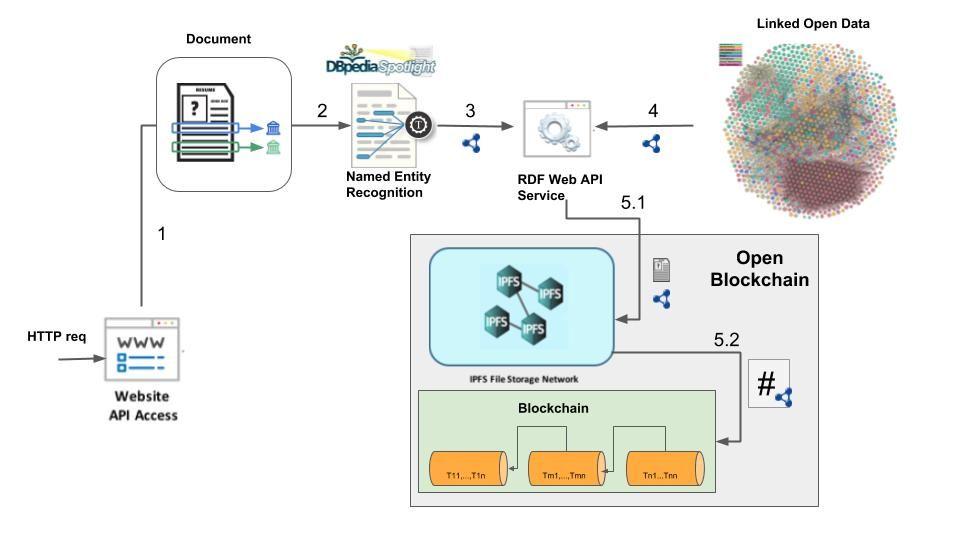}
\caption{Architecture Overview. Adapted from Domingue, J. (2018) Blockchains and Decentralised
Semantic Web Pill, ISWS 2018 Summer School, Bertinoro, Italy.}
\label{fig:hufflepuff_fig2}
\end{figure}

The prototype consists of multiple components which can be seen in the architecture overview in Figure ~\ref{fig:hufflepuff_fig2}. The user makes a HTTP request to the API, where he uploads the document that should be validated by the system (1). The Named Entity Recognition system extracts the semantic entities using natural language processing techniques (2). The entities are represented as RDF triples and combined together with information from the Linked Open Data cloud (4) put to the Open Blockchain network (5.1). In the network, the document and the RDF triples are stored in an InterPlanetary File System (IPFS) distributed file storage network and the retrieved hash is stored in the blockchain transaction(5.2) (\url{http://ipfs.io}). This information is then stored for validation. 

\subsubsection{Use Cases}
The proposed distributed validation approach can be used in multiple different use cases. 

\paragraph{Blockchain Dating}

The first suggested use case is storing dating data on a blockchain. In this case, the semantic triples from all personal dating-relevant data (e.g, interests, age, ex-partners, etc.) are extracted, encrypted, and put in IPFS. The retrieved hashes are stored on blockchain. Permissions are defined to allow and describe how and what parts of this personal data can be used by different services or dating websites. The description of data usage and permissions is written in a separate smart contract on the blockchain that is signed with each individual dating service provider. This ensures that the owner of the data is in full control of which platform uses what parts of data and how that data is used. Validation of the user data can be done by the peers in the blockchain network that have interacted with the user. As there is no trusted authority that can officially validate all the personal information like interests or events attended, the peers will (in)validate the presented information about the user. A trust system can be used to strengthen the validation system.

\paragraph{Distributed Career Validation}

Another possible use case is a distributed career validation system. The system should store and verify education, skill, and career information for individuals. The qualification documents are stored in distributed secure way and due to the qualities of blockchain can not be changed and will never disappear. The system saves the business resources and effort for recruiting and validation of the job applications. The authorities in this case are universities, online schools, and previous employees.   

Splitting a document such as a curriculum vitae (CV) in small easily verifiable pieces of information and fulfilling the missing information using semantic inferences can help authorities, such as universities, former employers, etc., easily prove the validity of the information the candidate has provided. And the new employer can trust that the authority proved the data provided by the candidate and it was validated.

\paragraph{Blockchain Democracy}

Blockchain-based authentication systems provide a more secure mechanism than conventional identity tools since they remove the intermediaries and as they are decentralized, the records are retrievable, even after cases of disaster. In order to achieve a successful transition between a centralized government to a decentralized one, the data in all the official databases needs to be transferred on the blockchain. Whenever new data is to be added in the blockchain, the smart contract regulates the process of validation as a governmental official will confirm or not the truthness of the data. 

In the case of e-Estonia, the citizens are can identify themselves in a secure way and every transaction can be approved and stored on the blockchain. The communication between different departments of the government is shortened in time, which makes the institutions more efficient. In the case that a citizen needs a certificate from the government, they identify themselves in the system and send the request to an institution. The employees of the institution (miners) are competing for the task and the first that completes the task is rewarded inside the blockchain. As soon as the task is done, it is stored in the system and can be accessed by the citizens.

\section{Related Work}

Zyskind and others (2015) defined a protocol that turns a blockchain into an automated access-control manager without need to trust a third party. Their work use blockchain storage to construct a personal data management platform focused on privacy. However, the protocol does not use LD and has not been implemented. To the best of our knowledge, there has been no follow-up to this work.

Previous work on validating Linked Open Data with blockchains includes several researches at the Open University \cite{blockchainb}(Open BlockChain 2018b). Allan Third et al. \cite{third2017linkchains}, for instance, compares four approaches to Linked Data/Blockchain verification with the use of triple fragments. 

Third \& Domingue (2017) have implemented a semantic index to the Ethereum blockchain platform to expose distributed ledger data as LD. Their system indexes both blocks and transactions by using the BLONDiE ontology, and maps smart contracts to the Minimal Service Model ontology. Their proof of concept is presented as “a first steps towards connecting smart contracts with Semantic Web Services”.  This paper as well as the previous one focuses on the technological aspects of blockchain and does not describe case studies related to privacy issues on the Web.

Sharples \& Domingue \cite{sharples2016blockchain} propose a permanent distributed record of intellectual effort and associated reputational reward, based on the blockchain. In this context, Blockchain is used as a reputation management system, both as a “proof of intellectual work” as an “intellectual currency”. This proposal, however, concerns only educational records, while ours aims is to address a wider variety of private data.

\section{Conclusion and Discussion}

In the present work, we propose a novel approach for validating LD using the Blockchain technology. We achieved this by constructing a set of rules that describes two validation models that can be encoded inside smart contracts. The advantages of using Blockchain technology with Linked Data for distributed data validation are: 1) The user maintains full control over their data and how this data is used (i.e. no third party stores any personal information), 2) Sensitive data is stored in a distributed and secure manner that minimises the risk of data loss or data theft, 3) The data is immutable and therefore a complete history of the changes can be retrieved at any time, 4) RDF stores can be used for indexing and for searching for specific triples in Linked Data; 5) Using LD, information can be enriched with semantic inferences; 6) Using smart contracts means that the validation rules on the decentralised system are reinforced forever.

However, the framework presented in the paper has a few limitations: 1) It is vulnerable to all weaknesses that the Blockchain technology suffers from (e.g. smaller networks are vulnerable to 51\% attack); 2) It requires a certain degree of trust in government organisations for maintaining accurate information about the data (i.e. garbage-in-garbage-out), and 3) In our formalisation we proposed to use a time-independent smart contract consensus model (where the parameters of the function that produces the final response are fixed). The model suffers from a “time-loss’’ problem in time-lag cases. This model can be further improved by defining time-dependent parameters that ensure obtaining a response in the defined time-frames. 

Building a decentralized system that uses blockchain technology to support the validation of LD opens up the possibility for secure data storage, control and ownership. It enables a trusted, secure, distributed data validation and share the only explicitly required information with the third parties. In the future work, we plan implement the validation and verification workflow described in our approach and to improve the limitations mentioned above.

\chapter{Using The Force to Solve Linked Data Incompleteness}
\label{sec:jedis}
\chapterauthor{Valentina Anita Carriero, David Chaves Fraga, Arnaud Grall, Lars Heling, Subhi Issa, Thomas Minier, Alberto Moya Loustaunau, Maria-Esther Vidal}

\begin{figure}[t!]
\includegraphics[width=1.0\textwidth]{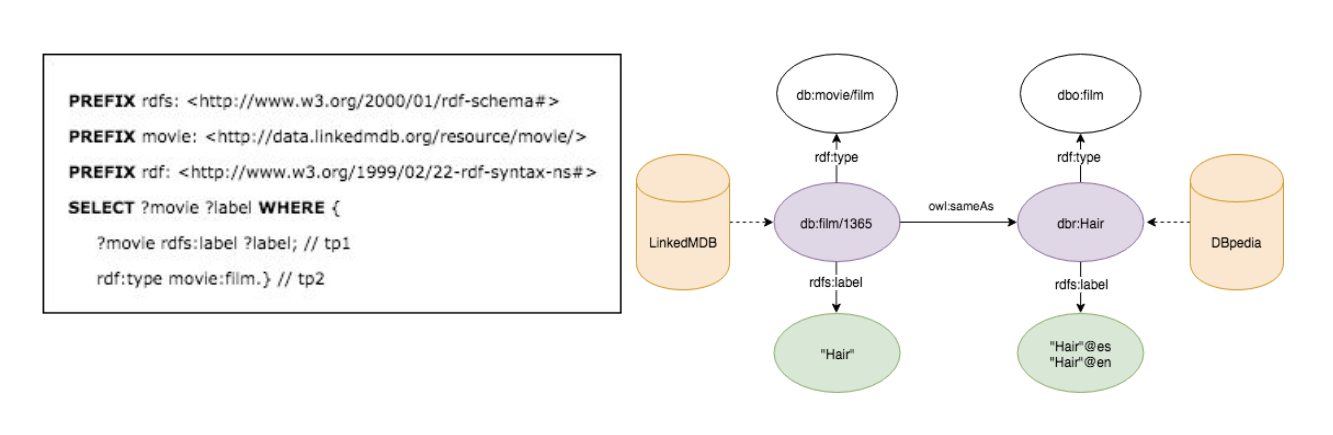}
\caption{\textbf{Motivating Example: incompleteness in SPARQL query results.} On the left, a query to
retrieve movies with their labels. On the right the property graph of the film "Hair" 1 with their respective
values for the LinkedMDB dataset and DBpedia dataset. In green, all labels related to the film "Hair" for
both datasets. LinkedMDB and DBpedia use different class names for movies resulting in incomplete
results when executing a federated query.}
\label{fig:jedis_fig1}
\end{figure}

Following the Linked Data principles, data providers have made available hundreds of RDF datasets \cite{schmachtenberg2014adoption}. The standardized approach to query this Linked Data is SPARQL, the W3C recommendation for querying RDF. Public SPARQL endpoints \cite{auer2007dbpedia,vrandevcic2014wikidata} allow any data consumers to query RDF datasets on the Web, and federated SPARQL query engines \cite{schwarte2011fedx,acosta2011anapsid,gorlitz2011splendid,gorlitz2011federated} allow to query multiple datasets at once. The majority of these datasets have been created by integrating multiple, typically heterogeneous sources and exhibit issues concerning Linked Data validity, including data incompleteness. To illustrate, consider the datasets LinkedMDB and DBpedia and query Q1 (c.f. Figure ~\ref{fig:jedis_fig1}) which retrieve all movies with their respective labels. Evaluating Q1 using a state-of-the-art federated SPARQL query engine over the federation only yields a single label for each movie. However, this result is considered incomplete, as not all relevant labels are provided, i.e., no labels from DBpedia are retrieved. This is due to the fact that these engines are not able to detect incomplete answers and leverage the description of the sources to enhance answer completeness.

In this work, we propose a new adaptive approach for federated SPARQL query processing which estimates the answer completeness and uses enhanced source descriptions to complete the answers by taking as few additional sources into account as possible. More precisely, we address the following research question: 
“Given a SPARQL query and a federation of SPARQL endpoints, how to minimize the number of sources to query during the execution while maximizing answer completeness?”. 
Our contributions are as follows:

\begin{itemize}
    \item We propose a framework, called {\bf extended RDF Molecule Template (eRDF-MTs)}, to describe an RDF dataset in terms of the RDF classes, their properties, and the similarity links between classes and properties across the federation. It also allows for detecting incompleteness.
    \item We propose a relevance-based {\bf cost-model} leveraging eRDF-MT to select sources in order to improve answer completeness without compromising on query execution time.
    \item We propose a new {\bf physical query operator}, the Jedi operator, which dynamically adds new sources during query execution according to the cost-model
\end{itemize}

The paper is organized as follows. Section 2 presents related work. Section 3 presents the problem statement, while Section 4 describes our main contributions. In Section 5 we experimentally study our approach. Finally, in Section 6, we conclude and outline future works. 

In Figure~\ref{fig:jedis_fig1}, on the left, a query to retrieve movies with their labels. On the right the property graph of the film "Hair"\footnote{\url{http://dbpedia.org/page/Hair_\%28film\%29 }} with their respective values for the LinkedMDB dataset and DBpedia dataset.  In green, all labels related to the film "Hair" for both datasets. LinkedMDB and DBpedia use different class names for movies resulting in incomplete results when executing a federated query. 

\section{Related Work}

In the following, we present the work related to our approach. First, we describe how a variety of federated SPARQL query engines select the relevant sources in the federation to minimize the execution time. Next, we present approaches addressing data incompleteness when querying Linked Data.

Federated SPARQL query engines \cite{schwarte2011fedx,acosta2011anapsid,gorlitz2011splendid,gorlitz2011federated} are able to evaluate SPARQL queries over a set of data sources. FedX \cite{schwarte2011fedx} is a federated SPARQL query engine introduced by Schwarte et al. It performs source selection by dynamically sending ASK queries to determine relevant sources and use bind joins to reduce data transfers during query execution. Anapsid \cite{acosta2011anapsid} is an adaptive approach for federated SPARQL query processing. It adapts query execution based on the information provided by the sources, e.g., their capabilities or the ontology used to describe datasets. Anapsid also proposes a set of novel adaptive physical operators for query processing, which are able to quickly produce answers while adapting to network conditions.

Endris et al. \cite{endris2017mulder} improve the performance of federated SPARQL query processing by describing RDF data sources in form of RDF molecule templates. RDF molecule templates (RDF-MTs) describe properties associated with entities of the same class available in a remote RDF dataset. RDF-MTs are computed for a dataset accessible via a specific web service. They can be linked to the same data set or across datasets accessible via other web services. MULDER \cite{endris2017mulder} is a federated SPARQL query engine that leverages these RDF-MTs in order to improve source selection and reduce query execution time while increasing the answer completeness. MULDER decomposes a query into star-shaped subqueries and associates them with the RDF-MTs to produce an efficient query execution plan.

Finally, Fedra \cite{montoya2015federated} and Lilac \cite{montoya2017decomposing} leverages replicated RDF data in the context of a federated process. They describe RDF datasets using fragments, which indicates which RDF triples can be fetched from which data source. Using this information, they compute a replication-aware source selection and decompose SPARQL queries in order to reduce redundant data transfers due to data replication.

However, neither of these approaches are able to detect data incompleteness in a federation. Furthermore, the presented source selection approaches will not be able to overcome semantic heterogeneity to improve answer completeness, as outlined in Section 1.

Acosta et al. \cite{acosta2017enhancing} propose HARE, a hybrid SPARQL engine which is able to enhance the completeness of query answers using crowdsourcing. It uses a model to estimate the completeness of the RDF dataset. HARE can automatically identify parts of queries that yield incomplete results and retrieves missing values via microtask crowdsourcing. A microtask manager proposes questions to provide specific values to complete the missing results. Thus, HARE relies on the crowd to improve answer completeness and is not able to leverage linked RDF datasets.

We conclude that, to the best of our knowledge, no federated SPARQL query engine is able to tackle the issue of data incompleteness in the presented context.

\section{Proposed Approach}

\begin{figure}[t!]
\includegraphics[width=1.0\textwidth]{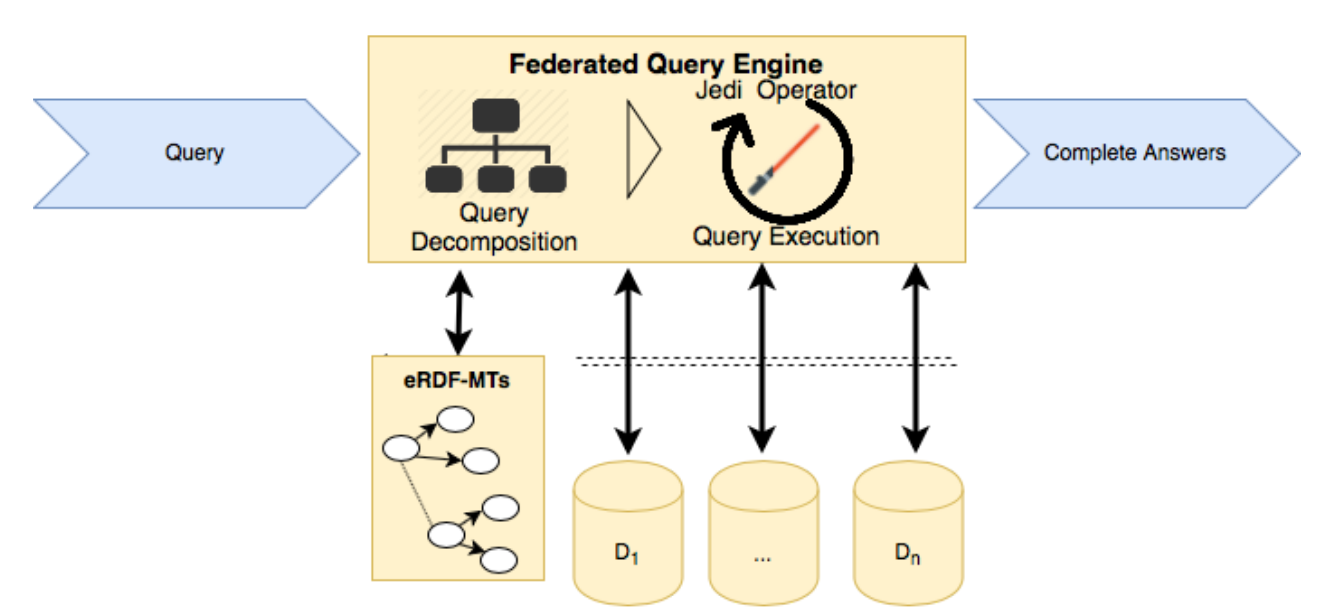}
\caption{\textbf{Overview of the approach.} The figure depicts the query processing model. The engine gets a
query as the input. During query execution, the Jedi operator leverages the eRDF-MTs of the data
sources in the federation to increase answer completeness. Finally, the complete answers are returned.}
\label{fig:jedis_fig2}
\end{figure}

In our work, we rely on the assumptions that the descriptions of RDF datasets are computed and provided by data providers and that Linked RDF datasets are correct but potentially incomplete.  Our approach is based on three keys contributions: (1) an extension of the RDF molecule template to detect data incompleteness, (2) a cost model to determine the relevancy of a source, and (3) a physical query operator which leverages the previous contributions to enhance answer completeness during query execution. An overview of the approach is provided in Figure~\ref{fig:jedis_fig2}. The figure depicts the query processing model. The engine gets a query as the input. During query execution, the Jedi operator leverages the eRDF-MTs of the data sources in the federation to increase answer completeness. Finally, the complete answers are returned. 

\section{Problem Statement}
First, we formalize the problem of data incompleteness and provide the notion of an oracle as a reference point for our definition. 

Given a set of RDF datasets $F = {D_1, ..., D_n}$ and a SPARQL query $Q$ to be evaluated over $F$, i.e., $[[Q]]F$. Consider $O$, the oracle dataset that contains all the data about each entity in the federation. Answer completeness for $Q$, with respect to $O$, is defined as $[[Q]]F = [[Q]]O$.

The problem of evaluating a complete federated SPARQL query over F is:
$min(|[[Q]]O| - |[[Q]]F*|)~s.t.~F* \subseteq F~and~min(|F*|)$.

In other words, the problem is to find the minimal set of sources in $F$ to use during query execution in order to maximize answer completeness.

\subsection{Extended RDF Molecule template}

Next, to tackle the problem of detecting data incompleteness, we rely on the HARE \cite{endris2017mulder} RDF completeness model. We now introduce key notions from this model that we are going to use. HARE is able to estimate that answers to a SPARQL query might be incomplete by leveraging the multiplicity of resources.

\begin{definition}{Predicate Multiplicity of an RDF Resource \cite{acosta2017enhancing}}
Given an RDF resource occurring in the data set $D$, the multiplicity of the predicate $p$ for the resource $s \in D$, denoted $MD(s|p)$, is $MD(s|p) := |{o | (s,p,o) \in D}|$.
\end{definition}

\begin{example}
Consider the RDF dataset from Figure~\ref{fig:jedis_fig1}. The predicate multiplicity of the predicate {\tt rdfs:label} for the resource {\tt dbr:Hair} is $MD(dbr:Hair | rdfs:label) = 2$, because the resource is connected to two labels.
\end{example}

Next, using resource multiplicity, HARE computes the aggregated multiplicity for each RDF class in the dataset.

\begin{definition}{Aggregated Predicate Multiplicity of a Class [2]}
For each class $C$ occurring in the RDF data set $D$, the aggregated multiplicity of $C$ over the predicate $p$, denoted $AMD(C|p)$, is: $AMD(C|p) := f({MD(s|p)|(s,p,o) \in D \wedge (s,a,C) \in D})$ where:
$f(s,a,C)$ corresponds to the triple $(s,rdf:type,C)$, which means that the subject $s$ belongs to the class $C$, and $f(.)$ is an aggregation function.
\end{definition}

\begin{example}
Consider again the RDF dataset from Figure \ref{fig:jedis_fig1}, and an aggregation function $f$ that computes the median. The aggregated predicate multiplicity of the class {\tt dbo:film} over the predicate {\tt rdfs:label} is $AMD(dbo:film | rdfs:label) = 2$.
\end{example}

However, HARE’s completeness model is not designed to be used in a federated scenario, as it can only be computed on a single dataset. To address this issue, we introduce a novel source description, called extended RDF Molecule template (eRDF-MT), based on RDF-MTs \cite{acosta2017enhancing}. An eRDF-MT, defined in Definition 3, describes each dataset of the federation as the set of properties that are associated with each RDF class. It also performs the interlinking of RDF class between datasets, to be able to find equivalent entities across the federation. Finally, eRDF-MTs also capture the equivalence between properties, in order to capture the semantic heterogeneity of datasets.

\begin{definition}{Extended RDF Molecule Template (eRDF-MT)}
An Extended RDF Molecule Template is a 7-tuple = <W, C, f, DTP, IntraC, InterC, InterP> where:
\begin{itemize}
    \item {\bf W} is a Web service API that provides access to an RDF dataset G via SPARQL protocol;
    \item {\bf C} is an RDF class such that the triple pattern (?s rdf:type C) is true in G;
    \item {\bf f} is an aggregation function;
    \item {\bf DTP} is a set of pairs (p, T, f(p)) such that p is a property with domain C and range T, and the triple patterns (?s p ?o), (?o rdf:type T) and (?s rdf:type C) are true in G. f(p) is the aggregated multiplicity of predicate p for class C;
    \item {\bf IntraC} is a set of pairs (p, Cj ) such that p is an object property with domain C and range Cj, and the triple patterns (?s p ?o) and (?o rdf:type Cj) and (?s rdf:type C) are true in G;
    \item {\bf InterC} is a set of 3-tuples (p, Ck, SW) such that p is an object property with domain C and range Ck; SW is a Web service API that provides access to an RDF dataset K, and the triple patterns (?s p ?o) and (?s rdf:type C) are true in G, and the triple pattern (?o rdf:type Ck) is true in K.
    \item {\bf InterP} is a set of 3-tuples (p, p’, SW) such that p is a property with domain C and range T, SW is a Web service API that provides access to an RDF dataset K and p’ is a property with domain C’ and range T’ such as the triples (p owl:sameAs p’) or (p’ owl:sameAs p) exists in G or K.
\end{itemize}
\end{definition}

\begin{figure}[ht!]
\includegraphics[width=0.8\textwidth]{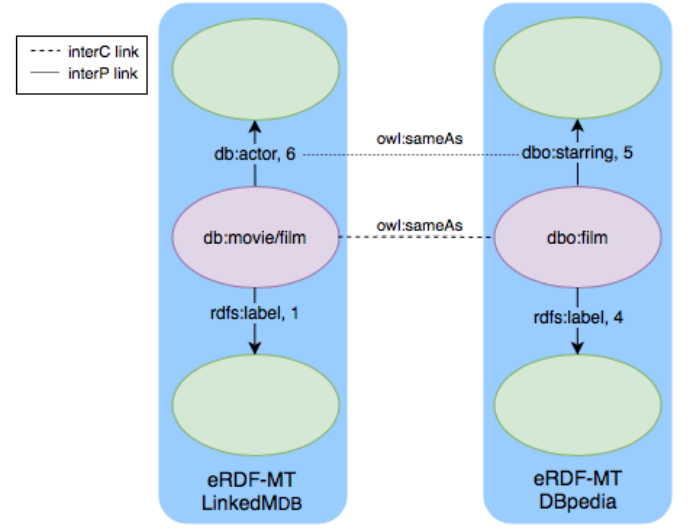}
\caption{An example of two interlinked eRDF-MT for the data sources LinkedMDB (left) and DBpedia
(right). InterC and InterP provide links between the classes and properties in the different data sources.
Additionally, the aggregated multiplicity of each predicate is displayed next to the predicates.}
\label{fig:jedis_fig3}
\end{figure}

The idea is to estimate the expected cardinalities of each property for each class in the data set. Thus, if the query engine finds fewer results for an entity of that class and a property than estimated by the eRDF-MT, it would consider the results to be incomplete. In this case, we assume that connected datasets in the eRDF-MT can be used to complete the missing values. Figure ~\ref{fig:jedis_fig3} provides an example of two eRDF-MTs.

\subsection{The Jedi Cost model}

We introduce a cost-model which relies on the eRDF-MTs to detect RDF datasets that can be used to complete query results, and estimate the relevance of these RDF datasets. This cost-model aims to solve our research problem, by selecting the minimal number of sources to contact. First, we formalize in Definitions 4 and 5 how to compute the relevant eRDF-MT that can be used to enhance the results when evaluating a given triple pattern in the federation.

\begin{definition}
Given a triple pattern $tp = (s, p, o)$, a root eRDF-MT $r = <W,C,f,DTP,IntraC,InterC,InterP>$ and a set of eRDF-MTs $M = {m_1, …, m_n }$, where $m_i$ summarizes the dataset $D_i$.
The set of relevant eRDF-MTs for tp and r are defined as $R(tp, r) =  m_i | \forall m_i \in M$, $m_i = <W’,C’,f’,DTP’,IntraC’,InterC’,InterP’>$ such as there exists $(p'',C,W') \in InterC$ and there exists $(p, W', p') \in InterP$.
\end{definition}

In other word, an eRDF-MT is considered to be relevant with respect to the root eRDF-MT, if it contains the same class (potentially with a different identifier) and the class has the same predicate (potentially also with a different identifier) as the triple tp.

\begin{definition}{Relevance of eRDF-MT}
Given a triple pattern $tp = (s, p, o)$ and a eRDF-MT $m = <W,C,f,DTP,IntraC,InterC,InterP>$, the relevance $\tau$ of m for tp is $\tau(tp, m) = f(p)$ if there exists a $(p, T, f(p))$ in DTP.

\end{definition}

Using these relevant eRDF-MTs, we next devise a strategy to minimize the number of relevant sources to select by ranking sources according to their relevance, formalized in the following definition.

\begin{definition}{Ranking relevant eRDF-MTs}
Given a triple pattern $tp = (s, p, o)$, a root eRDF-MT $r = <W,C,f,DTP,IntraC,InterC,InterP>$ and the set of relevant eRDF-MTs $R(tp, r) = {m_1, …, m_k }$. The ranking of $R(tp, r)$ is $R(tp, r)$ where eRDF-MTs are sorted by descending relevance.
\end{definition}

\paragraph{The Jedi operator for Triple Pattern evaluation}
Federated SPARQL query engine evaluates SPARQL query for building a plan of physical query operators \cite{gorlitz2011federated}. We choose to implement our approach as a physical query operator for triple pattern evaluation, named Jedi operator, in order to ease the integration of this operator in an existing federated SPARQL query engine. Thus, it can be used with state of the art physical operator, like Symmetric Hash Join \cite{gorlitz2011splendid} or Bind Join \cite{schwarte2011fedx,haas1997optimizing}, to handle query execution.

\begin{figure}[t!]
\centering
\includegraphics[width=0.8\textwidth]{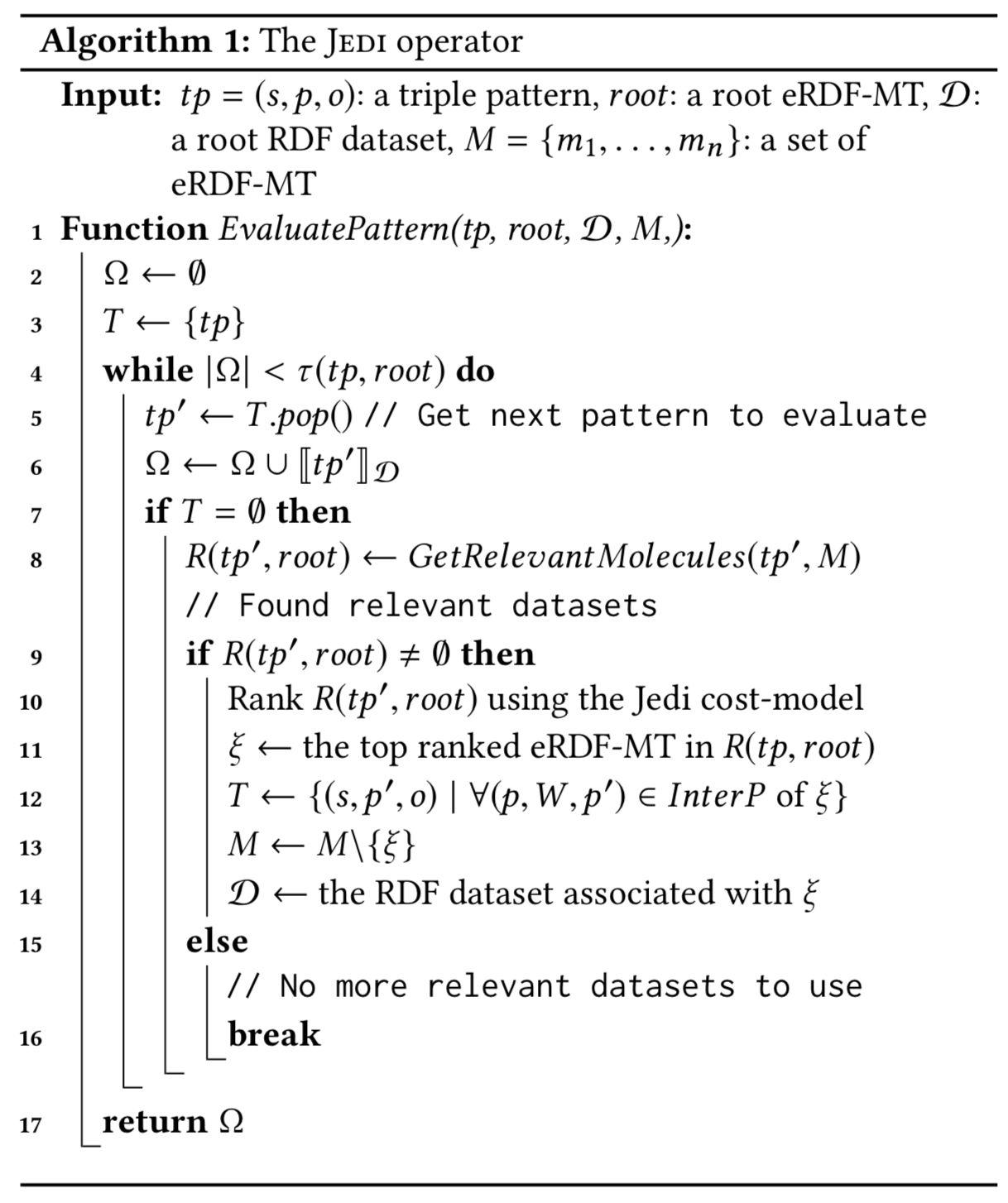}
\caption{The Jedi operator algorithm evaluates a triple pattern using eRDF-MTs}
\label{fig:jedis_fig4}
\end{figure}

The Jedi operator follows interlinking between eRDF-MTs using a breadth-first approach to find additional data during query execution. The algorithm of the operator is shown in Figure ~\ref{fig:jedis_fig4}. The inputs are a triple pattern, a root eRDF-MT (from which the computation will start) and a set of eRDF-MTs for the data sources in the federation. Starting with the root eRDF-MT, the Jedi operator first evaluates the triple pattern at the associated data source (Line 1-6). Then, if the results are incomplete according to the aggregated multiplicity, it uses the Jedi cost-model to find relevant datasets to use (Lines 7-8) and selects the more relevant one to continue query execution (Line 14). Next, it performs a triple pattern mapping (Line 12) using the property interlinks of eRDF-MTs, to maps the triple pattern to the schema used by the newly found dataset. The operator terminates if there the results are considered complete regarding the expected aggregated multiplicity, or if no more relevant eRDF-MTs to use to improve answer completeness.

\section{Evaluation and Results}

In the evaluation, we consider five queries evaluate over two data sets in order to determine the impact of our approach on answer completeness. Each query is associated with a certain domain in order to show that completeness issues are distributed over different parts of the data. The federation contains the data sets DBpedia and Wikidata and we assume Wikidata as a mirror data set of DBpedia. This means that, according to our cost-model, Wikidata is queried only in case the results from DBpedia are estimated to be incomplete. The original queries and the rewritten queries are provided in Appendix A of this work. 

For the sake of brevity, we discuss how the evaluation query q1 in the following. In the query, we want to determine the position, date of birth and the team for soccer players. When evaluating the query over DBpedia, we retrieve no results. However, the results are incomplete when considering Wikidata as well. Rewriting the query according to our proposed approach and executing it over the federation of both data sets, we find that there are 42 results.  As shown in Table 1, similar results can be observed for the other queries as well. 
The results of this first evaluation clearly indicates the potential of our approach to increase answer completeness over a federation of data sets. We expect similar results in other domains and for other data sets as well.

\begin{table}[htp!]
\centering
\begin{tabular}{|c|c|c|c|} \hline
Domain & Query & DBpedia  &  DBpedia + Wikidata \\ \hline
Sport & q1 & 0 & 42 \\ \hline
Movies & q2 & 3 & 6 \\ \hline
Culture & q3 & 0 & 31 \\ \hline
Drugs & q4 & 0 & 482 \\ \hline
Life Sciences & q5 & 0 & 9 \\ \hline

\end{tabular}
\caption{Results of our preliminary evaluation. The table shows the number of answers for 5 queries evaluated over the data set DBpedia and the corresponding rewritten queries evaluated over the federation of DBpedia and Wikidata.} \label{tab:jedis_1}
\end{table}

\section{Conclusion and Discussion}

In this paper, we proposed Jedi, a new adaptive approach for federated SPARQL query processing, which is able to estimate data incompleteness and uses links between classes and properties in different RDF datasets to improve answer completeness. It relies on extended RDF Molecule Templates, which describe the classes, properties as well as the links between data sources. Furthermore, by including the aggregated predicate multiplicity of entities, they allow for detecting incompleteness during query execution. Using these RDF-MTs and a cost-model, the Jedi operator is able to discover new data sources to improve answer completeness.

The results of our evaluation shows that answer incompleteness is presented in various domains of the well-known data sets DBpedia. Furthermore, we show that using our approach to rewritten according to the presented approach will increase the completeness of the results. 

Our approach suffers from one main limitation: it assumes that eRDF-MTs are pre-computed and published by data providers. We also suppose that data providers are aware of the interlinking between their datasets. One perspective is to research how these eRDF-MTs can be computed by data consumers instead, in order to reduce the dependence on data providers. 

In the future, we also aim to integrate the Jedi operator in a state-of-the-art federated SPARQL query engines, like FedX \cite{schwarte2011fedx}, MULDER \cite{acosta2017enhancing} or Anapsid \cite{acosta2011anapsid}, in order to conduct a more elaborate experimental study of our approach. According to this study, we will then improve on our approach to maximize the answer completeness.



\chapter*{Acknowledgement}
We would like to thank everyone who contributed to the organisation of ISWS, the students who are its soul and motivating engine, and the sponsors. Please visit http://www.semanticwebschool.org

\bibliographystyle{splncs04}
\bibliography{bibliographies/deloreans.bib,bibliographies/42s.bib,bibliographies/mordor.bib,bibliographies/dragons.bib,bibliographies/gryffinder.bib,bibliographies/hufflepuff.bib,bibliographies/ravenclaw.bib,bibliographies/hobbits.bib,bibliographies/jedis.bib}

%
%
%

%

\end{document}